\documentclass[a4paper,11pt]{article}
\pdfoutput=1 

\usepackage{adjustbox}
\usepackage{amsfonts}
\usepackage{amssymb, amscd}
\usepackage{graphicx}
\usepackage{tikz}
\usepackage{pgfplots}
\usepackage[mathscr]{euscript}
\usepackage{mathtools}
\DeclareSymbolFont{rsfs}{U}{rsfs}{m}{n}
\DeclareSymbolFontAlphabet{\mathscrsfs}{rsfs}
\usetikzlibrary{positioning,calc}
\pgfplotsset{major grid style={dashed,black,thick}}
\pgfmathdeclarefunction{gauss}{2}{%
	\pgfmathparse{1/(#2*sqrt(2*pi))*exp(-((x-#1)^2)/(2*#2^2))}%
}

\usepackage{mathrsfs}
\usepackage{slashed}

\usepackage{tikz} 
\usetikzlibrary{calc,decorations.markings}
\usetikzlibrary{decorations.pathmorphing}
\tikzset{snake it/.style={decorate, decoration=snake}}
\tikzset{
    set arrow inside/.code={\pgfqkeys{/tikz/arrow inside}{#1}},
    set arrow inside={end/.initial=>, opt/.initial=},
    /pgf/decoration/Mark/.style={
        mark/.expanded=at position #1 with
        {
            \noexpand\arrow[\pgfkeysvalueof{/tikz/arrow inside/opt}]{\pgfkeysvalueof{/tikz/arrow inside/end}}
        }
    },
    arrow inside/.style 2 args={
        set arrow inside={#1},
        postaction={
            decorate,decoration={
                markings,Mark/.list={#2}
            }
        }
    },
}

\DeclareSymbolFont{rsfs}{U}{rsfs}{m}{n}
\DeclareSymbolFontAlphabet{\mathscrsfs}{rsfs}

\usepackage{pdfpages}
\usepackage{braket}
\usepackage{sidecap}
\usepackage{arydshln}
\usepackage{floatrow}
\usepackage[font=small,labelsep=none]{caption}
\usepackage{hyperref}
\usetikzlibrary{arrows,decorations.markings}
\usepackage{subcaption}
\usepackage[left=2cm,right=2cm,top=3.5cm,bottom=3.5cm]{geometry}

\hypersetup{
colorlinks=true,
linkcolor=ceruleanblue,
filecolor=ceruleanblue,      
urlcolor=ceruleanblue,
citecolor=ceruleanblue,
}

\definecolor{cobalt}{rgb}{0.0, 0.28, 0.67}
\definecolor{myblue}{RGB}{174, 198, 219}
\definecolor{myred}{RGB}{157,31,68}
\definecolor{ceruleanblue}{rgb}{0.0, 0.2, 0.6}

\def\MP{M_{\rm Pl}}
\newcommand*{\Cc}{\mathcal}%
\newcommand{\vect}[1]{\boldsymbol{#1}}

\newcommand{\be}{\begin{equation}}      
\newcommand{\ee}{\end{equation}}

\numberwithin{equation}{section}

\newcommand{\rmS}[1]{\rm \scriptscriptstyle #1 }

\date{\today}

\begin{document}
\begin{center}
\LARGE{\bf Beyond Perturbation Theory in Inflation}
\\[1cm] 

\large{Marco Celoria$^{\,\rm a, \rm b}$, Paolo Creminelli$^{\,\rm a, \rm b}$, Giovanni Tambalo$^{\,{\rm c}}$, and Vicharit Yingcharoenrat$^{{\,\rm d }, {\rm e}}$}
\\[0.5cm]

\small{
\textit{$^{\rm a}$
ICTP, International Centre for Theoretical Physics\\ Strada Costiera 11, 34151, Trieste, Italy}}
\vspace{.2cm}

\small{
\textit{$^{\rm b}$
IFPU - Institute for Fundamental Physics of the Universe,\\ Via Beirut 2, 34014, Trieste, Italy }}
\vspace{.2cm}

\small{
\textit{$^{\rm c}$ Max Planck Institute for Gravitational Physics (Albert Einstein Institute)\\ Am M\"uhlenberg 1, D-14476 Potsdam-Golm, Germany}}
\vspace{.2cm}

\small{
\textit{$^{\rm d}$ SISSA, via Bonomea 265, 34136, Trieste, Italy}}
\vspace{.2cm}

\small{
\textit{$^{\rm e}$ INFN, National Institute for Nuclear Physics \\  Via Valerio 2, 34127 Trieste, Italy}}
\vspace{.2cm}

\end{center}

\vspace{0.3cm} 

\begin{abstract}\normalsize
Inflationary perturbations are approximately Gaussian and deviations from Gaussianity are usually calculated using in-in perturbation theory. This method, however, fails for unlikely events on the tail of the probability distribution: in this regime non-Gaussianities are important and perturbation theory breaks down for $|\zeta| \gtrsim |f_{\rmS NL}|^{-1}$. In this paper we show that this regime is amenable to a semiclassical treatment, $\hbar \to 0$. In this limit the wavefunction of the Universe can be calculated in saddle-point, corresponding to a resummation of all the tree-level Witten diagrams. The saddle can be found by solving numerically the classical (Euclidean) non-linear equations of motion, with prescribed boundary conditions.  We apply these ideas to a model with an inflaton self-interaction $\propto \lambda \dot\zeta^4$. Numerical and analytical methods show that the tail of the probability distribution of $\zeta$ goes as $\exp(-\lambda^{-1/4}\zeta^{3/2})$, with a clear non-perturbative dependence on the coupling. Our results are relevant for the calculation of the abundance of primordial black holes.

\end{abstract}

\vspace{0.3cm}

\vspace{2cm}


\newpage
{
\hypersetup{linkcolor=black}
\tableofcontents
}

\flushbottom

\vspace{1cm}


\section{Introduction and main ideas}
Primordial fluctuations generated during inflation are approximately Gaussian \cite{Akrami:2019izv} and deviations from Gaussianity are calculated in perturbation theory \cite{Maldacena:2002vr}. In this paper we point out that there are physically interesting questions whose answer lies beyond perturbation theory and we explain how to get non-perturbative results using semiclassical methods. 

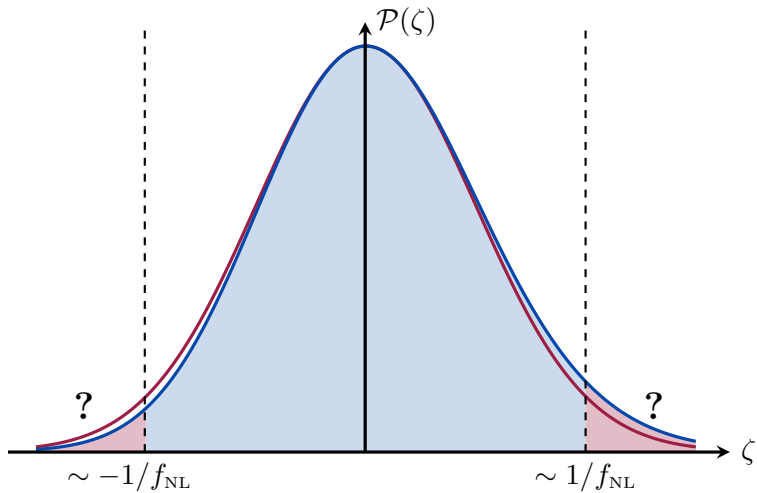
\begin{figure}[t!]
	\begin{center}
	\begin{tikzpicture}[every axis/.style={thick,black}, 
	]
	\begin{axis}[
	no markers,
	every axis plot/.append style={very thick},
	domain=-6:6,
	samples=200,
	axis x line=none,
	axis y line=none,
	height=8cm, width=12cm,
	]
	
	\addplot [fill=cobalt!20, draw=none, domain=-2:2] {1/(sqrt(2*pi))*exp(-(x^2)/(2) + 3*10^(-2)*x^3)} \closedcycle;
	\addplot [fill=myred!30, draw=none, domain=-3:-2] {1/(sqrt(2*pi))*exp(-((x)^2)/(2) + 3*10^(-2)*x^3)} \closedcycle;
	\addplot [fill=myred!30, draw=none, domain=2:3] {1/(sqrt(2*pi))*exp(-((x)^2)/(2) + 3*10^(-2)*x^3)} \closedcycle;
	\addplot [myred, domain=-3:3] {gauss(0,1)};
	\addplot [cobalt, domain=-3:3] {1/(sqrt(2*pi))*exp(-(x^2)/(2) + 3*10^(-2)*x^3)};

	\end{axis}
	
	\node[above] at (1.5,0.8) {\Large \textbf{?}};
	\node[above] at (9,0.8) {\Large \textbf{?}};
	\node[below] at (2.1,0.5) {$\sim -1/f_{\rmS NL}$};
	\node[below] at (8.1,0.5) {$\sim 1/f_{\rmS NL}$};
	\draw[->,>=stealth,very thick] (5.2,0.5)--(5.2,6.2) node[right]{$\mathcal{P}(\zeta)$};
	\draw[->,>=stealth,very thick] (0.5,0.5)--(10,0.5) node[right]  {$\zeta$};
	\draw[-,dashed,thick] (2.3,0.5)--(2.3,6.1) ;
	\draw[-,dashed,thick] (8.1,0.5)--(8.1,6.1) ;
	\end{tikzpicture}
		\caption{~Gaussian distribution (red curve) compared to a non-Gaussian one (blue curve). Close to the center the two distributions are close to each other and the difference can be studied in perturbation theory. On the tails the difference is large and one has to use non-perturbative methods.}  
		\label{fig:distribution}
	\end{center}
\end{figure}

Let us focus for concreteness on a particular question: the calculation of the Primordial Black Hole (PBH) abundance. Roughly, the probability of forming a PBH corresponds to the probability that the primordial curvature perturbation $\zeta(\vect{x})$, smoothed with a typical scale that depends on the mass of the PBH we are interested in, exceeds a certain threshold of order unity, $\zeta \gtrsim 1$ (for a recent discussion see \cite{Musco:2020jjb} and references therein). The formation of a PBH is a very unlikely event on the tail of the probability distribution. (To get a sizeable amount of PBHs one considers models of inflation with a power spectrum $P_\zeta$ on short scales that is much larger than the one measured on CMB scales, but still the formation of a PBH remains a very unlikely event.)  Let us see what happens in the presence of some primordial non-Gaussianity, characterised by a bispectrum $\langle\zeta\zeta\zeta\rangle$, a trispectrum $\langle\zeta\zeta\zeta\zeta \rangle$ and so on. These correlators imply that the probability distribution of $\zeta$ is, very schematically, of the form
\be
{\cal{P}}[\zeta] \sim \exp\left[-\frac{\zeta^2}{2 P_\zeta} + \frac{\langle\zeta\zeta\zeta\rangle}{P_\zeta^3} \zeta^3 + \frac{\langle\zeta\zeta\zeta\zeta\rangle}{P_\zeta^4} \zeta^4 + \ldots\right] \sim  \exp\left[-\frac{\zeta^2}{2 P_\zeta} \left(1 + \frac{\langle\zeta\zeta\zeta\rangle}{P_\zeta^2} \zeta + \frac{\langle\zeta\zeta\zeta\zeta\rangle}{P_\zeta^3} \zeta^2 + \ldots\right)\right] \;.
\ee
The corrections to the Gaussian result are thus
\be \label{eq:expansion_parameter}
\frac{\langle\zeta\zeta\zeta\rangle}{P_\zeta^2} \zeta \sim f_{\rmS NL} \zeta \quad , \quad \frac{\langle\zeta\zeta\zeta\zeta\rangle}{P_\zeta^3} \zeta^2 \sim g_{\rmS NL} \zeta^2 \;.
\ee 
For typical values of $\zeta$, $\zeta \sim P_\zeta^{1/2}$, these are small corrections, compatible with the experimental bounds on non-Gaussianity \cite{Akrami:2019izv} and amenable to a perturbative calculation. However, if we are interested in $\zeta \sim 1$, corrections are large if $|f_{\rmS NL}| \gtrsim 1$ or $|g_{\rmS NL}| \gtrsim 1$ (see Figure~\ref{fig:distribution}). (See for example \cite{Franciolini:2018vbk,Atal:2018neu} and references therein.) For instance in a single-field model of inflation with reduced speed of sound $c_s$, 
$f_{\rmS NL} \sim c_s^{-2}-1$ and $g_{\rmS NL} \sim (c_s^{-2}-1)^2$. Therefore, in these models the calculation of the PBH abundance cannot be done in perturbation theory, unless $c_s$ is close to unity\footnote{For a minimal slow-roll model the non-Gaussian parameters are slow-roll suppressed $f_{\rmS NL} \ll 1$ and $g_{\rmS NL} \ll 1$, so that Gaussainity is a good approximation even for $\zeta \sim 1$. Actually, even if the statistics of the inflaton perturbations can be taken as Gaussian, one needs to take into account the non-linear relation between inflaton perturbations and $\zeta$ and may need to resum the out-of-the-horizon evolution with a stochastic approach a la Starobinsky \cite{Starobinsky:1986fx} (for a recent rigorous derivation see \cite{Gorbenko:2019rza}), see \cite{Pattison_2017,Ezquiaga:2019ftu} and references therein. }. (Non-Gaussianity that cannot be represented by a finite number of $n$-point functions was studied in multifield models of inflation, see for example \cite{Chen:2018uul,Chen:2018brw,Panagopoulos:2019ail,Panagopoulos:2020sxp}.)

The breaking of perturbation theory on the tails of the distribution can be studied in a simple toy model (see Section~\ref{sec:QM}): a quantum mechanical oscillator in the ground state, characterised by a small anharmonicity. In general, one can treat the small anharmonicity in perturbation theory. However, if one is interested in exploring the tail of the ground-state wavefunction, very far from the origin, at a certain point the anharmonic correction to the potential will be large. This quantum-mechanical example suggests a possible approach: the tail of the wavefunction is very suppressed and one expects this regime to be amenable to a semiclassical treatment. Instead of using the WKB approximation (this is done in Appendix~\ref{app:WBK_approximation}), one can obtain the semiclassical wavefunction using the path integral formulation in the limit $\hbar \to 0$. This formulation can be generalized to the case of interest of cosmological inflation. 

In the limit $\hbar \to 0$, inflationary perturbations go to zero. Intuitively this limit should describe rare events, i.e.~events that exceed a given large ``threshold": sending this threshold to infinity with $\hbar$ constant is equivalent to send $\hbar \to 0$. Therefore, {\em the rare-event limit of inflationary perturbations is semiclassical}. 
Let us make this more concrete. The wavefunction of the Universe (WFU) is given by
\be\label{eq:WFU}
\Psi[\zeta(\vect x)]= \int^{\zeta_0 (\vect x)}_{\rm BD} {\mathcal D} \zeta e^{i S[\zeta]/\hbar}\;.
\ee
The functional integral has to be performed with Bunch-Davies boundary conditions at early times and a given configuration $\zeta_0(\vect x)$ at late times. (For simplicity we stick to a single-field model of inflation and neglect tensor modes.) To specify what one means with ``rare event", let us filter $\zeta_0(\vect x)$ with an appropriate window function:
\be
\hat\zeta_0(\vect x) = \int \frac{d^3 k}{(2\pi)^3} \;W(k) \zeta_0(\vect k) e^{i \vect k \cdot \vect x} \;.
\ee
The window function will select a certain range $\Delta k$, so that in real space the field $\hat \zeta_0$ is convoluted with an appropriate filter. A filtered field $\hat\zeta_0$ is relevant to describe the probability of an overdensity (or underdensity) in a certain region of the Universe, or the probability of forming a PBH of a given size. The CMB temperature in each pixel of a map is also a filtered map $\hat\zeta_0$ (but projected in 2 dimensions).

By translational invariance $\hat\zeta_0(\vect x)$ has the same probability ${\cal P}(\hat\zeta_0)$ at any point. The claim is that ${\cal P}(\hat\zeta_0 = \zeta_0)$ can be calculated semiclassically in the limit $|\zeta_0| \to \infty$.  Indeed in this limit we are imposing boundary conditions on the integral of eq.~\eqref{eq:WFU} that make the action large compared to $\hbar$. In this limit the functional integral can be calculated in saddle point approximation
\be\label{eq:WFUsemi}
\Psi[\zeta_0(\vect x)] \sim e^{i S[\zeta_{\rm cl}]/\hbar}\;.
\ee
The action is evaluated on-shell, i.e.~on the classical trajectory $\zeta_{\rm cl}$ that satisfies the boundary condition $\zeta_{\rm cl} = \zeta_0(\vect x)$ at late times and the Bunch-Davies conditions at early times. Notice that we are keeping the full non-linear action and not expanding in perturbation theory: the semiclassical expression \eqref{eq:WFUsemi} resums all non-linearities that are enhanced by the large $\zeta_0$. Corrections to this result come from looking at perturbations around this classical action and evaluating the functional integral over them. These fluctuations are of order $P_\zeta^{1/2}$ and are not enhanced by $\zeta_0$. They give a subleading contribution provided inflation is a weakly coupled EFT.

Before getting to a more realistic scenario of inflation, in Section~\ref{sec:2field} we study a simple toy model to appreciate the difference between the usual in-in perturbation theory and our semiclassical expansion. The model consists of two fields, $\chi$ and $\sigma$, with a cubic interaction $\lambda H \chi \sigma^2$ (with $H$ the Hubble scale during inflation). We will be interested in the regime in which the modes of $\chi$ have a very large amplitude (the unlikely tail of the distribution) so that the expansion in $\lambda$ is not reliable. Moreover, we are going to focus on configurations in which the modes of $\chi$ are much longer than the ones of $\sigma$. In this regime $\chi$ acts as a background for the modes of $\sigma$ and its effect can be easily calculated exactly since it simply corresponds to a change in the $\sigma$ mass. 

The main point of the paper is described in Section~\ref{sec:zeta'^4}, where the methods outlined above are applied to a particular interaction in single-field inflation: $\propto \lambda \dot\zeta^4$. This is a particular limit of inflation, which is consistent and technically natural, as we will discuss. The full evaluation of the wavefunction requires the numerical solution of a PDE: this is done in Section~\ref{subsec_zeta'^4:PDE}, while a test of the numerical code against perturbation theory is the subject of Appendix~\ref{sec:perturb}. One is able to understand the qualitative behaviour in $\lambda$ reducing the PDE to an ODE, which basically corresponds to looking at a single Fourier mode, instead of a realistic real-space profile. This ODE is numerically studied in Section~\ref{subsec_zeta'^4:ODE_approximation}, while an analytic understanding, based on a scaling argument is presented in Section~\ref{subsec_zeta'^4:ODE_analytic_approx}. The conclusion of all these different approaches is that the tail of the distribution goes as $\exp(-\lambda^{-1/4}\zeta^{3/2})$, a result which is clearly non-perturbative in the coupling $\lambda$. The numerical approach can only be performed after an analytic continuation of time to Euclidean time, to avoid integrals with fast oscillations. The possibility of doing this analytic continuation is studied in Section~\ref{sec:proof_anlyticity}. 

This paper is just a first step in the understanding of inflation beyond perturbation theory. Besides the interest in PBHs, there are many conceptual issues in being able to calculate (potentially) observable quantities in our Universe. Many directions remain open and some of them are listed in Section~\ref{sec:conclusions} together with the conclusions. 


\section{\label{sec:QM}Anharmonic oscillator} 

Let us consider an anharmonic oscillator with potential
\begin{align}\label{eq:QM_AHO_potential}
V(x) = \hbar \omega \left[\frac{1}{2}\left(\frac{x}{d}\right)^2 + \lambda\left(\frac{x}{d}\right)^4\right] \;,
\end{align}
and $d \equiv \sqrt{\hbar/m\omega}$. As usual, perturbation theory works provided that the dimensionless parameter $\lambda$ is much smaller than unity and that the particle remains close to the origin ($ x/d \sim1 $). Within the validity of perturbation theory one can perform the standard computations e.g. determine the first order corrections to the ground-state wavefunction and its energy level. The same thing happens in this example when $x/d \gg 1$ (far away from the origin), while $\lambda$ is kept small (and positive). Indeed, as we shall see more in detail, the expansion parameter involves the value of the position i.e. $\lambda (x/d)^2$ to which an analogy can be made with the case of inflation where the expansion parameter was given by (\ref{eq:expansion_parameter}). 

We are now going to study the ground-state wavefunction $\Psi_0(x)$ using functional methods (see e.g.~\cite{rattazzi2009path} for an introduction to path-integral methods in QM). Let us consider a particle evolving under the Hamiltonian $\hat{\rm H}$. The real-time (Lorentzian) action $S[x(t)]$ is
\begin{align}\label{eq:QM_AHO_action_Lorentzian}
S[x(t)] = \int_{t_{\rm i}}^{t_{\rm f}} dt \left[\frac{1}{2}m \left(\frac{dx}{dt}\right)^2 - V(x) \right] \;.
\end{align}
The propagator $K(x_{\rm f}, t_{\rm f}; x_{\rm i}, t_{\rm i})$ for going from some initial position $x_{\rm i}$ at time $t_{\rm i}$ to the position $x_{\rm f}$ at time $t_{\rm f}$ can be written in both operator and path-integral languages
\begin{equation}\label{eq:QM_propagator}
K(x_{\rm f}, t_{\rm f}; x_{\rm i}, t_{\rm i}) = \braket{x_{\rm f} | e^{- i \, \hat{\rm{H}} (t_{\rm f} -t_{\rm i}) / \hbar}| x_{\rm i}} = \int_{x(t_{\rm i}) = x_{\rm i}}^{x(t_{\rm f}) = x_{\rm f}} \mathcal{D} x(t) \, e^{i S[x(t)] / \hbar}\;.
\end{equation}
We can insert in the propagator a complete set of eigenstates $\ket{n}$ of $\hat{\rm H}$ with eigenvalues $E_{n}$ that we assume positive:
\begin{equation}\label{eq:QM_propagator_sum}
\braket{x_{\rm f} | e^{- i \, \hat{\rm H} (t_{\rm f} - t_{\rm i} ) / \hbar} | x_{\rm i}} = \sum_n e^{-i E_{n} (t_{\rm f} - t_{\rm i}) / \hbar} \, \Psi_n(x_{\rm f}) \Psi^*_n(x_{\rm i}) \;,
\end{equation}
where $\Psi_n(x) \equiv \braket{x | n}$ and $\Psi^*_n(x)$ is its complex conjugate.

The ground state can then be extracted by performing a Wick rotation $t \rightarrow - i \tau$ and by then taking the limit of $T \equiv \tau_{\rm f}-\tau_{\rm i}$ large. In this way, \eqref{eq:QM_propagator_sum} is dominated by the ground state and we obtain
\begin{equation}\label{eq:QM_ground_state_wavefunction_from_PI}
\Psi_0(x_{\rm f}) \Psi^*_0(x_{\rm i}) \,  e^{- E_0 T / \hbar}  = \lim_{T \rightarrow \infty}~ \int_{x(\tau_{\rm i}) = x_{\rm i}}^{x(\tau_{\rm f}) = x_{\rm f}} \mathcal{D}x(\tau) \, e^{-S_{\rm E}[x(\tau)] / \hbar}\;,
\end{equation}
where $S_{\rm E}$ is the Euclidean action obtained after Wick rotation and $\tau$ is the imaginary time. Notice that the point $x_{\rm i}$ can be chosen arbitrarily if our goal is to extract $\Psi_0(x_{\rm f})$ (the dependence on $x_{\rm i}$ will end up in a normalization factor).

Let $y(\tau)$ be a fluctuation around the classical path $x_{\rm cl}$: $x(\tau) = x_{\rm cl}(\tau) + y(\tau)$. $x_{\rm cl}(\tau)$ satisfies the Euclidean equation of motion (without any expansion in $\lambda$). The path integral in (\ref{eq:QM_ground_state_wavefunction_from_PI}) then becomes
\begin{align}\label{eq:QM_path_integral_semiclassics}
\int_{x(\tau_{\rm i}) = x_{\rm i}}^{x(\tau_{\rm f}) = x_{\rm f}} \mathcal{D}x(\tau) \, e^{-S_{\rm E}[x(\tau)] / \hbar} = e^{-S_{\rm E}[x_{\rm cl}(\tau)]/\hbar} \int_{y(\tau_{\rm i}) = 0}^{y(\tau_{\rm f}) = 0} \mathcal{D}y(\tau) \, e^{-\frac{1}{\hbar} \left(\frac{1}{2}\frac{\delta^2 S_{\rm E}}{\delta x^2}y^2 + \frac{1}{3!}\frac{\delta^3 S_{\rm E}}{\delta x^3}y^3 + \ldots \right)} \;. 
\end{align}
Neglecting the higher-order terms which capture the interactions of perturbations around $ x_{\rm cl}(\tau)$, we obtain the semiclassical approximation for the ground-state wavefunction $\Psi_0(x_{\rm f})$,
\begin{equation}\label{eq:QM_semiclassics_formula}
\Psi_0(x_{\rm f}) =  \mathcal{I}(x_{\rm f}) e^{-S_{\rm E}[x_{\rm cl}(\tau)] / \hbar  }\;,
\end{equation}
where the path integral of the quadratic action of $y(\tau)$ gives rise to the prefactor $\mathcal{I}(x_{\rm f})$. Let us emphasize that the higher-order terms we have neglected in (\ref{eq:QM_path_integral_semiclassics}) correspond to higher-order corrections in $\lambda$ in perturbation theory, which are equivalent to loop diagrams, see~\cite{rattazzi2009path}. The on-shell action in (\ref{eq:QM_semiclassics_formula}) only captures all the tree-level diagrams with many external legs $x$. Moreover, following the standard derivation in \cite{rattazzi2009path} one arrives to the VanVleck-Pauli-Morette formula of the prefactor $\mathcal{I}(x_{\rm f})$,
\begin{equation}\label{eq:QM_prefactor}
\mathcal{I}(x_{\rm f}) = \EuScript{N}\sqrt{\frac{m}{2 \pi i \hbar v_i v_f \int_{x_{\rm i}}^{x_{\rm f}} \frac{d x'}{v^3(x')}}}\;,
\end{equation}
where we defined $v_i$ and $v_f$ as the initial and final velocities on the classical trajectory and $\EuScript{N}$ is a normalization factor. Notice that the expression \eqref{eq:QM_semiclassics_formula} is correct up to corrections $\mathcal{O}(\hbar)$ and will be a good approximation in regions where $S_{\rm E} \gg \hbar$.

Now let us get back to the case of the anharmonic oscillator. From the formula (\ref{eq:QM_semiclassics_formula}) it is convenient to write down the action (\ref{eq:QM_AHO_action_Lorentzian}) in Euclidean space. We now have

\begin{align}\label{eq:QM_AHO_action_Euclidean}
S_{\rm E}[x(\tau)] = \int_{\tau_{\rm i}}^{\tau_{\rm f}} d \tau \left( \frac{1}{2}m\dot{x}^2 + V(x) \right) \;,
\end{align}
where dot denotes $d/d\tau$.

Let us first anticipate the semi-classical scaling of the wavefunction $\Psi_0(x_{\rm f})$ as a function of $\lambda$ and the final position $x_{\rm f}$. From the formula (\ref{eq:QM_path_integral_semiclassics}), the leading exponent $S_{\rm E}[x_{\rm cl}(\tau)] / \hbar $ scales as\footnote{This can be easily realized by performing $x \rightarrow (\sqrt{\hbar/\lambda})x$.}
\begin{align}\label{qm:scaling_action}
\frac{S_{\rm E}[x_{\rm cl}(\tau)]}{\hbar} \sim \frac{1}{\lambda}F(\lambda x^2_f/d^2) \;, 
\end{align}
where $F$ is a function to be determined explicitly later on. Having said that, the on-shell action resums all the tree-level diagrams. The prefactor instead goes as $\lambda^0 G(\lambda x^2_f/d^2)$ with $G$ being an arbitrary function of $\lambda x^2_f/d^2$ and it captures all the 1-loop diagrams. The terms we have neglected in (\ref{eq:QM_path_integral_semiclassics}) are associated to the higher-loop diagrams. 

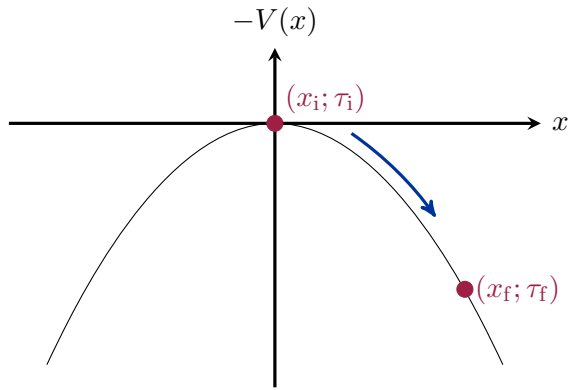
\begin{figure}[t!]
	\begin{center}
		\begin{tikzpicture}
		
		\draw[->,>=stealth,very thick] (-3.5,0)--(3.5,0) node[right]  {$x$};
		\draw[->,>=stealth,very thick] (0,-3.5)--(0,1) node[above]{$-V(x)$};
		
		\draw (0,0) parabola (3,-3.2);  
		\draw (0,0) parabola (-3,-3.2);
		\draw[->,>=stealth',very thick,ceruleanblue] ([shift=(55:3.5cm)]-1,-3) arc (55:35:4.5cm);
		
		\filldraw[myred] (0,0) circle (3pt) node[above right] {$(x_{\rm i};\tau_{\rm i})$};
		\filldraw[myred] (2.5,-2.2) circle (3pt) node[right] {$(x_{\rm f};\tau_{\rm f})$};
		
		\end{tikzpicture}
	\end{center}
	\caption{~The inverted potential for the case of anharmonic oscillator. } \label{fig:inverted_potential_anharmonic}
\end{figure}

Let us now use the formula (\ref{eq:QM_semiclassics_formula}) to calculate the ground-state wavefunction. First notice that from the action (\ref{eq:QM_AHO_action_Euclidean}) it is practically convenient to think of a particle moving in an inverted potential shown in Figure~\ref{fig:inverted_potential_anharmonic}. Without loss of generality, we set $x_{\rm i} = 0$\footnote{Notice that this choice has nothing to do with the choice of the ground state of the Hamiltonian. Also, if one keeps $x_{\rm i}$ finite and non-zero, the solution that does not run to infinity is the one with zero energy (it spends an infinite amount of time around the origin).}.
Another thing one should bear in mind is that since in eq.~\eqref{eq:QM_ground_state_wavefunction_from_PI} $\tau_{\rm f}-\tau_{\rm i}$ is taken to be very large, this means that the only real solution that exists corresponds to the zero-energy configuration (with a finite energy the particle would reach $x_{\rm f}$ from the origin in a finite time). Exploiting the conservation of energy the classical trajectory $x(\tau)$ satisfying the boundary conditions $x_{\rm cl}(\tau_{\rm i}) = x_{\rm i}$ and $x_{\rm cl}(\tau_{\rm f}) = x_{\rm f}$ is then determined by 
\begin{align}
\frac{d x}{d \tau} = \sqrt{\frac{2V(x)}{m}} \;,
\end{align}
which gives
\begin{align}
\tau -\tau_0 = \int_{\infty}^x \frac{d x'}{\sqrt{2V(x')/m}} = -\frac{1}{\omega} \text{arcsinh}\left(\frac{d}{\sqrt{2\lambda}~x}\right) \;,
\end{align}
where the integration constant $\tau_0$ corresponds to the lower limit of $x$ going to infinity. Inverting the expression above one gets
\begin{align}\label{eq:classicalPath_QM}
x(\tau) = -\frac{d}{\sqrt{2\lambda}\sinh(\omega\tau)} \;, \quad \tau < 0\;,
\end{align}
where $\tau_0$ has been absorbed into the variable $\tau$. 

Now let us calculate the exponent of (\ref{eq:QM_semiclassics_formula}). The action evaluated on the classical path is 
\begin{align}\label{eq:QM_AHO_action_E}
\frac{S_{\rm E}[x_{\rm cl}(\tau)]}{\hbar} &= \frac{1}{\hbar}\int_{\tau_{\rm i}}^{\tau_{\rm f}} d\tau~m\dot{x}^2 \nonumber \\
&= \frac{1}{\hbar}\int_{x_{\rm i}}^{x_{\rm f}} d x~\sqrt{2m V(x)} \nonumber \\
&= \frac{1}{6\lambda}\bigg[(1 + \bar{x}^2)^{3/2} - 1\bigg] \;,
\end{align}
where in the first line we have used the fact that the total energy vanishes, and in the second line we have changed the integration variable from time to position. Here we define $\bar{x}^2 \equiv 2 \lambda x_{\rm f}^2 / d^2$. Notice that to evaluate the action we did not need the explicit trajectory \eqref{eq:classicalPath_QM}. Eq.~(\ref{eq:QM_AHO_action_E}) agrees with the scaling argument (\ref{qm:scaling_action}).

At this point, using the formula (\ref{eq:QM_prefactor}) and the classical path (\ref{eq:classicalPath_QM}) one can easily compute the prefactor. Changing the integration variable to $\tau$ we obtain 
\begin{align}
v_i v_f \int_{\tau_{\rm i}}^{\tau_{\rm f}} \frac{d \tau}{v^2} =  \frac{e^{\omega T}}{4 \omega}(1 + \sqrt{1 + \bar{x}^2}) \sqrt{1 + \bar{x}^2} \;,
\end{align}
where $T = \tau_{\rm f} - \tau_{\rm i}$ which is taken to be very large. Therefore, the prefactor is 
\begin{align}\label{eq:QM_AHO_prefactor_final1}
\mathcal I(x_{\rm f}) = \EuScript{N}\frac{e^{-\omega T/2}}{(1 + \bar{x}^2)^{1/4}(1 + \sqrt{1 + \bar{x}^2})^{1/2}} \;,
\end{align} 
where we have absorbed all the $x_{\rm f}$-independent factors into the normalization factor $\EuScript{N}$. Again, this prefactor (\ref{eq:QM_AHO_prefactor_final1}) is only a function of $\lambda x_{\rm f}^2/d^2$ as anticipated from the scaling argument. The expressions for the Euclidean action \eqref{eq:QM_AHO_action_E} and for the prefactor \eqref{eq:QM_AHO_prefactor_final1} can now be inserted in eq.~\eqref{eq:QM_semiclassics_formula} to obtain the ground-state wavefunction as
\begin{equation}\label{eq:QM_AHO_ground_state_final}
\Psi_0(\bar{x}) = \EuScript{N} \frac{\exp \left\{-\frac{1}{6\lambda}\left[ \left(1 + \bar{x}^2 \right)^{3/2} - 1\right]  \right\}}{\left(1 + \bar{x}^2\right)^{1/4} \left( 1 + \sqrt{1 + \bar{x}^2} \right)^{1/2}}\bigg(1 +   {\cal O}(\lambda) f(\bar x)\bigg)\;.
\end{equation} 
This expression does not contain all $\lambda$ corrections to the ground-state wavefunction, but it resums all the leading corrections $(\lambda x_{\rm f}^2/d^2)^n$, the ones enhanced by $x_{\rm f}^2/d^2$ (for $\lambda =0$ one gets back to the harmonic ground-state $ \sim \exp[-x_{\rm f}^2/(2d^2)]$). Also, from (\ref{eq:QM_AHO_prefactor_final1}) we can read off the energy $E_0 = \hbar \omega/2$, which is the ground-state energy of the harmonic oscillator. This is consistent with the fact that $\lambda$ corrections to $E_0$ appear only at order $\hbar^2$ (corresponding to a two-loop effect, which we neglected). The wavefunction eq.~\eqref{eq:QM_AHO_ground_state_final} was obtained in \cite{Escobar-Ruiz:2017uhx} using periodic boundary condition $x_{\rm i} = x_{\rm f}$: the large $T$ limit corresponds in this case to the limit of zero temperature\footnote{An observable that is sensitive to the tail of the probability distribution is the moment $\langle x^N \rangle$ for large $N$. In the Gaussian case one finds that the leading contribution to the integral comes from $x \sim \sqrt{N}$. In standard perturbation theory the ground-state wavefunction gets corrections of order $\lambda x^4$, so that the perturbative calculation of $\langle  x^N \rangle$ is reliable for $\lambda x^4 \sim \lambda N^2 \lesssim 1$. The ``resummed" wavefunction eq.~\eqref{eq:QM_AHO_ground_state_final} allows to calculate $\langle x^N \rangle$ in saddle-point approximation for large $N$. In this case one only gets corrections ${\cal O}(\lambda)$ due to subleading corrections to the wavefunction  \eqref{eq:QM_AHO_ground_state_final} and ${\cal O}(1/N)$ due to the saddle-point approximation.}.

In the limit of large $\bar{x}$ keeping $\lambda$ small one obtains 
\begin{equation}
\Psi_0(\bar{x}) \sim \exp\bigg(-\lambda^{1/2}\frac{x_{\rm f}^3}{d^3} \bigg) \;.
\end{equation}
This expression shows how the tails of the distribution for $x_{\rm f}$ get modified. Moreover, it makes manifest the non-perturbative nature of the semiclassical approximation, since we obtain a non-analytic expression in the coupling $\lambda$. 

This result for the ground-state wavefunction can be obtained also in the more standard WKB approximation. As a consistency check for our procedure, in Appendix~\ref{app:WBK_approximation} we show that indeed the WKB wavefunction matches with eq.~\eqref{eq:QM_AHO_ground_state_final}.


\section{Two fields in dS}\label{sec:2field}
We are now going to consider an inflationary toy model in which one is able to analytically calculate the leading effect in the semiclassical expansion, effectively resumming an infinite set of diagrams of the perturbative series. Let us consider the action for two fields $\sigma$ and $\chi$: 
\begin{align}
S = \int d\eta d^3x \bigg[\frac{1}{2\eta^2H^2}(\sigma'^{2} - (\partial_i\sigma)^2) + \frac{1}{2\eta^2H^2}(\chi'^{2} - (\partial_i \chi)^2) - \frac{\lambda}{\eta^4 H^3}\chi \sigma^2\bigg] \;.  
\end{align}
The two fields interact through the cubic term and $\lambda \ll 1$ is the dimensionless parameter of the standard perturbative expansion. We want to calculate the WFU in a particular regime: the modes of $\chi$ have a much longer wavelength compared to the ones of $\sigma$ ($k_\chi \ll k_\sigma$), and $\chi$ is much larger than its typical fluctuation, $|\chi| \gg H $. Therefore, we do not want to assume that $\lambda \chi/H$ is small, while we are going to neglect all corrections suppressed by $\lambda$ only. (Notice that we assume $\sigma$ to have a typical fluctuation: $|\sigma| \sim H$.)

\begin{figure}[t!]
	\begin{subfigure}{1.0\textwidth}
		\centering
		\begin{tikzpicture}
		\draw[-,>=stealth,very thick] (-2,0)--(2,0);
		\draw[-,>=stealth,thick,dashed] (1.5,0) arc (0:-180:1.5cm);
		\node[above] at (1.5,0) {$\textcolor{myred}{\sigma_0}$}; 
		\node[above] at (-1.5,0) {$\textcolor{myred}{\sigma_0}$};
		\node at (2.5,-1) {$\textcolor{black}{+}$};
		\end{tikzpicture}
		\begin{tikzpicture}
		\draw[-,>=stealth,very thick] (-2,0)--(2,0);
		\draw[-,>=stealth,thick,dashed] (1.5,0) arc (0:-180:1.5cm);
		\draw[-,>=stealth,thick] (0.5,0)--(1.4,-0.5);
		\node[above] at (1.5,0) {$\textcolor{myred}{\sigma_0}$}; 
		\node[above] at (-1.5,0) {$\textcolor{myred}{\sigma_0}$};
		\node[above] at (0.5,0) {$\textcolor{myred}{\bar{\chi}}$};
		\node[right] at (1.4,-0.5) {$\textcolor{ceruleanblue}{\lambda}$};
		\node at (2.5,-1) {$\textcolor{black}{+}$};
		\end{tikzpicture}
		\begin{tikzpicture}
		\draw[-,>=stealth,very thick] (-2,0)--(2,0);
		\draw[-,>=stealth,thick,dashed] (1.5,0) arc (0:-180:1.5cm);
		\draw[-,>=stealth,thick] (0.5,0)--(1.4,-0.5);
		\draw[-,>=stealth,thick] (-0.2,0)--(1.2,-0.9);
		\node[above] at (1.5,0) {$\textcolor{myred}{\sigma_0}$}; 
		\node[above] at (-1.5,0) {$\textcolor{myred}{\sigma_0}$};
		\node[above] at (0.5,0) {$\textcolor{myred}{\bar{\chi}}$};
		\node[above] at (-0.2,0) {$\textcolor{myred}{\bar{\chi}}$};
		\node[right] at (1.4,-0.5) {$\textcolor{ceruleanblue}{\lambda}$};
		\node[right] at (1.2,-0.9) {$\textcolor{ceruleanblue}{\lambda}$};
		\node at (2.5,-1) {$\textcolor{black}{+}$};
		\node at (3.3,-1) {$\textcolor{black}{\ldots}$};
		\end{tikzpicture}
		\caption{~}
		\label{fig:TreeWitten_resum_chi}
	\end{subfigure}
	
    \bigskip
	
	\begin{subfigure}{1.0\textwidth}
		\centering
		\begin{tikzpicture}
		\draw[-,>=stealth,very thick] (-2,0)--(2,0);
		\draw[-,>=stealth,thick,dashed] (1.5,0) sin (1.2,-1.5) cos (0.9,0) ;
		\draw[-,>=stealth,thick,dashed] (-1.5,0) sin (-1.2,-1.5) cos (-0.9,0);
		\draw[-,>=stealth,thick] (-1.2,-1.5)--(1.2,-1.5) ;
		\node[above] at (1.5,0) {$\textcolor{myred}{\sigma_0}$}; 
		\node[above] at (-1.5,0) {$\textcolor{myred}{\sigma_0}$};
		\node[above] at (0.9,0) {$\textcolor{myred}{\sigma_0}$};
		\node[above] at (-0.9,0) {$\textcolor{myred}{\sigma_0}$};
		\node[below] at (-1.3,-1.4) {$\textcolor{ceruleanblue}{\lambda}$};
		\node[below] at (1.3,-1.4) {$\textcolor{ceruleanblue}{\lambda}$};
		\node at (2.5,-1) {$\textcolor{black}{+}$};
		\end{tikzpicture}
		\begin{tikzpicture}
		\draw[-,>=stealth,very thick] (-2,0)--(2,0);
		\draw[-,>=stealth,thick,dashed] (1.5,0) sin (1.2,-1.5) cos (0.9,0) ;
		\draw[-,>=stealth,thick,dashed] (-1.5,0) sin (-1.2,-1.5) cos (-0.9,0);
		\draw[-,>=stealth,thick] (-1.2,-1.5)--(1.2,-1.5) ;
		\draw[-,>=stealth,thick] (0.2,0)--(1,-0.75);
		\node[above] at (0.2,0) {$\textcolor{myred}{\bar{\chi}}$};
		\node[above] at (1.5,0) {$\textcolor{myred}{\sigma_0}$}; 
		\node[above] at (-1.5,0) {$\textcolor{myred}{\sigma_0}$};
		\node[above] at (0.9,0) {$\textcolor{myred}{\sigma_0}$};
		\node[above] at (-0.9,0) {$\textcolor{myred}{\sigma_0}$};
		\node[below] at (0.86,-0.6) {$\textcolor{ceruleanblue}{\lambda}$};
		\node[below] at (-1.3,-1.4) {$\textcolor{ceruleanblue}{\lambda}$};
		\node[below] at (1.3,-1.4) {$\textcolor{ceruleanblue}{\lambda}$};
		\node at (2.5,-1) {$\textcolor{black}{+}$};
		\node at (3.3,-1) {$\textcolor{black}{\ldots}$};
		\end{tikzpicture}
		\caption{~}
     \label{fig:Diagrams_not_resum}
	\end{subfigure}

    \bigskip
    
    \begin{subfigure}{1.0\textwidth}
    	\centering
    	\begin{tikzpicture}
    \draw[-,>=stealth,very thick] (-2,0)--(2,0);
    \draw[-,>=stealth,thick,dashed] (1.5,0) arc (0:-180:1.5cm);
    \draw[-,>=stealth,thick] (-1,-1.1) sin (0,-0.5) cos (1,-1.1);
    \node[above] at (1.5,0) {$\textcolor{myred}{\sigma_0}$}; 
    \node[above] at (-1.5,0) {$\textcolor{myred}{\sigma_0}$};
    \node[below] at (-1.1,-1) {$\textcolor{ceruleanblue}{\lambda}$};
    \node[below] at (1.1,-1) {$\textcolor{ceruleanblue}{\lambda}$};
    \node at (2.5,-1) {$\textcolor{black}{+}$};
    \end{tikzpicture}
    \begin{tikzpicture}
    \draw[-,>=stealth,very thick] (-2,0)--(2,0);
    \draw[-,>=stealth,thick,dashed] (1.5,0) arc (0:-180:1.5cm);
    \draw[-,>=stealth,thick,dashed] (-0.3,-0.8) sin (0,-0.5) cos (0.3,-0.8);
    \draw[-,>=stealth,thick,dashed] (-0.3,-0.8) sin (0,-1.1) cos (0.3,-0.8);
    \draw[-,>=stealth,thick] (-1,-1.1) sin (-0.3,-0.8);
    \draw[-,>=stealth,thick] (0.3,-0.8) cos (1,-1.1) ;
    \node[above] at (1.5,0) {$\textcolor{myred}{\sigma_0}$}; 
    \node[above] at (-1.5,0) {$\textcolor{myred}{\sigma_0}$};
    \node[below] at (-1.1,-1) {$\textcolor{ceruleanblue}{\lambda}$};
    \node[below] at (1.1,-1) {$\textcolor{ceruleanblue}{\lambda}$};
    \node[above] at (0.45,-0.85) {$\textcolor{ceruleanblue}{\lambda}$};
    \node[above] at (-0.45,-0.85) {$\textcolor{ceruleanblue}{\lambda}$};
    \node at (2.5,-1) {$\textcolor{black}{+}$};
    \node at (3.3,-1) {$\textcolor{black}{\ldots}$};
    \end{tikzpicture}
    \caption{~}
 \label{fig:LoopWitten_resum_chi}
    \end{subfigure}
\caption{~In the first row (Figure~\ref{fig:TreeWitten_resum_chi}) tree-level Witten diagrams that are enhanced by $\bar\chi$ and resummed. In the second (Figure~\ref{fig:Diagrams_not_resum}) tree-level diagrams with fewer powers of $\bar\chi$. In the third row (Figure~\ref{fig:LoopWitten_resum_chi}) loop-level diagrams, which are subleading in $\lambda$ and not captured in the semiclassical limit.}
\end{figure}
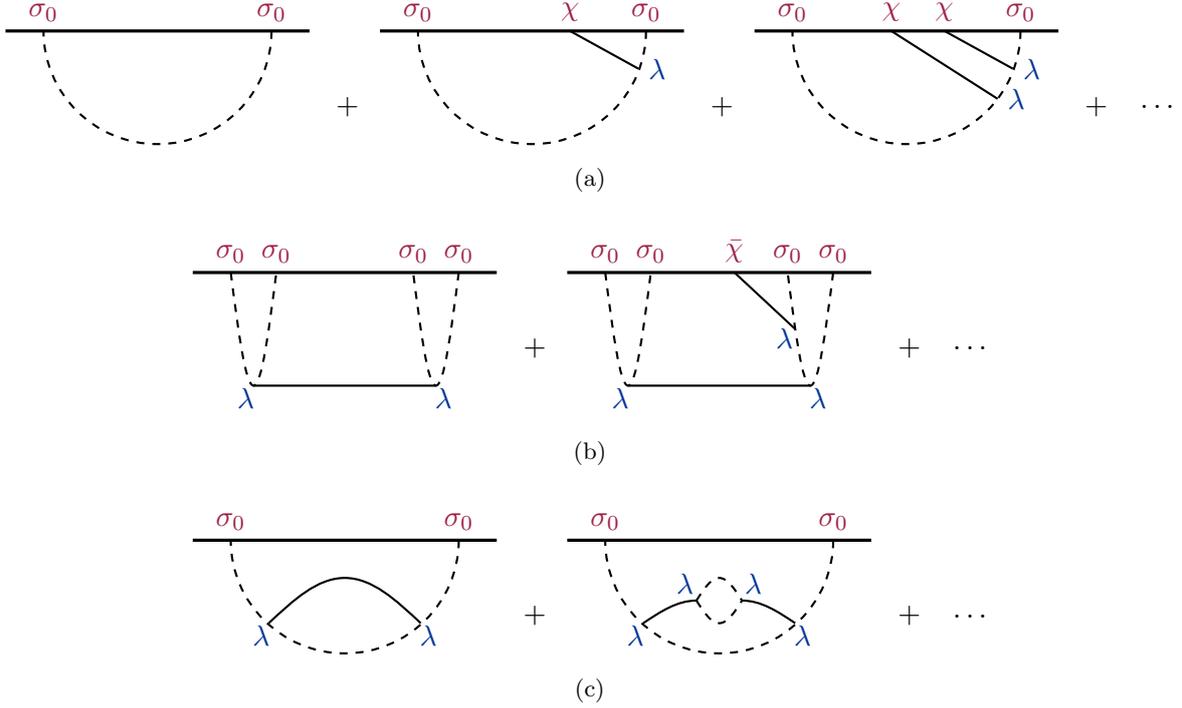

Loop corrections are suppressed by $\lambda$, so that the WFU can be calculated evaluating the classical action on-shell, as in eq.~\eqref{eq:WFUsemi}. The classical equations of motion in Fourier space read
\begin{align}
\sigma'' - \frac{2}{\eta}\sigma' + k_\sigma^2 \sigma + \frac{2\lambda}{\eta^2 H}\chi * \sigma &= 0 \;, \label{eq:sigma}\\ 
\chi'' - \frac{2}{\eta}\chi' + k_\chi^2 \chi + \frac{\lambda }{\eta^2 H}\sigma * \sigma &= 0 \;, \label{eq:chi}
\end{align}
where $*$ indicates a convolution in Fourier space.
The last term on the LHS of (\ref{eq:chi}) is negligible because it is of the order $\lambda$. Therefore, $\chi$ is just a free wave in de Sitter\footnote{For simplicity, we assume that there is a single Fourier mode of $\chi$, but the results would not change considering many modes, all much longer than the ones of $\sigma$, and giving $\bar\chi$ as the late-time value in the region of interest.},
\begin{align}
\chi_{\rm cl}(\vect k, \eta) = \bar{\chi}(1 - ik_\chi \eta)e^{ik_\chi \eta} \;,
\end{align}
with $\bar{\chi}$ its asymptotic value at late times. We need to keep, on the other hand, the last term on the LHS of (\ref{eq:sigma}) since $\lambda \bar{\chi}$ need not be small.  Plugging $\chi_{\rm cl}$ back into (\ref{eq:sigma}) we have
\begin{align}
\sigma'' - \frac{2}{\eta}\sigma' + k_\sigma^2 \sigma + \frac{2\lambda}{\eta^2 H}\bar{\chi}(1-ik_\chi \eta)e^{ik_\chi \eta} \sigma = 0 \;.
\end{align}
The last term becomes relevant compared to the gradient term only at late times when $|\eta| \lesssim k_\sigma^{-1}$. In this regime, since $k_\chi \ll k_\sigma$, one can treat $\chi$ as a constant, equal to its asymptotic value $\bar\chi$. The calculation reduces to the one of a massive scalar field in dS with the mass that depends on $\bar\chi$:
\begin{align}
S_\sigma = \int d\eta d^3x \bigg[\frac{1}{2\eta^2H^2}(\sigma'^{2} - (\partial_i\sigma)^2) - \frac{\alpha H^2}{2\eta^4}\sigma^2 \bigg] \;, \label{action:sigma_barchi}
\end{align}
where the dimensionless coupling $\alpha$ is defined by $\alpha \equiv 2\lambda \bar{\chi}/H$. Of course, one is able to solve exactly in $\alpha$ and there is no need of a perturbative expansion in this parameter. This corresponds to resumming the tree-level Witten diagrams shown in Figure~\ref{fig:TreeWitten_resum_chi}. The tree-level diagrams of Figure~\ref{fig:Diagrams_not_resum} are not enhanced by $\bar\chi$ (or less enhanced than the ones of Figure~\ref{fig:TreeWitten_resum_chi}) and are thus neglected, together with all loop diagrams, Figure~\ref{fig:LoopWitten_resum_chi}.
The power spectrum of $\sigma$ for $\alpha < 9/4$ reads at late times (prime means $(2\pi)^3 \delta(\vect k + \vect k')$ was dropped)
\begin{align}
\langle \sigma_{\vect{k}} \sigma_{-\vect{k}} \rangle' \simeq \frac{H^2}{2k^{3 - \frac{2}{3}\alpha}} = \frac{H^2}{2k^{3 - \frac{4}{3} \lambda \bar{\chi}/H}}\;. 
\end{align}
This shows we have resummed all powers of $\lambda \bar\chi$.


As an aside, one may wonder whether the exact power spectrum as a function of $\alpha$ coincides with the result of summing the perturbative series, or there are non-perturbative effects one cannot capture in perturbation theory. It turns out that the power spectrum as a function of the complex variable $\alpha$ is an entire function, without singularities at any finite point. Therefore, it coincides with the perturbative series for any $\alpha$. Let us verify this. Following the standard calculation for a massive field in dS (see e.g~\cite{Arkani-Hamed:2015bza}), the mode function $\sigma_{\rm cl}(\vect k, \eta)$ that multiplies the operator $\hat a^\dagger$, with the correct behaviour at early times reads
\begin{align}\label{Two field:sigma_nu_real}
\sigma_{\rm cl}(\vect k, \eta) = H \frac{\sqrt{\pi}}{2}e^{-i\nu\pi/2}(-\eta)^{3/2}H_\nu^{(2)}(-k\eta) \;,\qquad \nu \equiv \sqrt{\frac{9}{4} - \alpha} \;.
\end{align}
This expression is even in $\nu$ so that there is no ambiguity when the square root becomes imaginary. 
To calculate the power spectrum one needs the complex conjugate of this. Using the properties of the Hankel function this can be written as
\be
\sigma_{\rm cl}(\vect k, \eta)^* = H \frac{\sqrt{\pi}}{2}e^{i\nu\pi/2}(-\eta)^{3/2}H_\nu^{(1)}(-k\eta) \;,
\ee
where the equality holds for any real $\alpha$. Therefore, for any real $\alpha$ one has
\be
\langle \sigma_{\vect{k}} \sigma_{-\vect{k}} \rangle' =  |\sigma_{\rm cl}(\vect k, \eta)|^2 = H^2 \frac{\pi}{4} (-\eta)^3 H_\nu^{(1)}(-k\eta) H_\nu^{(2)}(-k\eta)  \;.
\ee
Using the properties of the Hankel functions, one can see that the RHS is an entire function of $\nu$ on the complex plane and moreover it is even in $\nu$. Therefore, it is an entire function of the complex variable $\alpha$. The analytic extension of the power spectrum as a function of $\alpha$ is entire and this implies that it coincides with its series expansion calculated around any point. (For a related discussion about analyticity of de Sitter propagators see \cite{Goodhew:2020hob}.)

In general, one cannot hope to find an analytical solution as in the simple case above. One has to approach the problem numerically and in this case it is necessary to analytically continue the problem to Euclidean time $\tau$ defined as $\eta = - i \tau$. The Bunch-Davies condition is that fields decay in the limit $\eta \to -\infty(1 - i \epsilon)$ and after analytic continuation to $\tau$, this condition becomes the requirement of decay for $\tau \to -\infty$. The advantage is that free fields exponentially decay for $\tau \to -\infty$, while in Lorentzian one has to deal with oscillating solutions. In order to perform the rotation, one has to assume (or prove) analyticity of the solution in the upper-left quadrant of the complex $\eta$ plane. We are going to come back to this issue in Section~\ref{sec:proof_anlyticity}. For the time being, let us notice that the solution \eqref{Two field:sigma_nu_real} is analytic in the required quadrant and this holds for any value of $\alpha$. This can also be seen as a consequence of the analyticity of the differential equation from the action \eqref{action:sigma_barchi}. Another advantage of the Euclidean rotation is that solutions are real, since both the differential equation and the boundary conditions are real. On the other hand, in the Lorentzian case, the Bunch-Davies boundary condition can only be satisfied by complex solutions.

In the following we concentrate on single-field models of inflation where there is no evolution outside the horizon. The non-perturbative results we are going to get are therefore unrelated with the stochastic approach, which resums the classical long-wavelength effects. We will study non-perturbative effects {\em at horizon crossing}, and these are fully quantum mechanical. In the future, it would be nice to explore the connection between the two approaches.


\section{Single-field inflation with \texorpdfstring{$\dot\zeta^4$}{zeta'4} interaction}\label{sec:zeta'^4}

Let us now apply our methods to a realistic scenario. We focus on a specific model of single-field inflation with a large quartic interaction $\dot\zeta^4$ \cite{Senatore:2010jy}. With a single interaction it will be easier and more transparent to explore the semiclassical limit and derive analytical estimates.  We leave to future work the generalization to other interactions. In the next Subsection we will review this model in the context of the Effective Field Theory of Inflation (EFTI). We will explain why it is consistent to focus on the non-linearities induced by the single operator $\dot\zeta^4$ and treat the geometry as an unperturbed de Sitter space.  After that, we will concentrate on the calculations of the $\zeta$ probability distribution for large values of $\zeta$, using both analytical and numerical methods. 

\subsection{Single-field inflation with large 4-point function}\label{subsec:EFT_for_zeta'^4}
The model we would like to discuss is naturally described within the EFTI \cite{Cheung:2007st}, which we briefly review below. 
In single-field inflation, the rolling of the inflaton $\phi(t)$ in a quasi-$\rm dS$ background leads to the spontaneous breaking of time diffeomorphisms. In unitary gauge, $\delta \phi(x) = 0$, the scalar mode is hidden inside the metric and the effective action for perturbations can be written as (see \cite{Cheung:2007st})
\begin{equation}\label{eq_sec_zeta'^4:EFT_action}
\begin{aligned}
S = \int d^{4} x \sqrt{-g}&\bigg[  \frac{1}{2}\MP^{2} R + \MP^{2} \dot{H} g^{00}-\MP^{2} (3 H^{2}+\dot{H}) ~ + \\
&+\frac{1}{2} M_{2}(t)^{4}(\delta g^{00})^{2}+\frac{1}{3 !} M_{3}(t)^{4}(\delta g^{00})^{3}+\frac{1}{4 !} M_{4}(t)^{4}(\delta g^{00})^{4}+ \ldots \bigg]\;,
\end{aligned}
\end{equation}
where $g_{\mu\nu}$ is the metric, $R$ is the Ricci scalar, $\delta g^{00} \equiv g^{00} + 1$ and $M_i(t)$ are functions of time with dimensions of a mass. The operators in the first line, expanded around the inflationary background, start linear in perturbations while those in the second line start at second and higher order. The dots stand for operators starting at even higher order in perturbations or containing more derivatives. 

The scalar mode $\pi$ can be reintroduced by performing a broken time diffeomorphism $t \rightarrow t + \xi^0(x)$ and then promoting $\xi^0$ to a field, $-\pi$, that transforms non-linearly under the broken time diffs. $\pi(x) \rightarrow \tilde \pi(\tilde x(x)) = \pi(x) - \xi^0(x)$. In this way the resulting action is fully diff-invariant. As an example, under this Stueckelberg procedure the $g^{00}$ component of the metric transforms (neglecting the mixing with metric perturbations) as
\begin{equation}\label{eq_zeta'^4:stueckelberg_g00}
g^{00} \rightarrow -1 -2 \dot \pi + (\partial_\mu \pi)^2\;.
\end{equation}
This will be the only transformation we will need in our discussion.  If one further assumes an approximate shift symmetry for $\pi$, then operators without at least one derivative acting on $\pi$ will be suppressed. This assumption allows us to neglect terms coming, for instance, from the time dependence of the functions $M_i(t)$ in the action \eqref{eq_sec_zeta'^4:EFT_action}. Notice that the Goldstone boson $\pi$ is related to the curvature perturbation $\zeta$ through the relation $\zeta = - H \pi$.

We want to explore a region of parameters where the $\pi$ non-linearities are dominated by a single quartic operator. Following \cite{Senatore:2010jy} let us start  with $M_4 \neq 0$ while all the other $M_i$'s in the action \eqref{eq_sec_zeta'^4:EFT_action} are zero. We are going to come back to discuss the radiative stability of this choice momentarily. The Stueckelberg procedure eq.~\eqref{eq_zeta'^4:stueckelberg_g00} then gives
\begin{equation}\label{eq_zeta'^4:action_g00^4}
\begin{aligned}
S_{\pi} = \int d^4 x \sqrt{-g} &\bigg[ -\dot H \MP^{2} \left( \dot \pi^2 - \frac{(\partial_i \pi)^2}{a^2}\right) ~ +\\
& + \frac{M_4^4}{4!}\left( 16 \dot{\pi}^{4}-32 \dot{\pi}^{3}\left(\partial_{\mu} \pi\right)^{2}+24 \dot{\pi}^{2}\left(\partial_{\mu} \pi\right)^{4}-8 \dot{\pi}\left(\partial_{\mu} \pi\right)^{6}+\left(\partial_{\mu} \pi\right)^{8} \right) \bigg] \;.
\end{aligned}
\end{equation}
The operator $M_4$ contains a whole slew of non-linearities, but we want to argue that there is a regime in which only the first term, $\dot\pi^4$, is relevant. In perturbation theory this operator contributes to the 4-point function as 
\begin{equation}\label{eq_zeta'^4:tau_NL_estimate}
g_{\rmS NL} \sim \frac{\braket{\zeta^4}}{ \braket{\zeta^2}^3} \sim \frac{\mathcal L_4}{\mathcal L_2} \frac{1}{P_\zeta} \sim \frac{M_4^4}{|\dot H| \MP^2}\;,
\end{equation}
where, in estimating the quadratic and quartic Lagrangians ${\mathcal L_2}$ and ${\mathcal L_4}$, derivatives are taken to be of order $H$. In the following we will focus on the limit $g_{\rmS NL} \gg 1 $. (The Planck experimental constraint on this parameter is $|g_{\rmS NL}| < 2 \cdot 10^6$ at $1 \sigma$ \cite{Akrami:2019izv}.)

After going to canonical normalization, $\pi_c \equiv \sqrt{- 2 \dot H \MP^2} ~\pi$, the interactions in eq.~\eqref{eq_zeta'^4:action_g00^4} read
\begin{equation}\label{eq_zeta'^4:L_4&L_5}
\mathcal L_4 \simeq \frac{1}{\Lambda_{\rmS U}^4}\dot \pi_c^4\;, \quad \mathcal L_5 \simeq \frac{1}{g_{\rmS NL}^{1/2} \Lambda_{\rmS U}^6 } \dot \pi_c^3 (\partial_i \pi_c)^2\;, \ldots
\end{equation}
where we defined the scale $\Lambda_{\rmS U}^4 \equiv (\dot H \MP^2)^2/M_4^4$ and dropped factors of order unity. The quantum mechanical expansion parameter is $\lambda \equiv H^4/\Lambda_{\rmS U}^4$, the analogue of the quartic coupling in the anharmonic oscillator example discussed above. We always assume $\lambda \ll 1$, since this is the regime of validity of the EFT: powers of $\lambda$ weight higher loops in calculating observables and in this paper we only look at the leading semiclassical approximation. Notice that $\lambda \simeq g_{\rmS NL} P_\zeta  $, so that the regime $g_{\rmS NL} \gg 1$ is compatible with $\lambda \ll 1$. For large $g_{\rmS NL} $ eq.~\eqref{eq_zeta'^4:L_4&L_5} shows that the additional operators inside $(\delta g^{00})^4$ are suppressed by a higher scale compared to $\dot\pi_c^4$. This separation of scales implies, as we are going to show,  that there is a regime of large values of $\zeta$ when the non-linearities associated with $\dot\pi^4$ are large, while the additional operators can be neglected.

Written in terms of $\zeta = - H\pi$ the Lagrangian is schematically of the form
\be\label{Szeta}
S_{\zeta} = \int d^4 x \sqrt{-g} \;\frac{|\dot H| \MP^2}{H^2} \left[(\partial_\mu\zeta)^2 + g_{\rmS NL} \frac{1}{H^2} \dot\zeta^4 + g_{\rmS NL}  \frac{1}{H^3} \dot\zeta^3 (\partial_\mu\zeta)^2 + \ldots\right] \;.
\ee
Since all derivatives are of order $H$, classical non-linearities associated with the quartic operator are of the same order as the kinetic term for $g_{\rmS NL} \zeta^2 \sim 1$. In this regime, since $g_{\rmS NL}  \gg 1$, the quintic term gives a contribution $g_{\rmS NL} \zeta^3 \ll 1$. Of course, the additional terms will become relevant if $g_{\rmS NL} \zeta^2$ becomes even larger, of order $g_{\rmS NL}^{1/3}$. In our Universe the experimental constraints impose $g_{\rmS NL}^{1/3} \lesssim 10^2$; however, since in this paper we are mostly interested in presenting the general method and not in applying to phenomenology, in the following we are going to disregard this upper limit and explore the effect of the quartic term for arbitrarily large $g_{\rmS NL} \zeta^2$, neglecting the other operators. (Notice that if one is interested in the PBH abundance, $\zeta \sim 1$, one is actually sensitive to all the terms inside a given operator $(\delta g^{00})^n$.)

Let us now come back to the issue of whether the choice of setting to zero all operators except $M_4$ is stable under radiative corrections. We start with the operators $M_2$ and $M_3$, following \cite{Senatore:2010jy}, and show that these operators are automatically suppressed by an approximate symmetry in the setup we are studying. Since the quintic operator in eq.~\eqref{eq_zeta'^4:L_4&L_5} is suppressed for large $g_{\rmS NL}$, the action \eqref{eq_zeta'^4:action_g00^4} acquires an approximate $\mathbb{Z}_2$ symmetry $\pi \rightarrow - \pi$: odd operators are suppressed by $g_{\rmS NL}^{1/2}$. This observation guarantees that loop corrections to $(\delta g^{00})^2$ and $(\delta g^{00})^3$ are not sizeable. To see this notice, using  eq.~\eqref{eq_zeta'^4:stueckelberg_g00}, that the leading interactions arising from these operators are odd in $\pi$. Thus, they are generated radiatively by loops with insertions of terms odd in $\pi$ hence suppressed by $g_{\rmS NL}$. As an example, we can estimate the scale at which the operator $\dot \pi_c (\partial_i \pi_c )^2$, contained in $(\delta g^{00})^2$, is generated. A loop with the interaction $\mathcal L_5$ of eq.~\eqref{eq_zeta'^4:L_4&L_5}  generates the cubic operator
\begin{equation}
\mathcal L_3 \sim \frac{1}{\Lambda_{\rmS U}^2 g_{\rmS NL}^{1/2}} \dot \pi_c (\partial_i \pi_c )^2\;,
\end{equation}
where the loop integral was cut off at the highest possible scale $\Lambda_{\rmS U}$. A similar estimate for the operator $\dot \pi_c^3$, contained in $(\delta g^{00})^3$, gives the same suppression scale. This corresponds to $M_2^4$, $M_3^4 \sim |\dot H| \MP^2 \ll M_4^4$: this model features $f_{\rmS NL} \lesssim 1$ while $g_{\rmS NL}$ can be arbitrarily large  \cite{Senatore:2010jy}. These radiatively generated operators would contribute terms of order $(\partial_i\zeta)^2 \dot\zeta/H$ and $\dot\zeta^3/H$ inside the brackets of eq.~\eqref{Szeta} and they are thus negligible for large $g_{\rmS NL}$.

Let us now come to the operators $(\delta g^{00})^n$ with $n\geq 5$. The radiative generation of the odd ones will be suppressed by the aforementioned approximate symmetry. For the even ones, however, there is no suppression, so that if the loop integral is pushed up to the unitarity cut-off $\Lambda_{\rmS U}$, the first operator inside each $(\delta g^{00})^n$ will read in canonical normalization
\be
(\delta g^{00})^n \quad\rightarrow\quad \frac{\dot\pi_c^n}{\Lambda_{\rmS U}^{2n-4}} \qquad  n\;{\rm even\;.}
\ee
It is easy to see that all terms in these expressions will become relevant exactly when the operator $M_4$ becomes of the same order as the kinetic term. Going to even larger values of $\zeta$, the terms with larger $n$ will dominate the lower ones. The estimate however may be pessimistic, since in general the loop integral will be cut at a scale much lower than the unitarity cutoff $\Lambda_{\rmS U}$. For instance, if one considers the spontaneous breaking of a global $U(1)$ via a Higgs mechanism, the resulting EFT for the Goldstone boson is of the form $-\frac12 (\partial\pi)^2 + (\partial\pi)^4/\Lambda^4$, with all additional operators $(\partial\pi)^{2n}$ suppressed in the limit the Higgs field is weakly coupled. See, for instance, the discussion in Section~4 of \cite{Creminelli:2019kjy} and references therein.
In the following we are going to assume that these extra operators are sufficiently suppressed to be negligible in the regime of interest.

This discussion leads us to an important general point. The questions we are addressing are sensitive to the full non-linear structure of the EFT, including in principle the whole series of operators. One may worry that this does not make sense and goes beyond the regime of validity of the EFT itself. First of all, notice we are always in a regime of small energy: derivatives are of order $H$ and are suppressed with respect to the cut-off of the theory. Indeed, the quantum mechanical expansion parameter $\lambda$ is small. What is getting large is $\zeta$, i.e.~we are in the regime of  large number of particles, or large occupation number. In general, there is nothing wrong in exploring an EFT for large values of the fields: for instance we do it in General Relativity all the times, when we study the full Einstein equations to obtain for example the Schwartzschild solution. Of course, there is no guarantee that the solution remains healthy: perturbations around the solution may become pathological signalling that the EFT is actually breaking down (see Section~4 of \cite{Creminelli:2019kjy}). Thus one should always check that the non-linear solution remains healthy. Another point of concern is the knowledge of the EFT: to find a reliable solution one should have control of all the terms in the EFT with the minimum number of derivatives, but this looks challenging. In some cases the symmetries of the problem are such that the whole non-linear structure of the theory is fixed. Again GR is the prototypical example: the Ricci scalar contains an infinite series of non-linearities of the graviton, all terms with two derivatives. In the case of scalars, one can consider symmetries that enforce a complete non-linear structure. For instance the scalars that describe the embedding of a brane in an extra dimensional space have an action fixed by the (non-linear realization of) geometrical symmetries: the DBI action \cite{Alishahiha:2004eh}. Another example is the one of Galileons \cite{Nicolis:2008in}: at leading order in derivatives there are only three possible interaction terms (in 3+1 dimensions). Even in cases in which symmetries are not powerful enough, some assumptions about the UV completion may fix the full non-linear structure of the EFT.  We already gave above the example of the Abelian Higgs model, while another example is the Euler-Heisenberg Lagrangian obtained integrating out the electron from QED. The necessity to know the whole non-linear action is therefore a feature more than a pathology, not that different from the necessity of knowing the full scalar potential $V(\phi)$ to describe inflation from observable scales to reheating. 

Before moving to the actual calculation with the $\dot\zeta^4$ interaction, let us comment on another approximation: we are going to neglect metric perturbations, considering a scalar field in exact de Sitter space. 
This corresponds to the usual ``decoupling limit": the effect of $\pi$ perturbations on the metric is suppressed by the slow-roll parameter $\epsilon \equiv -\dot H/H^2$, which also describes the deviation of the unperturbed background from de Sitter.  This is not changed by the fact that we are taking large values of $\zeta$; the leading interaction can be read by looking at $\pi$ only and treating the metric as unperturbed.

\subsection{\texorpdfstring{$\dot\zeta^4$}{zeta'4} beyond perturbation theory}

We can now apply the main ideas of this paper to the model introduced in the previous Section, with the discussed approximations. The action for $\zeta$ using conformal time is
\begin{align}\label{lagrangian:lambdaphi4}
S = \int d ^3x d \eta \left\{\frac{1}{2\eta^2 P_\zeta}\bigg[\zeta'^2 - (\partial_i\zeta)^2\bigg] + \frac{\lambda \zeta'^4}{4! P_\zeta^2}\right\} \;, 
\end{align}
where $P_\zeta \equiv H^2 / (2 \epsilon \MP^2)$ and $\lambda \equiv \left( H / \Lambda_{\rmS U} \right)^{4} \ll 1$. The standard in-in perturbation theory for $\zeta$ corresponds to an expansion of the various correlators in powers of $\lambda$. From now on we call $\zeta_0$ the asymptotic late-time value of $\zeta$. Comparing the free action with the quartic interaction, one sees that the relevant expansion parameter is $\lambda \zeta_0^2/P_\zeta$. The semiclassical expansion corresponds to an expansion in $\lambda \ll 1$ keeping $\lambda \zeta_0^2/P_\zeta$ finite and not necessarily small. The wavefunction of the Universe is calculated evaluating the action on-shell
\be\label{eq:WFUsemiagain}
\Psi[\zeta_0(\vect x)] \sim e^{i S[\zeta_{\rm cl}]}\;.
\ee
From the expression of the action eq.~\eqref{lagrangian:lambdaphi4} one can see that the on-shell action scales as
\begin{equation}\label{eq_zeta'^4:semiclassical_action_scaling}
S[\zeta_{\rm cl}] =\frac{1}{\lambda}F\left(\lambda \zeta_0^2/P_\zeta\right)\;,
\end{equation}
where $F$ is a function to be determined (in analogy with the case of the anharmonic oscillator in eq.~\eqref{qm:scaling_action}).

The field $\zeta_{\rm cl}$ is a solution of the equation of motion one can derive from the action \eqref{lagrangian:lambdaphi4}. For analytical and numerical purposes it is better to consider the system in Euclidean time $\tau$ defined as $\eta = - i \tau$. The equation of motion reads
\begin{align}\label{eq:PDE phi_lambda4}
-\zeta'' + \frac{2}{\tau}\zeta' - \partial_i^2\zeta - \frac{\lambda}{2 P_\zeta}\tau^2\zeta'^2\zeta'' = 0\;.
\end{align} 
(With an abuse of notation we indicate with primes both derivatives with respect to the conformal time $\eta$ and the Euclidean time $\tau$. The appearance of $\eta$ or $\tau$ in the equation should help not creating confusion.) We are going to solve the PDE above with boundary conditions at early and late times. At early times $\zeta$ must go to zero, while at late times if must give the profile $\zeta_0(\vect x)$ we are interested in. The action in Euclidean time is given by
\begin{align}\label{lagrangian:lambdaphi4_euclidean}
S_{\rm E} \equiv -\int d^3x d \tau \left\{\frac{1}{2\tau^2 P_\zeta}\bigg[\zeta'^2 + (\partial_i\zeta)^2\bigg] + \frac{\lambda \zeta'^4}{4! P_\zeta^2}\right\} \;, 
\end{align}
with 
\be\label{eq:WFUsemiE}
\Psi[\zeta_0(\vect x)] \sim e^{-S_{\rm E}[\zeta_{\rm cl}]}\;.
\ee
Notice that we started with an integral in $\eta$ slightly deformed above the real axis to project in the vacuum. Now we are effectively integrating $\eta$ along the positive imaginary axis. The two procedures give the same result assuming analyticity of the Lagrangian as a function of complex $\eta$, in the quadrant of interest. For the time being, we assume this property and we will come back to this point in Section~\ref{sec:proof_anlyticity}.

Let us go through the calculation in the case of the free theory $\lambda =0$, following \cite{Maldacena:2002vr}. This is useful to understand the dependence of the WFU on time: indeed we have been sloppy so far and we should have written the WFU as $\Psi[\zeta_0(\vect x),\eta_{\rm f}] $, where $\eta_{\rm f}$ is the (late) time of interest. In Fourier space, the solution with prescribed boundary conditions at $\eta_{\rm f}$ that decays to zero when $\eta$ acquires a small positive imaginary part in the far past is
\be \label{sec4:classical_free_zeta}
\zeta_{\rm cl}(\vect k, \eta) = \zeta_0 (\vect k) \frac{(1-i k \eta)e^{i k \eta}}{(1-i k \eta_{\rm f})e^{i k \eta_{\rm f}}} \;.
\ee
One has then to evaluate the free action on these solutions. Integrating by parts the free action gives a term proportional to the equation of motion, which is zero on-shell, and a boundary term:
\be\label{sec4:onshell_free_action}
i S = \frac{i}{2 P_\zeta} \int \frac{d^3 k}{(2 \pi)^3} \left. \frac{1}{\eta_{\rm f}^2} \zeta_{\rm cl} (-\vect k, \eta) \partial_\eta \zeta_{\rm cl} (\vect k, \eta) \right|_{\eta = \eta_{\rm f}} \simeq  \int \frac{d^3 k}{(2 \pi)^3} \frac{1}{2 P_\zeta} \left(i \frac{k^2}{\eta_{\rm f}} - k^3 + \ldots \right) \zeta_0 (-\vect k) \zeta_0 (\vect k) \;,
\ee
where we dropped terms subleading for $\eta_{\rm f} \to 0^-$. The time-dependence of the WFU is a pure phase that does not affect the probability of $\zeta_0$, which is time-independent at late times (this justifies our sloppy notation).

It is useful to also do the same calculation in Euclidean time $\tau$ since this is what we are going to use for the interacting theory.  One has
\be
\zeta_{\rm cl}(\vect k, \tau) = \zeta_0 (\vect k) \frac{(1- k \tau)e^{k \tau}}{(1-k \tau_{\rm f})e^{k \tau_{\rm f}}} \;,
\ee
\be\label{eq:SEfree}
S_{\rm E} = -\frac{1}{2 P_\zeta} \int \frac{d^3 k}{(2 \pi)^3} \left. \frac{1}{\tau_{\rm f}^2} \zeta_{\rm cl} (-\vect k, \tau) \partial_\tau \zeta_{\rm cl} (\vect k, \tau) \right|_{\tau = \tau_{\rm f}} \simeq  \int \frac{d^3 k}{(2 \pi)^3} \frac{1}{2 P_\zeta} \left(\frac{k^2}{\tau_{\rm f}} + k^3 + \ldots \right) \zeta_0 (-\vect k) \zeta_0 (\vect k) \;.
\ee
This is the same as the Lorentzian result after the analytic continuation of $\tau_{\rm f}$. Notice that in the Euclidean calculation both the divergent part and the finite part are real. Notice also that since $\tau_{\rm f} \to 0^-$, $S_{\rm E} <0$ and indeed there is an overall minus sign in front of eq.~\eqref{lagrangian:lambdaphi4_euclidean}. However, after analytic continuation $1/\tau_{\rm f}$ becomes purely imaginary and the remaining piece is positive as it should.  In the interacting case, one has to deal with this divergence to make the problem numerically tractable. The crucial simplification is that the divergence, which in Lorentzian describes the phase of the wavefunction is a late-time effect and at late times the interaction is negligible, since it contains more derivatives than the free action. Therefore, as we will see, one can extract a finite result comparing the interacting case with the free one. 

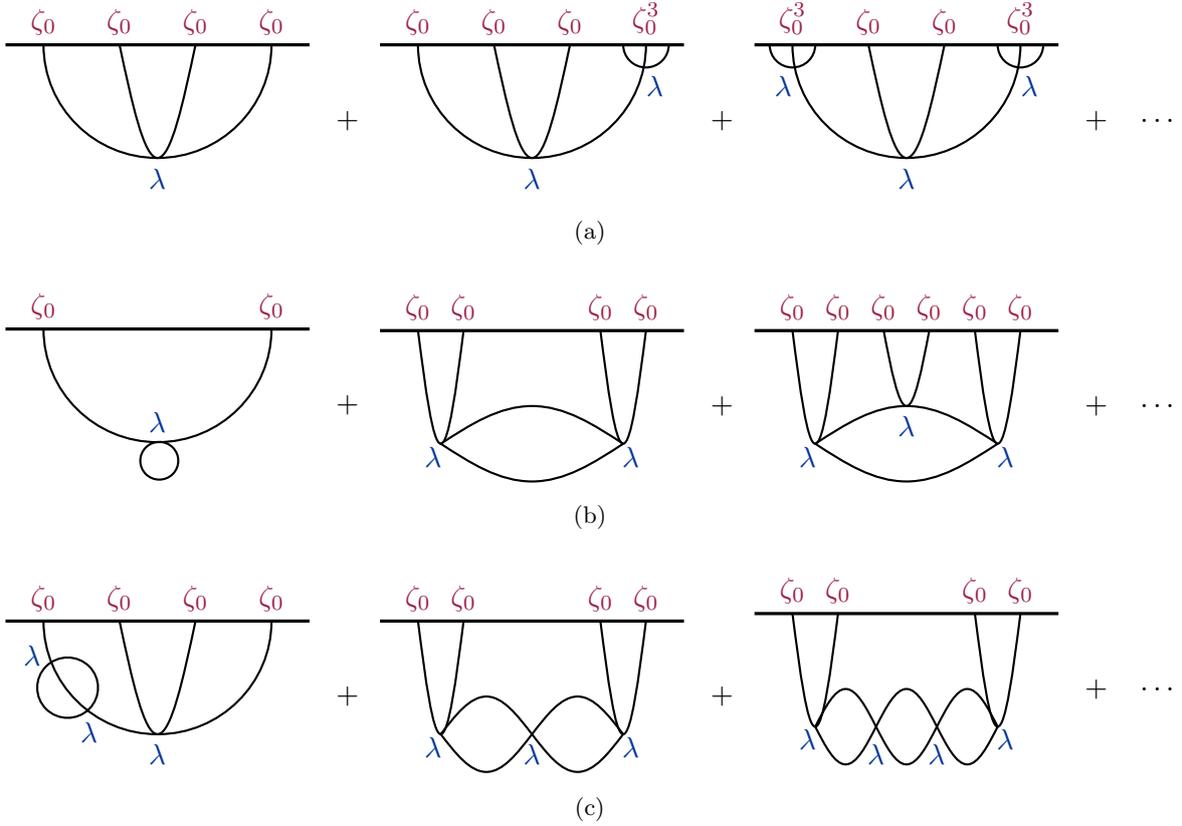
\begin{figure}[t!]
	\begin{subfigure}{1.0\textwidth}
		\centering
			\begin{tikzpicture}
		\draw[-,>=stealth,very thick] (-2,0)--(2,0);
		\draw[-,>=stealth,thick] (1.5,0) arc (0:-180:1.5cm);
		\draw[-,>=stealth,thick] (0.5,0) sin (0,-1.5) cos (-0.5,0);
		\node[above] at (1.5,0) {$\textcolor{myred}{\zeta_0}$}; 
		\node[above] at (-1.5,0) {$\textcolor{myred}{\zeta_0}$};
		\node[above] at (0.5,0) {$\textcolor{myred}{\zeta_0}$};
		\node[above] at (-0.5,0) {$\textcolor{myred}{\zeta_0}$};
		\node[below] at (0,-1.5) {$\textcolor{ceruleanblue}{\lambda}$};
		\node at (2.5,-1) {$\textcolor{black}{+}$};
		\end{tikzpicture}
		\begin{tikzpicture}
		\draw[-,>=stealth,very thick] (-2,0)--(2,0);
		\draw[-,>=stealth,thick] (1.5,0) arc (0:-180:1.5cm);
		\draw[-,>=stealth,thick] (1.8,0) arc (0:-180:0.3cm);
		\draw[-,>=stealth,thick] (0.5,0) sin (0,-1.5) cos (-0.5,0);
		\node[above] at (1.5,0) {$\textcolor{myred}{\zeta_0^3}$}; 
		\node[above] at (-1.5,0) {$\textcolor{myred}{\zeta_0}$};
		\node[above] at (0.5,0) {$\textcolor{myred}{\zeta_0}$};
		\node[above] at (-0.5,0) {$\textcolor{myred}{\zeta_0}$};
		\node[below] at (0,-1.5) {$\textcolor{ceruleanblue}{\lambda}$};
		\node[below] at (1.62,-0.28) {$\textcolor{ceruleanblue}{\lambda}$};
		\node at (2.5,-1) {$\textcolor{black}{+}$}; 
		\end{tikzpicture}
		\begin{tikzpicture}
		\draw[-,>=stealth,very thick] (-2,0)--(2,0);
		\draw[-,>=stealth,thick] (1.5,0) arc (0:-180:1.5cm);
		\draw[-,>=stealth,thick] (1.8,0) arc (0:-180:0.3cm);
		\draw[-,>=stealth,thick] (-1.2,0) arc (0:-180:0.3cm);
		\draw[-,>=stealth,thick] (0.5,0) sin (0,-1.5) cos (-0.5,0);
		
		\node[above] at (1.5,0) {$\textcolor{myred}{\zeta_0^3}$}; 
		\node[above] at (-1.5,0) {$\textcolor{myred}{\zeta_0^3}$};
		\node[above] at (0.5,0) {$\textcolor{myred}{\zeta_0}$};
		\node[above] at (-0.5,0) {$\textcolor{myred}{\zeta_0}$};
		\node[below] at (0,-1.5) {$\textcolor{ceruleanblue}{\lambda}$};
		\node[below] at (1.62,-0.28) {$\textcolor{ceruleanblue}{\lambda}$};
		\node[below] at (-1.62,-0.28) {$\textcolor{ceruleanblue}{\lambda}$};
		\node at (2.5,-1) {$\textcolor{black}{+}$}; 
		\node at (3.3,-1) {$\textcolor{black}{\ldots}$};
		\end{tikzpicture}
\caption{~} \label{fig:TreeWitten_resum}
	\end{subfigure}
	
	\bigskip
	
	\begin{subfigure}{1.0\textwidth}
		\centering
		\begin{tikzpicture}
		\draw[-,>=stealth,very thick] (-2,0)--(2,0);
		\draw[-,>=stealth,thick] (1.5,0) arc (0:-180:1.5cm);
		\draw[-,>=stealth,thick] (0.2,-1.57) arc (45:405:0.25);
		\node[above] at (1.5,0) {$\textcolor{myred}{\zeta_0}$}; 
		\node[above] at (-1.5,0) {$\textcolor{myred}{\zeta_0}$};
		\node[above] at (0,-1.5) {$\textcolor{ceruleanblue}{\lambda}$};
		\node at (2.5,-1) {$\textcolor{black}{+}$};
		\end{tikzpicture}
		\begin{tikzpicture}
		\draw[-,>=stealth,very thick] (-2,0)--(2,0);
		\draw[-,>=stealth,thick] (1.5,0) sin (1.2,-1.5) cos (0.9,0) ;
		\draw[-,>=stealth,thick] (-1.5,0) sin (-1.2,-1.5) cos (-0.9,0);
		\draw[-,>=stealth,thick] (-1.2,-1.5) sin (0,-1) cos (1.2,-1.5);
		\draw[-,>=stealth,thick] (-1.2,-1.5) sin (0,-2) cos (1.2,-1.5);
		\node[above] at (1.5,0) {$\textcolor{myred}{\zeta_0}$}; 
		\node[above] at (-1.5,0) {$\textcolor{myred}{\zeta_0}$};
		\node[above] at (0.9,0) {$\textcolor{myred}{\zeta_0}$};
		\node[above] at (-0.9,0) {$\textcolor{myred}{\zeta_0}$};
		\node[below] at (-1.3,-1.4) {$\textcolor{ceruleanblue}{\lambda}$};
		\node[below] at (1.3,-1.4) {$\textcolor{ceruleanblue}{\lambda}$};
		\node at (2.5,-1) {$\textcolor{black}{+}$};
		\end{tikzpicture}
		\begin{tikzpicture}
		\draw[-,>=stealth,very thick] (-2,0)--(2,0);
		\draw[-,>=stealth,thick] (1.5,0) sin (1.2,-1.5) cos (0.9,0) ;
		\draw[-,>=stealth,thick] (-1.5,0) sin (-1.2,-1.5) cos (-0.9,0);
		\draw[-,>=stealth,thick] (0.3,0) sin (0,-1) cos (-0.3,0); 
		\draw[-,>=stealth,thick] (-1.2,-1.5) sin (0,-1) cos (1.2,-1.5);
		\draw[-,>=stealth,thick] (-1.2,-1.5) sin (0,-2) cos (1.2,-1.5);
		\node[above] at (1.5,0) {$\textcolor{myred}{\zeta_0}$}; 
		\node[above] at (-1.5,0) {$\textcolor{myred}{\zeta_0}$};
		\node[above] at (0.9,0) {$\textcolor{myred}{\zeta_0}$};
		\node[above] at (-0.9,0) {$\textcolor{myred}{\zeta_0}$};
		\node[above] at (0.3,0) {$\textcolor{myred}{\zeta_0}$};
		\node[above] at (-0.3,0) {$\textcolor{myred}{\zeta_0}$};
		\node[below] at (-1.3,-1.4) {$\textcolor{ceruleanblue}{\lambda}$};
		\node[below] at (0,-1) {$\textcolor{ceruleanblue}{\lambda}$};
		\node[below] at (1.3,-1.4) {$\textcolor{ceruleanblue}{\lambda}$};
		\node at (2.5,-1) {$\textcolor{black}{+}$}; 
		\node at (3.3,-1) {$\textcolor{black}{\ldots}$};
		\end{tikzpicture}
\caption{~} \label{fig:LoopWitten_resum}
	\end{subfigure}
	
	\bigskip
	
	\begin{subfigure}{1.0\textwidth}
		\centering
		\begin{tikzpicture}
		\draw[-,>=stealth,very thick] (-2,0)--(2,0);
		\draw[-,>=stealth,thick] (1.5,0) arc (0:-180:1.5cm);
		\draw[-,>=stealth,thick] (0.5,0) sin (0,-1.5) cos (-0.5,0);
		\draw[-,>=stealth,thick] (-0.9,-0.6) arc (45:405:0.4);
		\node[above] at (1.5,0) {$\textcolor{myred}{\zeta_0}$}; 
		\node[above] at (-1.5,0) {$\textcolor{myred}{\zeta_0}$};
		\node[above] at (0.5,0) {$\textcolor{myred}{\zeta_0}$};
		\node[above] at (-0.5,0) {$\textcolor{myred}{\zeta_0}$};
		\node[below] at (0,-1.5) {$\textcolor{ceruleanblue}{\lambda}$};
		\node[below] at (-0.9,-1.2) {$\textcolor{ceruleanblue}{\lambda}$};
		\node[left] at (-1.4,-0.45) {$\textcolor{ceruleanblue}{\lambda}$};
		\node at (2.5,-1) {$\textcolor{black}{+}$}; 
		\end{tikzpicture}
		\begin{tikzpicture}
		\draw[-,>=stealth,very thick] (-2,0)--(2,0);
		\draw[-,>=stealth,thick] (1.5,0) sin (1.2,-1.5) cos (0.9,0) ;
		\draw[-,>=stealth,thick] (-1.5,0) sin (-1.2,-1.5) cos (-0.9,0);
		\draw[-,>=stealth,thick] (-1.2,-1.5) sin (-0.6,-1) cos (0,-1.5);
		\draw[-,>=stealth,thick] (-1.2,-1.5) sin (-0.6,-2) cos (0,-1.5);
		\draw[-,>=stealth,thick] (0,-1.5) sin (0.6,-1) cos (1.2,-1.5);
		\draw[-,>=stealth,thick] (0,-1.5) sin (0.6,-2) cos (1.2,-1.5);
		\node[above] at (1.5,0) {$\textcolor{myred}{\zeta_0}$}; 
		\node[above] at (-1.5,0) {$\textcolor{myred}{\zeta_0}$};
		\node[above] at (0.9,0) {$\textcolor{myred}{\zeta_0}$};
		\node[above] at (-0.9,0) {$\textcolor{myred}{\zeta_0}$};
		\node[below] at (-1.3,-1.4) {$\textcolor{ceruleanblue}{\lambda}$};
		\node[below] at (1.3,-1.4) {$\textcolor{ceruleanblue}{\lambda}$};
		\node[below] at (0,-1.5) {$\textcolor{ceruleanblue}{\lambda}$};
		\node at (2.5,-1) {$\textcolor{black}{+}$};
		\end{tikzpicture}
		\begin{tikzpicture}
		\draw[-,>=stealth,very thick] (-2,0)--(2,0);
		\draw[-,>=stealth,thick] (1.5,0) sin (1.2,-1.5) cos (0.9,0) ;
		\draw[-,>=stealth,thick] (-1.5,0) sin (-1.2,-1.5) cos (-0.9,0);
		\draw[-,>=stealth,thick] (-1.2,-1.5) sin (-0.8,-1) cos (-0.4,-1.5);
		\draw[-,>=stealth,thick] (-1.2,-1.5) sin (-0.8,-2) cos (-0.4,-1.5);
		\draw[-,>=stealth,thick] (-0.4,-1.5) sin (0,-1) cos (0.4,-1.5);
		\draw[-,>=stealth,thick] (-0.4,-1.5) sin (0,-2) cos (0.4,-1.5);
		\draw[-,>=stealth,thick] (0.4,-1.5) sin (0.8,-1) cos (1.2,-1.5);
		\draw[-,>=stealth,thick] (0.4,-1.5) sin (0.8,-2) cos (1.2,-1.5);
		\node[above] at (1.5,0) {$\textcolor{myred}{\zeta_0}$}; 
		\node[above] at (-1.5,0) {$\textcolor{myred}{\zeta_0}$};
		\node[above] at (0.9,0) {$\textcolor{myred}{\zeta_0}$};
		\node[above] at (-0.9,0) {$\textcolor{myred}{\zeta_0}$};
		\node[below] at (-1.3,-1.4) {$\textcolor{ceruleanblue}{\lambda}$};
		\node[below] at (1.3,-1.4) {$\textcolor{ceruleanblue}{\lambda}$};
		\node[below] at (-0.4,-1.6) {$\textcolor{ceruleanblue}{\lambda}$};
		\node[below] at (0.4,-1.6) {$\textcolor{ceruleanblue}{\lambda}$};
		\node at (2.5,-1) {$\textcolor{black}{+}$};
		\node at (3.3,-1) {$\textcolor{black}{\ldots}$};
		\end{tikzpicture}
\caption{~} \label{fig:2LoopWitten_resum}
	\end{subfigure}
	\caption{~In the first row (Figure~\ref{fig:TreeWitten_resum}) tree-level Witten diagrams; these are captured by the semi-classical method. In the second row (Figure~\ref{fig:LoopWitten_resum}) one-loop diagrams; these would be captured by the (one-loop) prefactor in the semi-classical method. In the third row (Figure~\ref{fig:2LoopWitten_resum}) higher-loop diagrams; these are only captured at subleading order in the semi-classical calculation.}
\end{figure}

The semiclassical approach effectively resums a subset of diagrams of the standard in-in perturbation theory.
In $\rm dS$ space, $\Psi[\zeta_0(\vect x)]$ is defined by the path integral of eq.~\eqref{eq:WFU} where one imposes Dirichlet boundary conditions for $\zeta$ at late times. This path integral can be conveniently computed in perturbation theory as a sum of Witten diagrams  (see for example \cite{Anninos:2014lwa}).
The tree-level Witten diagrams of Figure~\ref{fig:TreeWitten_resum} have the same scaling as the lowest-order term in the semiclassical expansion, which corresponds to the on-shell action \eqref{eq_zeta'^4:semiclassical_action_scaling}. This is immediate to realize since for any additional vertex we add we increase the number of boundary legs by two. Thus, a tree-level diagram with $V$ vertices scales as $\sim \frac{1}{\lambda} (\lambda \zeta_0^2)^{V+1}$. 
The subleading order in $\lambda$ in the semiclassical expansion is instead obtained through a one-loop calculation around a non-trivial background for $\zeta$ (this corresponds to the calculation of the prefactor in eq.~\eqref{eq:QM_prefactor} in our quantum mechanical example). The scaling of this factor is $\lambda^0 G(\lambda \zeta_0^2)$, which corresponds to the scaling of the one-loop Witten diagrams of Figure~\ref{fig:LoopWitten_resum}, while the diagrams of Figure~\ref{fig:2LoopWitten_resum} are computed by two- or higher- loop calculations around the semiclassical solution for $\zeta$. 

Before moving to the actual calculation of the action, it is useful to comment on the choice of the asymptotic value $\zeta_0(\vect x)$. To answer a concrete question, like the probability of producing a PBH, one would be interested in evaluating the WFU for all functions that are above a certain threshold\footnote{The threshold is of course an approximate concept: one should know the precise boundary in the functional space $\zeta_0(\vect x)$ that separates the configurations giving rise to a PBH from the ones that do not. }. More specifically, as we discussed in the introduction, one would consider a filtered field $\hat \zeta_0(\vect x)$ and require that this field exceeds a certain numerical threshold at a point of interest. In the limit of a very high threshold all configurations $\zeta_0(\vect x)$ that are above threshold have a ``large $\zeta_0$" and as such the WFU can be calculated semiclassically. Of course, to get to the final answer one should eventually sum over all $\zeta_0(\vect x)$ that are above threshold. This final integral can also be done in saddle-point approximation: since the probability of all interesting configurations is small, the integral will be dominated by the least unlikely. In this paper we do not want to commit to a very specific question, which would require the details of the window function and the threshold. We are going simply to choose a given $\zeta_0(\vect x)$ and take it large enough for our approach to be applicable. Since the question we are addressing is not completely specified, we will be mostly interested in the behaviour of the probability as a function of the parameter $\lambda \zeta_0^2/P_\zeta$, especially once this becomes large. We leave the actual implementation of these techniques to the calculation of the PBH abundance to future work.

\subsection{ODE approximation}\label{subsec_zeta'^4:ODE_approximation}

The qualitative behaviour of the action as a function of the boundary values of $\zeta$ can be understood focussing on a single Fourier mode and thus reducing the problem to an ordinary differential equation (ODE). In perturbation theory the interaction $\zeta'^4$ induces coupling mainly among modes with comparable wavelength: this is the reason why one gets non-Gaussianities of ``equilateral" kind \cite{Babich:2004gb}. For the same reason one expects that if the boundary condition at late times $\zeta_0(\vect x)$ has Fourier transform concentrated on a typical value $k$ (\footnote{For instance PBH with a certain mass will typically form at a given time and the modes of interest will be the ones with wavelength comparable with the Hubble radius at that moment.}) then only modes with similar wavelength will be relevant in the full solution $\zeta_{\rm cl}(\vect x, \tau)$.   Therefore, one can concentrate on a single Fourier mode and eq.~\eqref{eq:PDE phi_lambda4} reduces to the following ODE 
\begin{align}\label{eq:ODE phi_lambda4}
-\zeta'' + \frac{2}{\tau}\zeta' + H^2 \zeta - \frac{\lambda}{2 P_\zeta}\tau^2\zeta'^2\zeta'' = 0\;,
\end{align}
where we have set $ k / H = 1$, using scale-invariance. The boundary conditions we need to impose are 
\begin{align}
\zeta (\tau \to - \infty) = 0 \;, \quad \zeta(\tau_{\rm f}) = \zeta_0 \;,
\end{align}
where $\tau_{\rm f}$ is the final conformal time.\footnote{Since the field goes to zero at early times, the free theory becomes a good approximation. Numerically we implement the boundary condition at an early time $\tau_{\rm i}$ as
\be\label{sec4:condition_tau_i}
\zeta'(\tau_{\text{i}}) = \frac{ H \tau_{\text{i}}}{H \tau_{\text{i}} - 1} \zeta(\tau_{\text{i}}) \;,
\ee
which is the relation between the field and the derivative in the free theory.}
This approximation, as we are going to argue, is useful in order to obtain an analytic understanding of the scaling of $S$ as a function of $\lambda$ and $\zeta_0$. In Section~\ref{subsec_zeta'^4:PDE} we will instead solve the full PDE and we will compare with the results of the ODE approximation. 

In order to solve eq.~\eqref{eq:ODE phi_lambda4} numerically and compare different solutions, it is convenient to rescale $\zeta \rightarrow \zeta_0 \zeta$ so that one has $\zeta(\tau_{\rm f}) = 1$. Moreover, one can define 
\be
\tilde \lambda \equiv \lambda \zeta_0^2 / P_\zeta \;,
\ee
this is the parameter that quantifies the classical non-linearities, the analogue of $\bar x$ in the quantum mechanical example of Section~\ref{sec:QM}. In this way the EoM \eqref{eq:ODE phi_lambda4} keeps the same form with $\lambda$ replaced by $\tilde \lambda$ and $P_\zeta$ set to 1, while the action rescales as $S \rightarrow (\zeta_0^2/ P_\zeta) S$.  Our analysis will be exact in $\tilde\lambda$, but perturbative in $\lambda \ll 1$: this separation requires $\zeta_0^2/P_\zeta \gg 1$, corresponding to $\tilde\lambda \gg \lambda$. (For instance at first order in $\lambda$ we are keeping the first graph of Figure~\ref{fig:TreeWitten_resum}, but we are dropping the first diagram of Figure~\ref{fig:LoopWitten_resum}. The first is larger than the second by a factor $\zeta_0^2/P_\zeta$.)

\begin{figure}[t!]
	\begin{subfigure}{0.48\textwidth}
		\includegraphics[width=\linewidth]{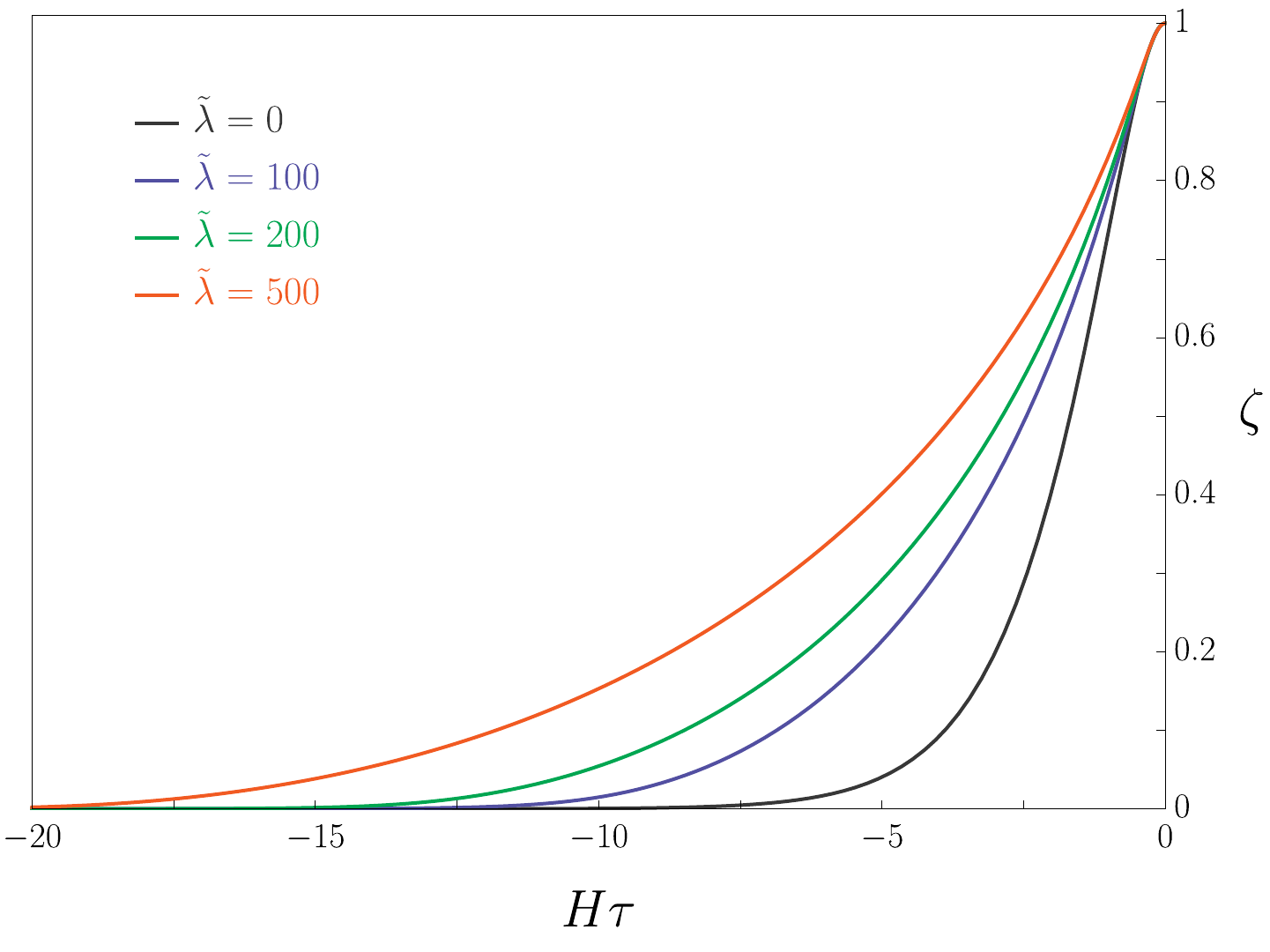}
		\caption{\label{fig:Sol_ODE_zeta4}} 
	\end{subfigure}
	\hspace*{\fill} 
	\begin{subfigure}{0.48\textwidth}
		\includegraphics[width=\linewidth]{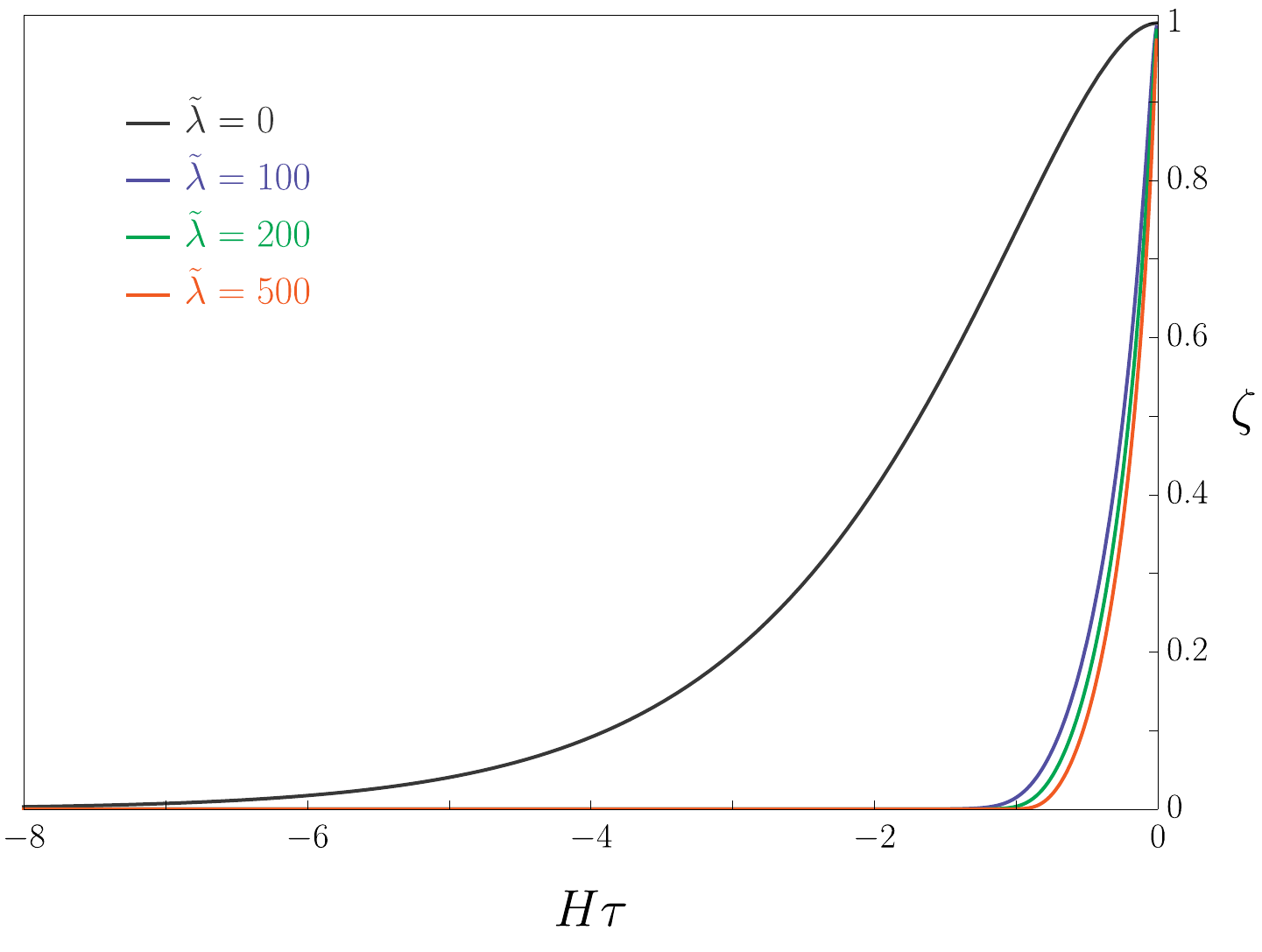}
		\caption{\label{fig:Sol_ODE_rescale_zeta4}} 
	\end{subfigure}
	\hspace*{\fill} 
	\caption{~Left panel (Figure~\ref{fig:Sol_ODE_zeta4}): the numerical solutions for $\tilde\lambda = \{0, 100, 200, 500\}$ and $\zeta_0 = 1$. Right panel (Figure~\ref{fig:Sol_ODE_rescale_zeta4}): the same solutions after $\tau \rightarrow \sqrt{\tilde \lambda}\tau$ (the solution for $\tilde\lambda =0$ is copied for comparison).} 
\end{figure}

The numerical solutions are shown in Figure~\ref{fig:Sol_ODE_zeta4} (we set $\zeta(\tau_{\rm f})= 1$ and $H \tau_{\rm f} = -0.001$).  Notice that the non-linear interaction acts as a sort of non-linear friction so that the solution varies more slowly as $\tilde\lambda$ increases. (Thus one has to correspondingly adjust the value of $H \tau_{\rm i}$ to earlier and earlier values.)
We notice that the solutions approach a universal behaviour for large $\tilde \lambda$ that can be obtained by rescaling $\tau \rightarrow \sqrt{\tilde \lambda}\tau$. The rescaled solutions are illustrated in Figure~\ref{fig:Sol_ODE_rescale_zeta4}. We will come back to this point in the next Section.

Now let us evaluate the on-shell action. First we note that, as in the case of a free field in ${\rm dS}$ discussed above, the action (in particular the free gradient energy) gives a singularity for $\tau \rightarrow 0$.  This can be completely removed since its contribution to the wavefunction is purely imaginary:
\begin{equation}\label{eq_zeta'^4:regularized_action_ODE}
\Delta S_{\rmS ODE} = -\frac{\zeta_0^2}{P_\zeta}\int_{\tau_{\rm i}}^{\tau_{\rm f}} d \tau \left\{ \frac{1}{2 \tau^{2}} \bigg[ \zeta'^{2} + H^2 ( \zeta^2 - 1 ) \bigg] + \frac{\tilde \lambda}{4!} \zeta'^{4}\right\} = \frac{1}{\lambda} F(\tilde\lambda) \;.
\end{equation}
We have subtracted $1$, the asymptotic value of $\zeta$, inside the innermost parentheses. This additional term gives a term proportional to $1/\tau_{\rm f}$ after integration and this becomes purely imaginary after rotation to $\eta$. Therefore, the extra term does not contribute to the probability of $\zeta$. The advantage is that, after this subtraction, the action is finite and can be treated numerically.

\begin{figure}[t!]
	\centering
	\includegraphics[width=0.6\linewidth]{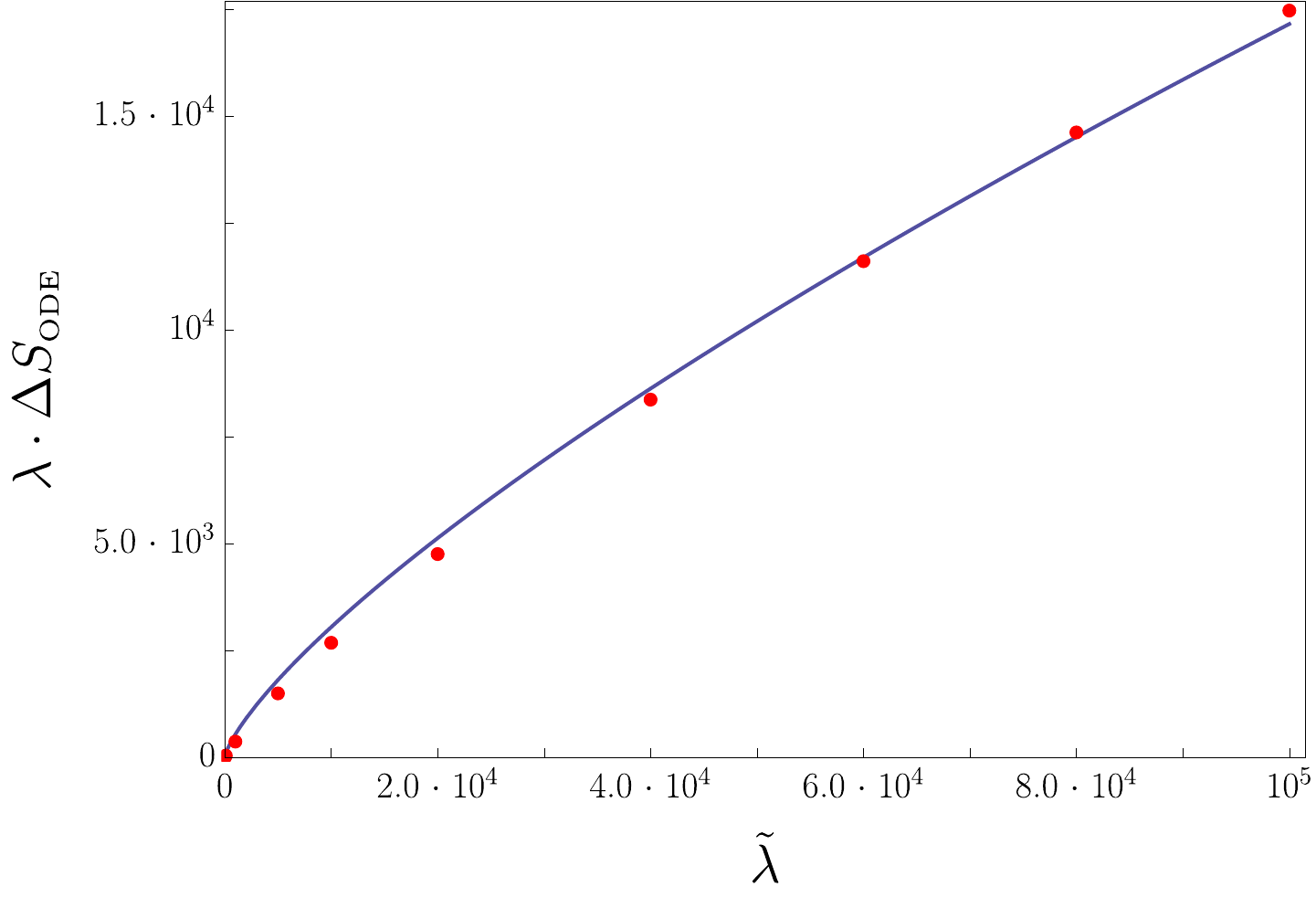}
	\caption{~The on-shell action as a function of the expansion parameter $\tilde{\lambda} = \lambda\zeta_0^2/P_\zeta$. The blue curve is the best fit curve proportional to $\tilde\lambda^{3/4}$. The red points indicate the numerical values of $\lambda\cdot\Delta S_{\rmS ODE}$.}  
	\label{fig:action_ODE_zeta4}
\end{figure}

The behaviour of the on-shell action evaluated on the numerical solutions is given in Figure~\ref{fig:action_ODE_zeta4}. It shows that the on-shell action $\Delta S_{\rmS ODE} \sim \frac{1}{\lambda}\tilde \lambda^{3/4}$ for large $\tilde \lambda$.  
The real part of the WFU therefore behaves as 
\begin{align}\label{sec4:WFU_ODE}
\Psi[\zeta_0] \sim \exp\left[-\frac{1}{\lambda}\tilde{\lambda}^{3/4}\right] \;,
\end{align}
with some unspecified order one coefficient in front of the exponent. The WFU is multiplied by a time-dependent phase, the same as in the free theory eq.~\eqref{sec4:onshell_free_action}, which does enter in the calculation of the correlation functions of $\zeta$.

\subsection{Analytic understanding of the ODE result}\label{subsec_zeta'^4:ODE_analytic_approx}

\begin{figure}[t!]
	\centering
	\begin{tikzpicture}[xscale=8]
	
	\fill[fill=myblue!100,opacity=0.8] (0.1,0) rectangle (0.5,1); 
	\fill[fill=myblue!100,opacity=0.8] (1.2,0) rectangle (1.6,1); 
	\fill[fill=ceruleanblue!100,opacity=0.65] (0.5,0) rectangle (1.2,1); 
	\draw[->,>=stealth][draw=black, very thick] (0,0) -- (1.8,0); 
	
	\node[above] at (0.3,0.2) {I: Free}; 
	\node[above] at (1.4,0.2) {III: Free}; 
	\node[above] at (0.85,0.2) {II: Non-linear}; 
	\draw [thick] (0.5,-.1) node[below]{\ \hspace{0.1cm} $\textcolor{myred}{H \tau_1}$} -- (0.5,0.1);
	\draw [thick] (1.2,-.1) node[below]{\ \hspace{0.5cm} $\textcolor{myred}{H \tau_2} \sim -1/\tilde{\lambda}^{1/4}$} -- (1.2,0.1);
	\draw [thick] (1.6,-.1) node[below]{$\tau = 0$} -- (1.6,0.1);
	\node[above] at (1.8,0.1) {$\tau$}; 
	
	\end{tikzpicture}
	\caption{:~Three regimes of the solution.} \label{fig:NLsolution}
\end{figure}
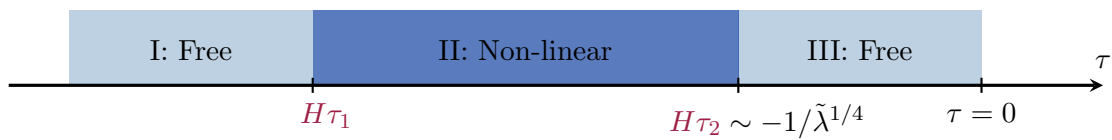

In this Section we show that the behaviour of the ODE for large $\tilde\lambda$ that we found numerically can also be understood analytically.  First we notice that there are three regimes for the ODE solution, summarized in Figure~\ref{fig:NLsolution}.
At very early times $\tau\to - \infty$, the field must approach the BD vacuum, so its amplitude is exponentially small. Therefore, in this regime the interaction term becomes negligible and we approach a free solution. We define $\tau_1$ as the earliest time at which the interaction term is comparable with the free time kinetic term. For $\tau < \tau_1$ the solution is approximatively free (region I of the Figure), while for $\tau > \tau_1$ we enter the non-linear regime (region II). 
\vspace{.3cm}

At very late times, $\tau \rightarrow 0^-$, the interaction term becomes subdominant once again since it contains more derivatives than the kinetic term. Thus the solution becomes free (region III) for times $\tau > \tau_2$. In region III the solution is approximatively given by
\begin{equation}\label{eq:approx_region_III}
\zeta_{\rmS III} \simeq (1-  H \tau ) e^{H \tau },
\end{equation}
since our boundary condition, after rescaling, is $\zeta(\tau_{\rm f}) = 1$. From this we can estimate $\tau_{2}$ as the time when $\zeta''$ and the non-linear term in eq.~\eqref{eq:ODE phi_lambda4} are of the same order
\begin{equation}
\tilde \lambda \tau_{2}^2 \zeta_{\rmS III}'(\tau_2)^2 \sim 1\;.
\end{equation}
Since this happens after horizon crossing, we can expand this equation at lowest order in $\tau_{2}$. We obtain
\begin{equation}
- H \tau_2 \sim \frac{1}{\tilde \lambda^{1/4}}\;,
\end{equation}
which for large $\tilde \lambda$ is consistent with the expansion we performed. In this approximation we can take $\zeta_{\rmS III} \sim 1$ at $\tau_{2}$. 

Now we can focus on region II. In this regime one expects the non-linear term to dominate over the kinetic term and Hubble friction, and to be compensated in the equation of motion by the spacial kinetic term. To see this let us consider eq.~\eqref{eq:ODE phi_lambda4} and rescale $\tau = \sqrt{\tilde \lambda} \tilde \tau$. We obtain 
\begin{align}
\frac{1}{\tilde \lambda}\bigg(-\ddot{\zeta} + \frac{2}{\tilde{\tau}}\dot{\zeta}\bigg) + H^2 \zeta - \frac{1}{2}\tilde{\tau}^2\dot{\zeta}^2\ddot{\zeta} = 0 \;,
\end{align} 
where $\dot{\zeta} \equiv d \zeta/d \tilde{\tau}$. We see that, when $\tilde \lambda$ is large, the first and the second terms can be neglected compared to the rest (one can check this in the numerical solutions).  Therefore, in this regime one has
\begin{align}
H^2 \zeta_{\rmS II} - \frac{\tilde \lambda}{2}{\tau}^2{\zeta'}_{\rmS II}^2{\zeta}_{\rmS II}'' = 0 \;. \label{eq:non_linear_regime_ODE}
\end{align}
This explains why in the previous Section we found a universal solution as a function of $\tau/\sqrt{\tilde \lambda}$.
This equation does not have an analytic solution. However, let us now assume that $\zeta \sim 1$ for small $\tau$ (as we argued from the behaviour in region III). Then, from (\ref{eq:non_linear_regime_ODE}) $\zeta_{\rmS II}$, $\zeta'_{\rmS II}$ and $\zeta''_{\rmS II}$ can be approximated, neglecting factors of order one, as 
\begin{align}\label{eq:approx_region_II}
\zeta_{\rmS II} \sim 1 - \frac{H^{2/3} |\tau|^{2/3}}{\tilde \lambda ^{1/3}} \;, \quad \zeta'_{\rmS II} \sim \frac{H^{2/3}}{(\tilde \lambda |\tau|)^{1/3}} \;, \quad \zeta_{\rmS II}'' \sim \frac{H^{2/3}}{\tilde \lambda^{1/3}{|\tau|}^{4/3}} \;. 
\end{align}
As we increase $\tilde \lambda$, the time dependence of the solution becomes milder. In comparison, the free solution decays exponentially when moving to earlier times. This analytic behaviour is in agreement with the numerical results of Figure~\ref{fig:Sol_ODE_zeta4}. Physically this is expected: for the model at hand the non-linearities have the effect of increasing the time kinetic term, hence reducing the forcing term. In Euclidean time this induces a slower decay at early times, and in the limit $\tilde \lambda \rightarrow \infty$ one recovers $\zeta \sim 1$ at any $\tau$. 

The solution in region I is again free, proportional to eq.~\eqref{eq:approx_region_III}, but with a different normalization. Time $\tau_1$ is approximately given by
\begin{equation}\label{eq:estimate_eta_1_ode}
\tilde \lambda \tau_1^2 \zeta'_{\rmS II}(\tau_1)^2 \sim 1\;.
\end{equation}
We do not know how to estimate $\tau_1$, since we do not have an analytic expression of $\zeta_{\rmS II}$. However, since for large enough $\tilde\lambda$, $\zeta_{\rmS II}$ is a function of $\tau/\sqrt{\tilde\lambda}$, one can argue that $|\tau_1|$ grows at least as fast as $\sqrt{\tilde\lambda}$, and this is sufficient to estimate the action. (Numerically one finds actually $|\tau_1| \sim \sqrt{\tilde\lambda}$.)

With these observations at hand we are now ready to estimate the on-shell action. We note that in the limit $\tilde \lambda \rightarrow \infty$ the regularized action \eqref{eq_zeta'^4:regularized_action_ODE} approaches zero (since we have $\zeta \sim 1$ and $\zeta' \sim 0$). On the other hand, for small $\tilde \lambda$ we need to recover the free result $\Delta S_{\rmS ODE}^{\rm free} = \zeta_0^2/(2 P_\zeta)$. Therefore, one expects $\Delta S_{\rmS ODE}$ to be a decreasing function of $\tilde \lambda$.

Let us assume momentarily that the contribution to the action of region I is negligible. Using the approximate solutions for regions II and III, eq.~\eqref{eq:approx_region_III} and \eqref{eq:approx_region_II}, we obtain
\begin{align}\label{eq:ODEfinal}
\Delta S_{\rmS ODE} & = -\frac{\zeta_0^2}{P_\zeta}\int_{-\infty}^{0} d \tau \left\{ \frac{1}{2 \tau^{2}} \bigg[ \zeta'^{2} + H^2 ( \zeta^2 - 1 ) \bigg] + \frac{\tilde \lambda}{4!} \zeta'^{4}\right\} \nonumber \\
& \!\!\!\!\!\sim -\frac{\zeta_0^2}{P_\zeta} \left\{ \int_{\tau_{\rm 1}}^{\tau_{\rm 2}} \!\!\!d \tau \left[ \frac{1}{2 \tau^{2}} \bigg( \zeta_{\rmS II}'^{2} + H^2( \zeta_{\rmS II}^2 - 1 ) \bigg) + \frac{\tilde \lambda}{4!} \zeta_{\rmS II}'^{4}\right]  + \int_{\tau_{\rm 2}}^{0} \!\!\!d \tau \left[ \frac{1}{2 \tau^{2}} \bigg( \zeta_{\rmS III}'^{2} + H^2( \zeta_{\rmS III}^2 - 1 ) \bigg)+ \frac{\tilde \lambda}{4!} \zeta_{\rmS III}'^{4}\right] \right\}\nonumber \\
& \!\!\!\!\sim \frac{\zeta_0^2}{P_\zeta} \frac{1}{\tilde \lambda^{1/4}}  = \frac{1}{\lambda } (\lambda \zeta_0^{2} / P_\zeta)^{3/4}\;.
\end{align}
Both integrals are dominated by the region around $\tau_2$. One can check that each single term in the action, $\zeta'^2$, $(\zeta^2-1)$ and $\tilde\lambda\zeta'^4$ contributes, both in region II and III, to a term of order $\tilde\lambda^{-1/4}$. This result confirms the numerical behaviour found in the previous Section.

To conclude, let us check that the contribution of region I is actually negligible. In this region we can use free modes (whose normalization, however, we do not know) to integrate the action. The integral of the free action can be written as
\be
-\frac{\zeta_0^2}{P_\zeta}\int_{-\infty}^{\tau_1} d \tau \left\{ \frac{1}{2 \tau^{2}} \bigg[ \zeta_{\rmS I}'^{2} + H^2 \zeta_{\rmS I}^2 \bigg] \right\} \sim -\frac{\zeta_0^2}{P_\zeta} \frac{H \zeta_{\rmS I}'^2(\tau_1)}{\tau_1^2}\;.
\ee
Using eq.~\eqref{eq:estimate_eta_1_ode} and that $|\tau_1|$ increases at least as fast as $\sqrt{\tilde\lambda}$, one sees that this term goes as $\tilde\lambda^{-3}$, and it is thus subleading compared with eq.~\eqref{eq:ODEfinal}. One gets the same estimate for the contribution of the term $\tilde\lambda\zeta'^4$. The integral of the $\zeta$-independent term in the action, $-H^2/(2\tau^2)$, gives a contribution of order $1/\tau_1 \sim \tilde\lambda^{-1/2}$, which is also subleading.


\subsection{PDE analysis}\label{subsec_zeta'^4:PDE}
In the last two Sections we have seen how the tail of the WFU can be estimated (as a function of the parameter $\tilde{\lambda}$) assuming that all the modes have comparable wavelength: the PDE was reduced to an ODE. The ODE is easy to treat numerically (Section~\ref{subsec_zeta'^4:ODE_approximation}) and one can also provide an analytic understanding of the numerical result (Section~\ref{subsec_zeta'^4:ODE_analytic_approx}). However, in this way one can only capture the qualitative dependence on $\tilde\lambda$ and not the constants of order unity: if one wants to use our semi-classical method to answer some specific questions, e.g.~compute the PBH abundance, the full PDE analysis is required. We are now going to study the PDE and check that the ODE treatment correctly captured the qualitative behaviour in $\tilde\lambda$ and that this result holds quite generally as we change the space dependence of the boundary condition $\zeta_0(\vect x)$ (\footnote{In the model we are studying there is a neat separation: the Euclidean action has an overall dependence on the amplitude of $\zeta_0(\vect x)$, i.e.~$\propto \zeta_0^{3/2}$, and a subleading dependence on the precise shape of $\zeta_0(\vect x)$. It is not obvious a priori that the same will hold for any possible interaction.}). In particular we are going to make two choices for $\zeta_0(\vect x)$: a sinusoidal wave and a spherically symmetric Gaussian profile.

Let us first start with the sinusoidal case. For simplicity, we impose the conditions 
\begin{align}\label{sec4:sine_profile}
\zeta(\tau_{\rm{f}},x) = \zeta_0 \sin(kx) \;,
\end{align}
with $kx \in [0,2\pi]$. Notice that the action we are considering is even in $\zeta$ so that it is consistent to impose that $\zeta$ vanishes at $0$ and $2 \pi$ for any $\tau$. Also in the PDE case we implement the initial condition (\ref{sec4:condition_tau_i}) and take $k =H$ using scale-invariance. Like the ODE Section, we redefine $\zeta \rightarrow \zeta_0 \zeta$ so that the condition (\ref{sec4:sine_profile}) becomes $\zeta(\tau_{\rm{f}},x) = \sin(kx)$.  

\begin{figure}[t!]
	\begin{subfigure}{0.48\textwidth}
		\includegraphics[width=\linewidth]{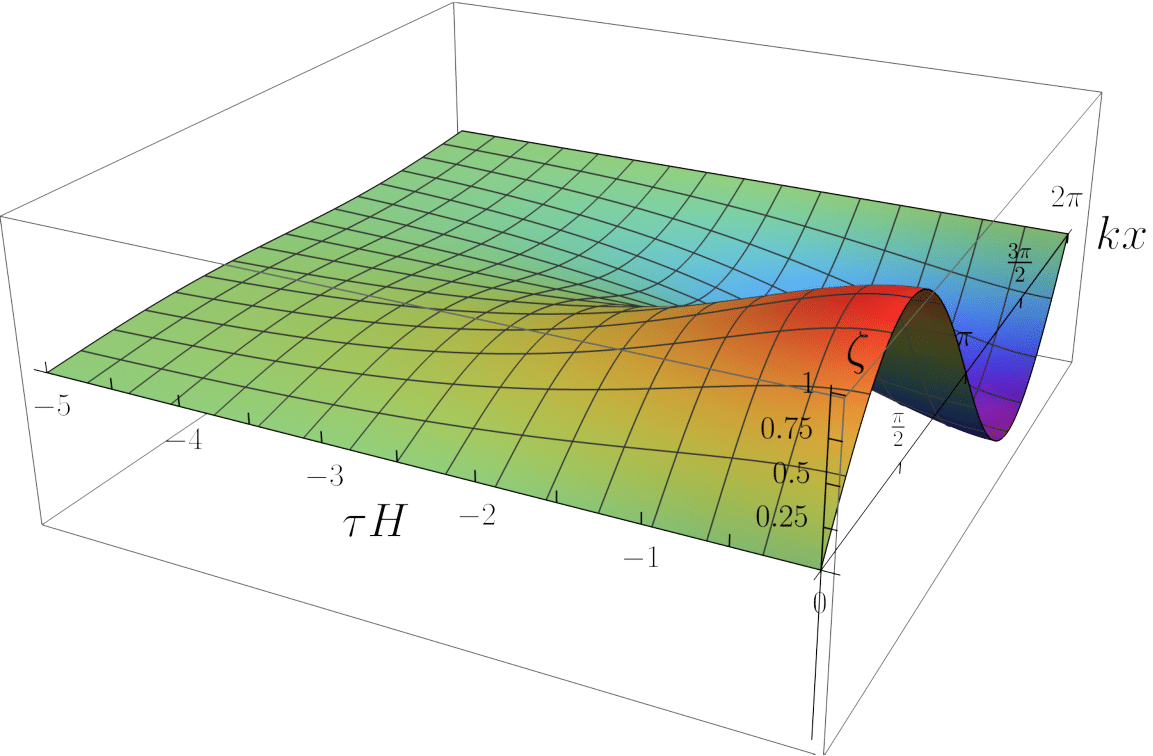}
		\caption{\label{fig:Sol_PDE_sine_lambda0_zeta4}} 
	\end{subfigure}
	\hspace*{\fill} 
	\begin{subfigure}{0.48\textwidth}
		\includegraphics[width=\linewidth]{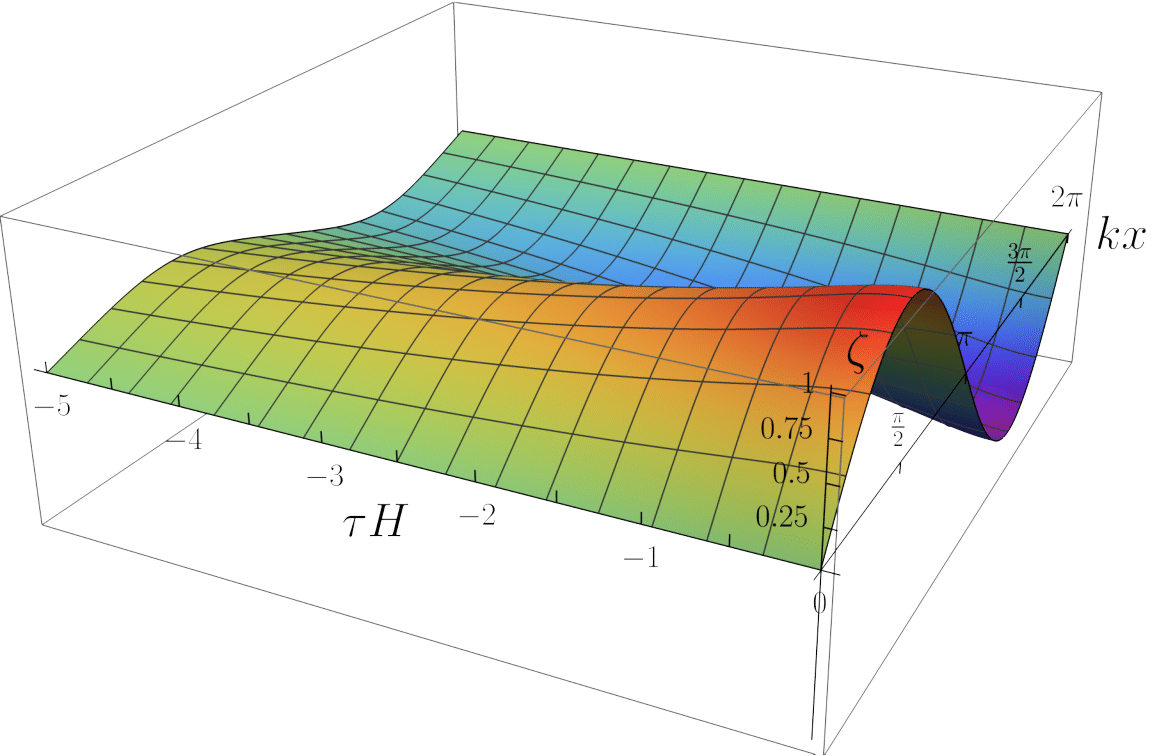}
		\caption{\label{fig:Sol_PDE_sine_lambda200_zeta4}} 
	\end{subfigure}
	\hspace*{\fill} 
	\caption{~Numerical solutions with sinusoidal boundary condition at late times with $\tilde{\lambda} = 0$ (left panel, Figure~\ref{fig:Sol_PDE_sine_lambda0_zeta4}) and $\tilde{\lambda} = 200$ (right panel, Figure~\ref{fig:Sol_PDE_sine_lambda200_zeta4}).} 
\end{figure}

The problem does not depend on $y$ and $z$, thus the PDE (\ref{eq:PDE phi_lambda4}) simplifies to a $1+1$-dimensional problem: 
\begin{align}\label{eq:PDE sine_1+1}
-\zeta'' + \frac{2}{\tau}\zeta' - \frac{\partial^2 \zeta}{\partial x^2} - \frac{\tilde{\lambda}}{2}\tau^2\zeta'^2\zeta'' = 0\;,
\end{align}     
where $\tilde{\lambda}$ was previously defined in Section~\ref{subsec_zeta'^4:ODE_approximation}.  The numerical solutions for $\tilde{\lambda} = 0$ and $\tilde{\lambda} = 200$ are given by Figures~\ref{fig:Sol_PDE_sine_lambda0_zeta4} and \ref{fig:Sol_PDE_sine_lambda200_zeta4}.  (We took $H \tau_{\rm{f}} = -0.001$ and $H \tau_{\rm{i}} = -80$. As in the ODE case the value of $|\tau_{\rm i}|$ must be taken larger and larger as $\tilde\lambda$ increases. We used {\texttt {Mathematica}} for all numerical analysis.)  The plots show that the solution remains smooth going to large $\tilde\lambda$, without generating large higher harmonics: this justifies the use of the ODE as an approximation to the full problem.

\begin{figure}[t!]
	\centering
	\includegraphics[width=0.6\linewidth]{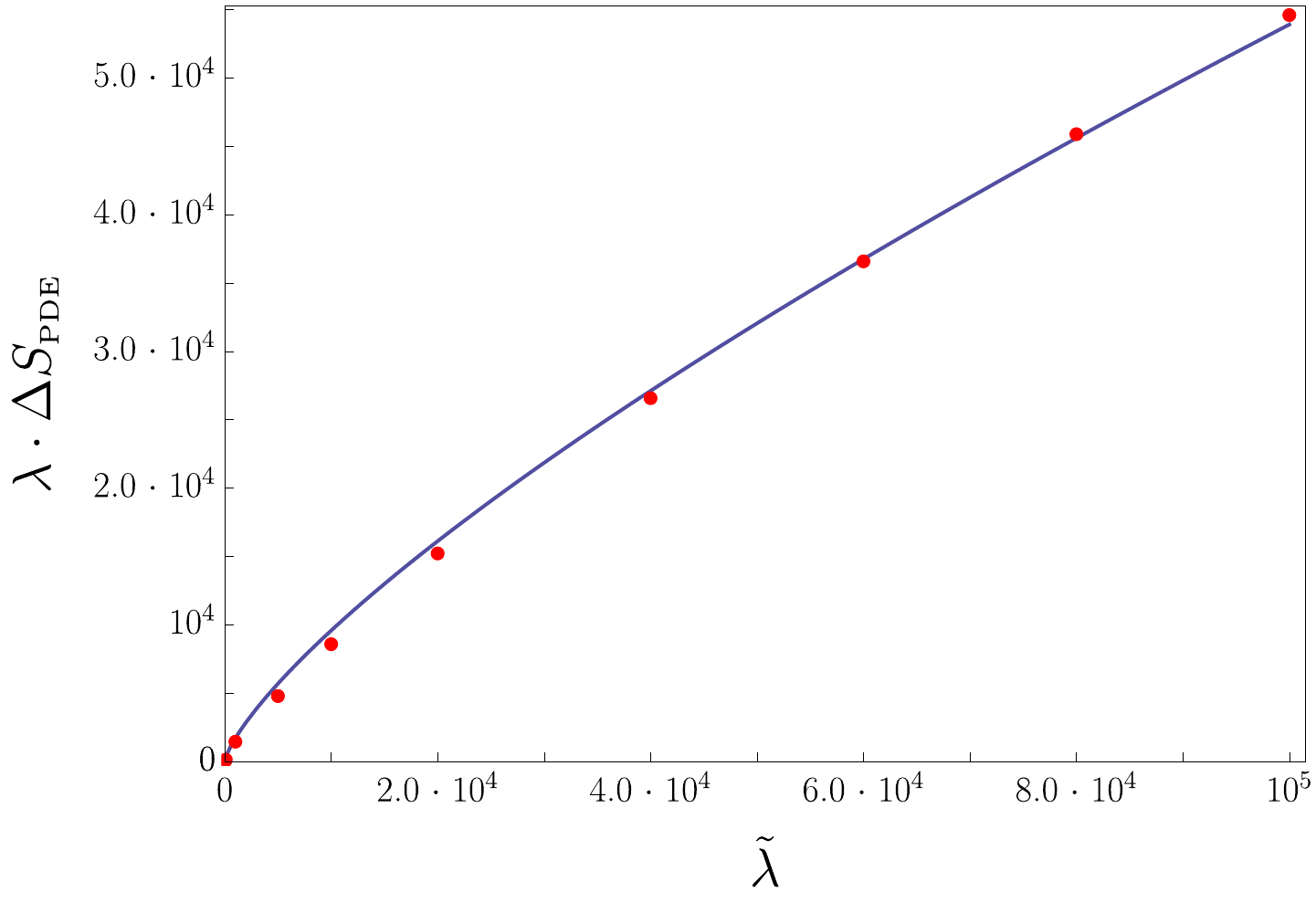}
	\caption{~The function $F(\tilde{\lambda}) = \lambda\cdot\Delta S_{\rmS PDE}$ for the sinusoidal case. The blue curve shows the best fit of $\lambda \cdot \Delta S_{\rmS PDE}$ (red points), proportional to $\tilde{\lambda}^{3/4}$.}  
	\label{fig:action_PDE_Sine_zeta4}
\end{figure}

Let us now compute the Euclidean action evaluated on the numerical solutions. Following the same procedure of the ODE Section, the finite part of the action $\Delta S_{\rmS PDE}$ is given by
\begin{align}\label{sec4:action_PDE_sine}
\Delta S_{\rmS PDE}  = -\frac{\zeta_0^2}{P_\zeta} \int_{\tau_{\rm{i}}}^{\tau_{\rm{f}}} d\tau \int_{x_{\rm i}}^{x_{\rm{f}}} dx \bigg\{\frac{1}{2\tau^2} \bigg[\zeta'^2 + (\partial_{x} \zeta)^2 - k^2 \cos^2(kx) \bigg] + \frac{\tilde{\lambda}}{4!}\zeta'^4\bigg\} = \frac{1}{\lambda}F(\tilde{\lambda})\;. 
\end{align}
The $k^2\cos^2(kx)$ term is added to remove the divergence of the free action at late times (at $\tau = \tau_{\rm{f}}$ one has $(\partial_{x}\zeta)^2 = k^2\cos^2(kx)$). We now evaluate the integral in (\ref{sec4:action_PDE_sine}) numerically on the solutions $\zeta(\tau,x)$ for different values of $\tilde{\lambda}$, starting from $\tilde{\lambda} = 0$ up to $\tilde{\lambda} = 10^5$. We then plot in Figure~\ref{fig:action_PDE_Sine_zeta4} the function $F(\tilde{\lambda}) = \lambda \cdot \Delta S_{\rmS PDE}$ against the parameter $\tilde{\lambda}$. For large $\tilde{\lambda}$ the function $F(\tilde{\lambda})$ approaches $ \tilde{\lambda}^{3/4}$ in agreement with the ODE result (\ref{sec4:WFU_ODE}). 

The advantage of considering a single Fourier mode is that it is easy to check the numerical result with perturbation theory in the limit of small $\tilde\lambda$: we leave this check to Appendix~\ref{sec:perturb}.

Let us now come to the study of the PDE with a Gaussian, spherically symmetric profile of $\zeta$ at late times. This is similar to what one should do for a proper calculation of PBH formation, where the assumption of spherical symmetry should be reasonably accurate. Notice that one should eventually sum over all the radial profiles exceeding a certain threshold. Here we simply choose a certain profile, leaving a proper investigation about PBH formation to future work.  

We simply impose the conditions 
\begin{align}\label{sec4:Gaussian_profile}
\zeta(\tau_{\rm{f}},r) = \zeta_0 \exp(-k^2r^2) \;,
\end{align}
and $\partial_r \zeta(\eta,r_{\rm{i}}) = 0 = \zeta(\eta,r_{\rm{f}})$ where $r \in [r_{\rm{i}},r_{\rm{f}}]$.  As usual the condition (\ref{sec4:condition_tau_i}) at early times has been imposed. Following the same rescaling procedure as before, we have $\zeta \rightarrow \zeta_0 \zeta$, so that the condition above becomes $\zeta(\tau_{\rm{f}},r) = \exp(-k^2r^2)$.   

\begin{figure}[t!]
	\begin{subfigure}{0.48\textwidth}
		\includegraphics[width=\linewidth]{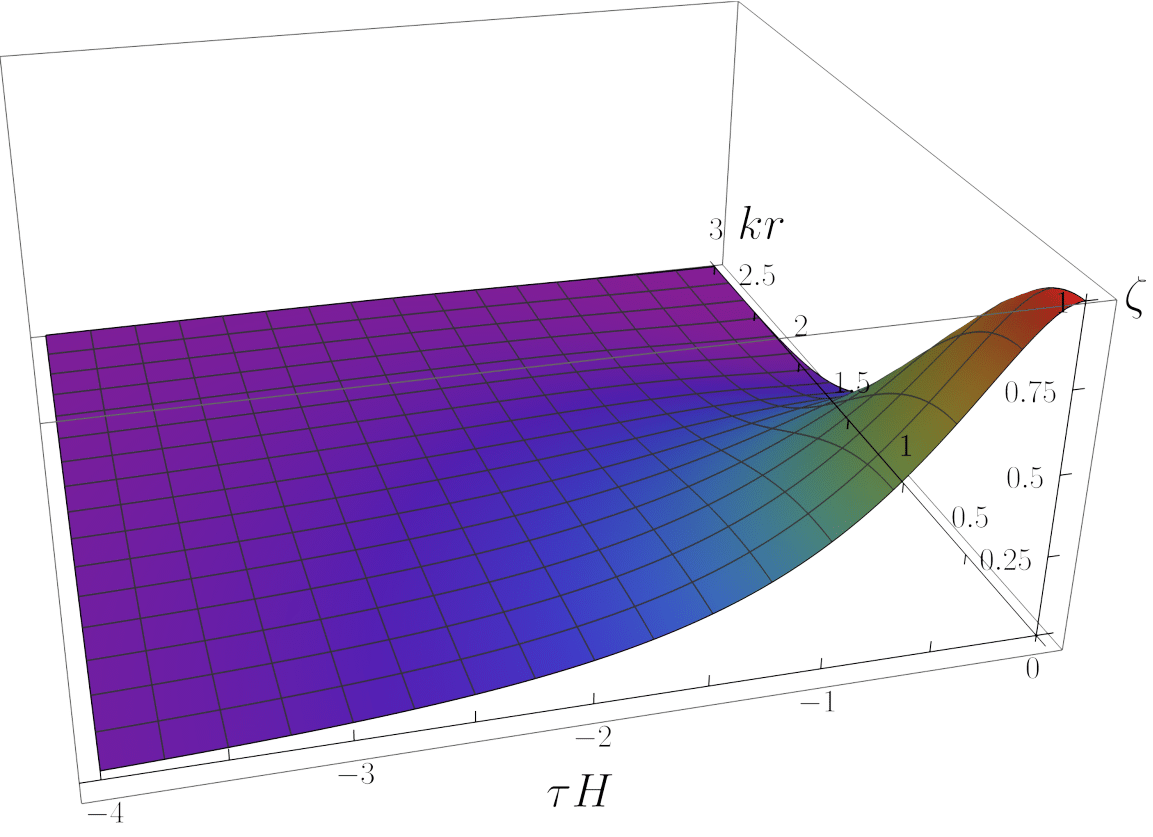}
		\caption{\label{fig:Sol_PDE_Gaussian_lambda0_zeta4}} 
	\end{subfigure}
	\hspace*{\fill} 
	\begin{subfigure}{0.48\textwidth}
		\includegraphics[width=\linewidth]{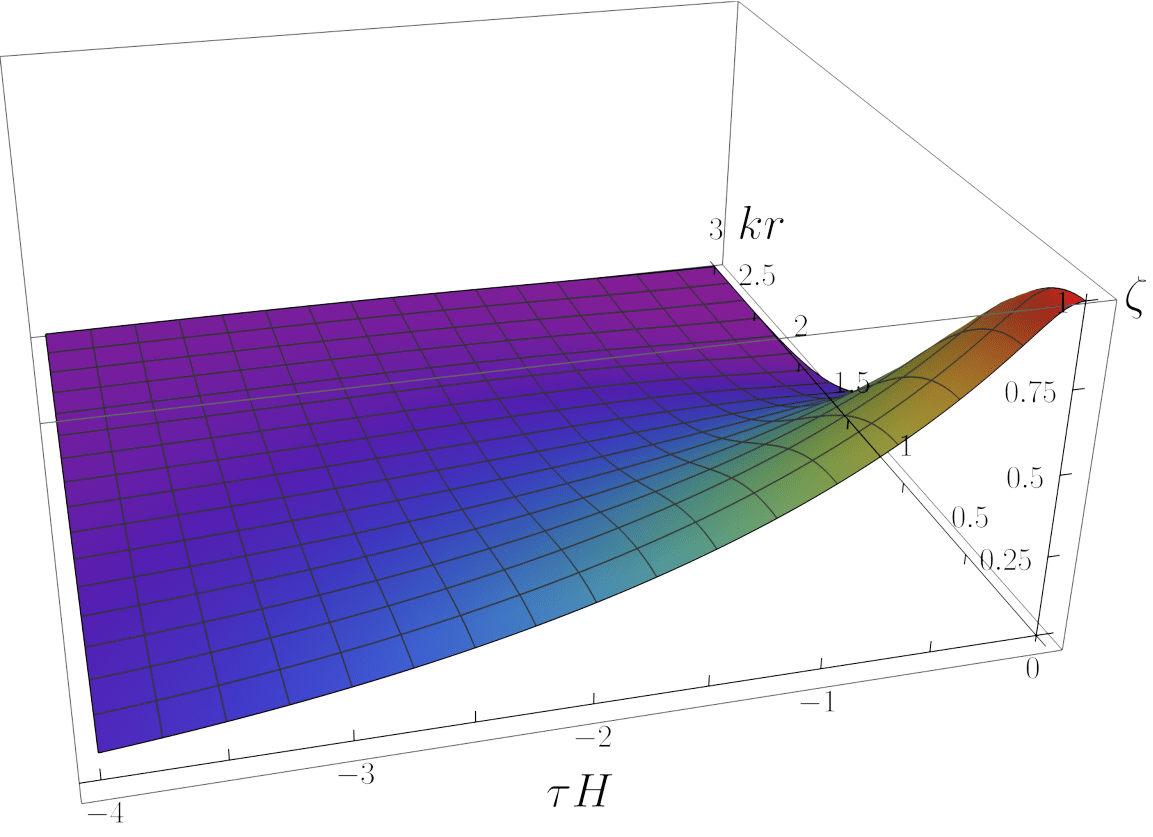}
		\caption{\label{fig:Sol_PDE_Gaussian_lambda200_zeta4}} 
	\end{subfigure}
	\hspace*{\fill} 
	\caption{~Numerical solutions with Gaussian boundary condition at late times for $\tilde{\lambda} = 0$ (left panel, figure~\ref{fig:Sol_PDE_Gaussian_lambda0_zeta4}) and $\tilde{\lambda} = 200$ (right panel, figure~\ref{fig:Sol_PDE_Gaussian_lambda200_zeta4}).} 
\end{figure}

Now let us proceed with the PDE. Given spherical symmetry eq.~(\ref{eq:PDE phi_lambda4}) takes the form,
\begin{align}\label{eq:PDE gaussian_1+1}
-\zeta'' + \frac{2}{\tau}\zeta' - \frac{1}{r^2}\frac{\partial}{\partial r}\left(r^2 \frac{\partial\zeta}{\partial r}\right) - \frac{\tilde{\lambda}}{2}\tau^2\zeta'^2\zeta'' = 0\;.
\end{align}  
The numerical solutions are shown in Figures~\ref{fig:Sol_PDE_Gaussian_lambda0_zeta4} and \ref{fig:Sol_PDE_Gaussian_lambda200_zeta4} for $\tilde{\lambda} = 0$ and $\tilde{\lambda} = 200$, respectively. (We chose $H \tau_{\rm{i}} = -80$, $H \tau_{\rm{f}} = -0.001$.  The value of $r_{\rm{f}}$ has to be sufficiently large to capture the decay of the Gaussian far from the center.) 

\begin{figure}[t!]
	\centering
	\includegraphics[width=0.6\linewidth]{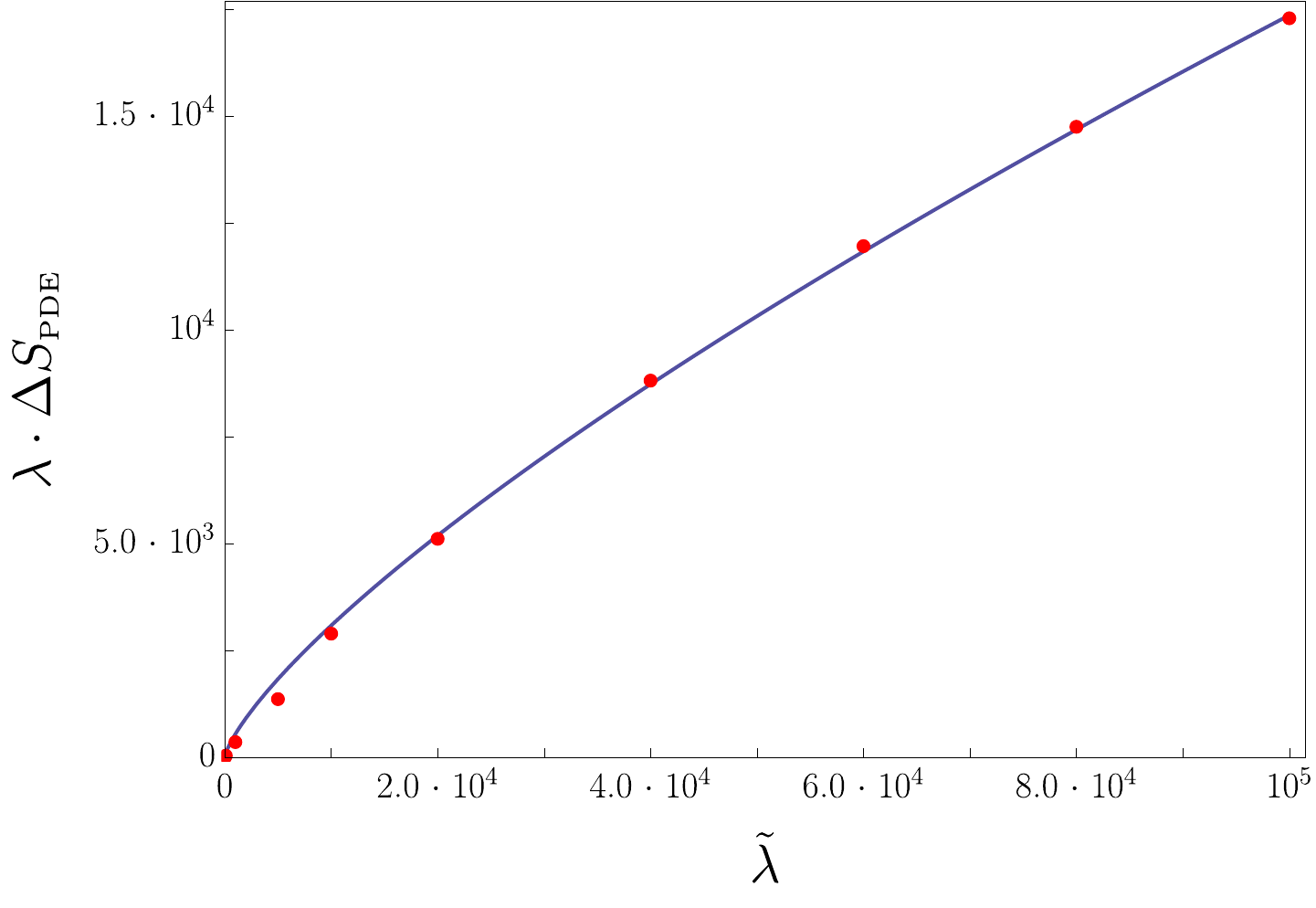}
	\caption{~The function $F(\tilde{\lambda}) = \lambda\cdot\Delta S_{\rmS PDE}$ for the Gaussian case. The blue curve shows the best fit of $\lambda \cdot \Delta S_{\rmS PDE}$ (red points), proportional to $\tilde{\lambda}^{3/4}$.}  
	\label{fig:action_PDE_Gaussian_zeta4}
\end{figure}

Like in the previous case, the finite part of the action $\Delta S_{\rmS PDE}$ reads
\begin{align}\label{sec4:action_gaussian}
\Delta S_{\rmS PDE} = -\frac{\zeta_0^2}{P_\zeta}\int_{\tau_{\rm{i}}}^{\tau_{\rm{f}}} d\tau \int_{r_{\rm{i}}}^{r_{\rm{f}}} dr~r^2\bigg\{\frac{1}{2\tau^2} \bigg[\zeta'^2 + (\partial_r \zeta)^2 - 4k^4r^2 e^{-2k^2r^2} \bigg] + \frac{\tilde\lambda}{4!}\zeta'^4\bigg\} = \frac{1}{\lambda}F(\tilde{\lambda})\;. 
\end{align}
As before, we subtracted the late-time value of $(\partial_{r}\zeta)^2$, i.e.~$4k^4r^2 e^{-2k^2r^2}$, from the full action to get rid of the divergent piece at late times. We now perform the integral in (\ref{sec4:action_gaussian}) numerically on the solutions for $\tilde{\lambda} = 0$ up to $\tilde{\lambda} = 10^5$. We then again plot in Figure~\ref{fig:action_PDE_Gaussian_zeta4} the numerical value of $F(\tilde{\lambda}) = \lambda \cdot \Delta S_{\rmS PDE}$ as a function of $\tilde{\lambda}$. As expected, the asymptotic behaviour of $F(\tilde{\lambda})$ fits very well with $\tilde{\lambda}^{3/4}$, in agreement with the ODE result (\ref{sec4:WFU_ODE}).

\section{Analytic continuation to Euclidean time}\label{sec:proof_anlyticity}

\begin{figure}[t!]
\centering
\begin{tikzpicture}[scale=0.8]
\def\gap{0.0}
\def\bigradius{3}
\def\larger{5.5}
\def\littleradius{0.7}

\draw [line width=0.7pt, black, ->,>=stealth] (-2*\bigradius, 0) -- (0.25*\bigradius,0);
\draw [line width=0.7pt, black, ->, >=stealth] (0, -0.25*\bigradius) -- (0, 2*\bigradius);

\draw  [line width=1pt, orange,
  postaction={decorate}]
  let
     \n1 = {asin(\gap/\bigradius)}
  in (-\littleradius, 0.08) arc (173.5:90:\littleradius)
  [arrow inside={end={stealth},opt={orange,scale=1.5}}{0.7}];

\draw [line width=1pt, orange,
  postaction={decorate}]
  let
     \n1 = {asin(\gap/\bigradius)}
  in (90:\larger) arc (90:177.1:\larger)
  [arrow inside={end={stealth},opt={orange,scale=1.5}}{0.55}];

\draw [line width=1pt, orange] plot [smooth, tension=0] coordinates { (-5.5, 0.3) (-0.7,0.1) } [arrow inside={end={stealth},opt={orange,scale=1.5}}{0.3}];

\draw [line width=1pt, orange] plot [smooth, tension=0] coordinates { (0,0.677) (0,5.522) } [arrow inside={end={stealth},opt={orange,scale=1.5}}{0.3}];


\node at (1.5,0){ $\text{Re}\, \eta$};
\node at (0,6.5) { $\text{Im}\,\eta$};
\node[text=orange, above left] at (-0.5,0.5) { $ \Gamma_\varepsilon$};
\node[text=orange, above left] at (-3.9, 3.8) { $\Gamma_\infty$};
\node[text=orange, right] at (0.1, 2.9) { $\Gamma_{\rm E}$};
\node[text=orange, below] at (-3.25,0) { $\Gamma_{\rm L}$};
\end{tikzpicture}
\caption{\label{fig:Complex_contour_rotation} Complex contour for the evaluation of the action. The Lorentzian action ($\rm dS_4$) is obtained integrating along $\Gamma_{\rm L}$ (notice the $i \epsilon$ prescription) while the Euclidean action ($\rm -EAdS_4$) along $-\Gamma_{\rm E}$. The large semicircle $\Gamma_{\infty}$ goes to zero for large radii because of the Bunch-Davies vacuum condition. The small circle $\Gamma_{\varepsilon}$ contains a singularity as $\eta_{\rm f}\rightarrow 0$.}
\end{figure}
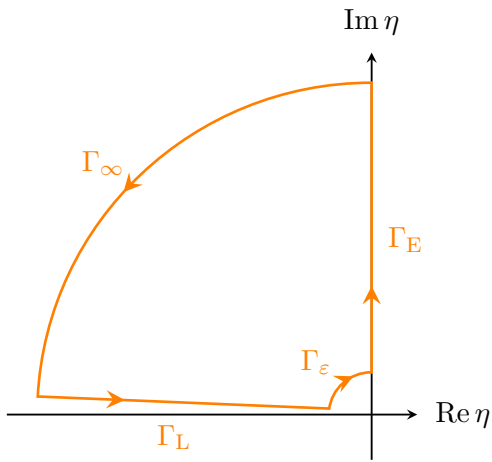

The analysis of the previous Sections relies on the analytic continuation of the action in $\eta$. This continuation corresponds to a rotation of the contour of integration as in Figure~\ref{fig:Complex_contour_rotation}. After this rotation one has a negative-definite metric
\be
ds^2 = - \frac{1}{H^2 \tau^2} (d \tau^2 + d \vect x^2) \;,
\ee
which we dub $\rm -EAdS_4$ (for a related discussion see \cite{Hertog:2011ky,Maldacena:2019cbz}). (We will only comment at the end about the possibility of further continuing this to Euclidean AdS, $\rm EAdS_4$, with an analytic continuation of the Hubble radius $H \rightarrow - i / L$, where $L$ is the $\rm EAdS_4$ radius.)

In this Section we would like to justify this analytic continuation by proving that the classical trajectory, and thus the Lagrangian, are analytic in the upper-left quadrant of the complex-$\eta$ plane. The proof holds at any order of tree-level perturbation theory, i.e.~for the diagrams we are resumming in the semiclassical expansion. As in the previous Sections we consider the geometry as unperturbed. This implies that the integral that gives the action and then the WFU can indeed be rotated from $\rm dS_4$ to its Euclidean version, $\rm -EAdS_4$, without encountering singularities. Towards the end of the Section we give a plausible non-perturbative argument for analyticity. 

As already emphasized, the Lagrangian has a pole at $\eta = 0$ so that there is a contribution of $\Gamma_\varepsilon$ in Figure \ref{fig:Complex_contour_rotation}. This pole is due to the quadratic part of $\mathcal L$ and gives a divergent contribution to the integral $\propto 1/\tau_{\rm f} = - i/\eta_{\rm f}$, see the discussion below eq.~\eqref{eq:SEfree}. This is only a phase in the WFU and it does not affect the statistical properties of $\zeta$. From now we assume that this divergent part is removed (see the discussion below eq.~\eqref{eq_zeta'^4:regularized_action_ODE}) and the integral can be extended to the origin. Provided that $\zeta$ is analytic for ${\rm Re}\; \eta < \eta_{\rm f} < 0$, ${\rm Im}\; \eta > 0$, the Lagrangian is also analytic in the same domain (we assume it is an analytic function of derivatives of $\zeta$). Hence, our goal is to show that, at any order in perturbation theory in some coupling $\lambda$, the classical solution with fixed boundary conditions at $\eta = \eta_{\rm f}$ for $\zeta$ remains analytic.

We start by writing the formal classical solution for $\zeta$ in ${\rm dS_4}$, with Bunch-Davies vacuum conditions for $\eta \rightarrow -\infty$ and Dirichlet boundary conditions at late times: $\zeta(\eta_{\rm f}, \vect k ) = \zeta_0(\vect k)$. Given a generic interaction term in the action $S_{\rm int}$, $\zeta(\eta, \vect k)$ reads
\begin{equation}\label{eq:formal_sol_zeta}
\zeta(\eta, \vect k) = K(\eta, \vect k) \zeta_0(\vect k) + \int_{-\infty(1 - i \epsilon)}^{\eta_{\rm f}}   G(\eta, \eta'; \vect k) \frac{\delta S_{\rm int}}{\delta \zeta(\eta', \vect k)}\, d \eta'\;, 
\end{equation}
where $K(\eta, \vect k)$ is the \emph{bulk-to-boundary} propagator and $G(\eta, \eta'; \vect k)$ is the \emph{bulk-to-bulk} propagator (see for instance \cite{Anninos:2014lwa}). For a massless scalar in $\rm dS_4$ they read
\begin{equation}\label{eq:bulk-to-boundary_dS}
K(\eta, \vect k)  = \frac{(1-i k \eta)}{(1-i k \eta_{\rm f})} e^{i k (\eta - \eta_{\rm f})}\;, 
\end{equation}
and
\begin{equation}
G(\eta, \eta'; \vect k) = \frac{-iH^2}{2k^3}\left[\theta(|\eta'| - |\eta|  ) \phi_+(\eta')\phi_-(\eta)  + \theta(|\eta| - |\eta'|) \phi_+(\eta)\phi_-(\eta')  - \frac{\phi_-(\eta_{\rm f})}{\phi_+(\eta_{\rm f})} \phi_+(\eta')\phi_+(\eta)\right]  \;,  \label{eq:bulk-to-bulk_dS}
\end{equation} 
where $\theta$ is the step function and $\phi_-(\eta)$, $\phi_+(\eta)$ are the wave-modes solving the free equation of motion
\begin{align}\label{eq:independent_wavemodes_dS}
\phi_-(\eta) \equiv (1 + ik\eta)e^{-ik\eta}\;, \quad \phi_+(\eta) \equiv (1 - ik\eta) e^{ik\eta} \;. 
\end{align}
Expression \eqref{eq:formal_sol_zeta} is a formal solution that can be used iteratively to obtain  corrections to $\zeta$ as a power series in the coupling $\lambda$. Indeed, by evaluating the source $S(\eta', \vect k) \equiv \delta S_{\rm int}/ \delta \zeta(\eta', \vect k)$ at order $n$ in $\lambda$, we can obtain the solution for $\zeta$ at order $n+1$ by evaluating the right-hand side of eq.~\eqref{eq:formal_sol_zeta}.
Such perturbative iteration corresponds to an expansion in tree-level Witten diagrams (with an increasing number of legs connected to the boundary). 

We can proceed by induction. We will start by assuming that the source term $S(\eta', \vect k)$ is analytic at order $n$ in the perturbative expansion in $\lambda$. Then, we will argue that the solution for $\zeta$ at order $n+1$ is also analytic.
Since the zeroth-order solution for $\zeta$, given by $\zeta^{(0)}(\eta, \vect k) = K(\eta, \vect k ) \zeta_0(\vect k)$, is manifestly analytic this will prove that $\zeta$ remains analytic at any order.

In order to show analyticity, we need to properly extend eq.~\eqref{eq:formal_sol_zeta} to complex-$\eta$ values. Assuming analyticity for the source $S(\eta', \vect k)$, the only difficulty resides in the propagator $G(\eta, \eta'; \vect k)$, which displays a discontinuity in the complex-$\eta'$ plane when $|\eta| = |\eta'|$ with ${\rm arg}\; \eta \neq {\rm arg}\; \eta'$. Note however that $G(\vect k ; \eta, \eta')$ is analytic in $\eta$ and $\eta'$ in the two regions 
$|\eta|>|\eta'|$ and $|\eta|<|\eta'|$.

\begin{figure}[t!]
\centering
\begin{tikzpicture}[scale=0.8]
\def\gap{0.0}
\def\bigradius{3}
\def\littleradius{0.5}

\draw [line width=0.7pt, black, ->,>=stealth] (-2*\bigradius, 0) -- (0.25*\bigradius,0);
\draw [line width=0.7pt, black, ->, >=stealth] (0, -0.25*\bigradius) -- (0, 2*\bigradius);

\draw[line width=1pt, myred,
  postaction={decorate}]
  let
     \n1 = {asin(\gap/\bigradius)}
  in (90:\bigradius) arc (90:180:\bigradius);

\draw [line width=1pt, orange] plot [smooth, tension=0.7] coordinates { (-5.3, 5.3) (-4.8,4.8) (-4.1,3)   (-\bigradius/ 1.41 , \bigradius/ 1.41) (-1.2,0.6) (0,0) } [arrow inside={end={stealth},opt={orange,scale=1.5}}{0.3,0.8}];

\filldraw[black] (-\bigradius/ 1.41 ,\bigradius/ 1.41) circle (2pt) ;
\node[text=black] at (-\bigradius/ 1.41 ,1.2* \bigradius/ 1.41) { $\eta$};

\node at (1.5,0){ $\text{Re}\, \eta'$};
\node at (0,6.5) { $\text{Im}\,\eta'$};
\node[text=orange] at (-0.8,1.2) { $\Cc C_2$};
\node[text=orange] at (-3.8,3.8) { $\Cc C_1$};
\end{tikzpicture}
\caption{\label{fig:Complex_eta1_contour} In orange, the complex contour for the integral in eq.~\eqref{eq:formal_sol_zeta}. The red line indicates the points where $G(\eta, \eta'; \vect k)$ is discontinuous.}
\end{figure}
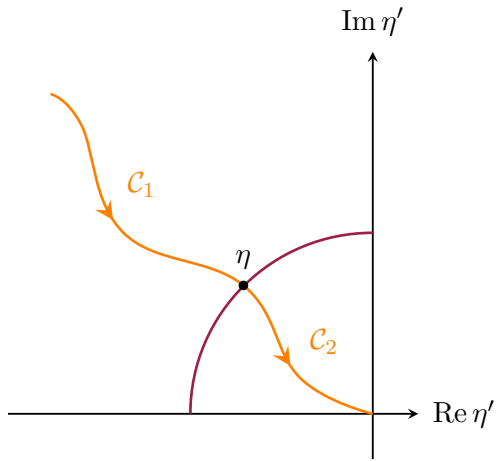 

Because of these properties, we can extend eq.~\eqref{eq:formal_sol_zeta} to complex $\eta$ by choosing a proper contour of integration in $\eta'$ for the integral on the right-hand side. As shown in Figure~\ref{fig:Complex_eta1_contour}, we pick a path $\mathcal C_1$ going from infinity (in the upper-left quadrant) to $\eta$, and then a second path $\mathcal C_2$ from $\eta$ to $0$. 
The explicit expression for $\zeta$ for complex $\eta$, in terms of the wave-modes, is then
\begin{equation}\label{eq:formal_sol_complex}
\begin{split}
\zeta(\eta, \vect k)  = K(\eta, \vect k) \zeta_0(\vect k) -  \frac{i H^2}{2 k^{3}} \bigg[ \phi_{-}(\eta)  \int_{\mathcal C_{1}} \phi_{+}(\eta') S(\eta', \vect k) d \eta' & +  \phi_{+}(\eta) \int_{\mathcal C_{2}} \phi_{-}(\eta') S(\eta', \vect k) d \eta' \\
&  - \phi_{+}(\eta) \int_{\mathcal C_1 \cup \, \mathcal C_2} \phi_{+}(\eta' )S(\eta', \vect k) d \eta '\bigg]\;.
\end{split}
\end{equation}

Because of the decaying properties of the Green functions in eqs.~\eqref{eq:bulk-to-boundary_dS} and \eqref{eq:bulk-to-bulk_dS}, the integrals are convergent and the solution is overall exponentially decaying at infinity, as expected. 
Due to Cauchy's theorem, the paths $\mathcal C_1$ and $\mathcal C_2$ can be chosen arbitrarily (as long as they do not cross the $|\eta'| = |\eta|$ and they remain in the upper-left quadrant) and therefore the integral over $\eta'$ will depend on $\eta$ but not on its complex conjugate $\eta^*$, as a consequence of the fundamental theorem of calculus. This means that the result is holomorphic ($\partial_{\eta^*} \zeta = 0$).
Finally, in order to prove analyticity, we only need to show that $\partial_\eta \zeta$ exists everywhere. This is the case, as we can see by direct inspection
\begin{equation}
\begin{split}
\partial_{\eta} \zeta(\eta, \vect k)  &=  \partial_{\eta} K(\eta, \vect k) \zeta_0(\vect k) -  \frac{i H^2}{2 k^{3}} \bigg[ \partial_{\eta}\phi_{-}(\eta)  \int_{\mathcal C_{1}} \phi_{+}(\eta') S(\eta', \vect k) d \eta'   \\ 
&+  \partial_{\eta} \phi_{+}(\eta) \int_{\mathcal C_{2}} \phi_{-}(\eta') S(\eta', \vect k) d \eta'  - \partial_{\eta} \phi_{+}(\eta) \int_{\mathcal C_1 \cup \, \mathcal C_2} \phi_{+}(\eta' )S(\eta', \vect k) d \eta '\bigg]\;.
\end{split}
\end{equation}
This shows that indeed $\zeta$ is analytic, as claimed.

Additionally, it is easy to realize that this choice of the contour is the correct one since $\zeta$ then satisfies its classical equation of motion, even for $\eta$ complex. Therefore, when $\eta$ is purely imaginary (with positive imaginary part), $\zeta$ reduces to the correct ${\rm -EAdS_4}$ classical solution. This can be checked by noticing that, in this case, one can take the two paths $\mathcal C_1$ and $\mathcal C_2$ to belong to the imaginary $\eta'$ axis. Hence, in eq.~\eqref{eq:formal_sol_complex} we can simply replace $\eta \rightarrow i z$, $\eta' \rightarrow i w$, with $z$, $w \in \mathbb{R}_{+}$. 
The bulk-to-boundary propagator and the two wave-modes of eqs.~\eqref{eq:bulk-to-boundary_dS} and \eqref{eq:independent_wavemodes_dS} map to their ${\rm -EAdS_4}$ counterparts ($K_{\rm dS}(i z, \vect k) = K_{\rm -EAdS}(z, \vect k)$, $\phi_{+}(iz) = \phi^{\rm -EAdS}_{+}(z)$ and $\phi_{-}(iz) = \phi^{\rm -EAdS}_{-}(z)$), whereas the bulk-to-bulk propagator picks up a phase ($G_{\rm dS}(iz, i w; \vect k) = i G_{\rm -EAdS}(z, w; \vect k)$), as we see from its expression eq.~\eqref{eq:bulk-to-bulk_dS}. This factor of $i$ then combines with the measure of integration in eq.~\eqref{eq:formal_sol_complex} so to obtain the correct formal solution for $\zeta$ in ${\rm -EAdS_4}$.\footnote{When going to Euclidean one should also consider the rotation of the background solution, which is time dependent. This will induce some extra phases in the coefficient of some of the $\pi$ interactions.} (Notice also that, by further rotating $H \rightarrow -i / L$, our expressions in $\rm -EAdS_4$ maps to the expressions to $\rm EAdS_4$. See \cite{Anninos:2014lwa} for a dictionary between $\rm dS_4$ and $\rm EAdS_4$. We will comment more about this point below.)

This concludes the proof of why we can analytically rotate to $\rm -EAdS$. At this point we want to give an argument about analyticity when we treat interactions non-perturbatively in the couplings. In particular, we focus on the case where the EoM can be approximated by an ODE. Formally speaking, our discussion only applies to QM but, as we saw for the model $\lambda \dot \zeta^4$, an ODE can be a good approximation for the behaviour of $\zeta$. It looks quite challenging to have a rigorous proof in the case of the PDE.
Further, we assume that a solution to the boundary-value problem for our ODE exists. Given this solution, we can think of our problem as an initial-value problem by computing the value of $\dot \zeta$ at early times and using it as an initial condition. By doing so, we are now allowed to apply standard results in the theory of ODEs. 

This ODE can be written as a first-order system of equations that we study in the full complex plane, schematically of the form $\dot x(\tau) = f(x, \tau; \lambda)$\footnote{In QM, the EoM $\ddot x = -V'(x)$ can be converted to a first order form by defining $y \equiv \dot x$. The equation then takes the form of a 2-dimensional system $\dot x = y$, $\dot y = - V'(x)$. For simplicity we schematically write the system as a single equation since our conclusions would not change.}, where $\tau$ is a complex-time variable and $\lambda$ is a generic coupling of the theory. Let us assume in the following that $f(x, \tau; \lambda)$ is an analytic function on all its three variables.\footnote{In the case of the ODE of Section~\ref{subsec_zeta'^4:ODE_approximation}, the function $f$ has a singularity around a time $\eta_0$ where $1+\tilde \lambda \eta_0^2 \zeta'^2(\eta_0) = 0$. We do not think however this is an obstacle to rotation: one can readily check that around $\eta_0$, $\zeta(\eta)$ admits a series expansions without singularities.} 
If this is the case, then it follows that around any point $\tau_0$, $x(\tau)$ can be written as an analytic series in $\tau- \tau_0$, with radius of convergence determined by $|f|$ (see e.g.~\cite{teschl2012ordinary}). This fact implies that the solution will be analytic at every point where $|f|$ is bounded.  From this observation we conclude that $x(\tau)$ is also analytic, except in the cases where this solution probes regions of the potential infinitely far away, where $|f|$ is expected to diverge. (Given that $f$ is analytic by assumption, it must be unbounded in some direction, otherwise it would be a constant.)
By gluing together these local solutions we expect that the overall solution is also analytic, with no obstacle for analytic continuation. In the case of the anharmonic oscillator the non-linear solution eq.~\eqref{eq:classicalPath_QM} is indeed analytic in the quadrant of interest, since the function $1/\sinh$ has poles only for imaginary $\tau$, i.e.~for real time $t$ and these are avoided in the Lorentzian calculation by the $i \epsilon$ prescription. Notice also that the solution decays for large radius in the quadrant of interest. 

Let us finally comment on the rotation to $\rm EAdS_4$. This continuation involves, on top of the rotation of $\eta$, also a rotation of Hubble $H \rightarrow - i / L$. It seems difficult that this rotation can be done at full non-perturbative level. Indeed many couplings of the theory depend on $H$ so that analyticity in this parameter is similar to analyticity in the couplings $\lambda$ of the theory.  A standard result in ODEs shows that $x(\tau)$ is an entire function of $\lambda$ for any fixed $\tau$, provided that the initial conditions for $x$ and $\dot x$ are $\lambda$-independent.  However we have here a boundary value problem and this may create non-analyticity. This happens for instance in the case of the anharmonic oscillator discussed above. Let us choose the origin of $\tau$ in eq.~\eqref{eq:classicalPath_QM} in such a way that the solution sits at $x_{\rm f}$ at $\tau =0$ (we remind the definition $\bar x^2 \equiv 2 \lambda x_{\rm f}^2/d^2$):
\begin{equation}
\begin{split}
x(\tau) & = - \frac{d}{\sqrt{2 \lambda}} \frac{1}{\sinh(\omega\tau-{\rm arcsinh}(1/\bar x) )}  = \frac{x_{\rm f}}{\cosh(\omega\tau)} \frac{1}{1-\sqrt{1+\bar x^2} \tanh(\omega\tau)}\;.
\end{split}
\end{equation}
(Notice that for $\bar x \ll 1$ the solution reduces to $x_{\rm f} e^{\omega\tau}$.) We see that one can rotate the solution from positive to negative $\lambda$ only if $|\lambda|$ is small enough. When $2\lambda x_{\rm f}^2/d^2 < -1$ the square root becomes imaginary and indeed it is easy to realize one cannot find a solution in this regime. The point of transition corresponds to $\bar x = - 1$, which is the point where the action eq.~\eqref{eq:QM_AHO_action_E} reaches a branch point. This shows that in general one cannot expect analyticity in $\lambda$. Actually the full series in $\lambda$ has zero radius of convergence, following Dyson's argument \cite{dyson1952divergence}, while the series we are resumming, $(\lambda x_{\rm f}^2/d^2)^n$, has a finite radius. 

The rotation to $\rm EAdS_4$ surely works at tree-level \cite{Harlow:2011ke}: this can be seen using the perturbative argument we gave in the first part of this Section upon continuation of $H$.  However it is probably not correct at the non-perturbative level as already suggested in \cite{Harlow:2011ke}.\footnote{We thank V. Gorbenko and L.~Di Pietro for discussions about this point.}


\section{\label{sec:conclusions}Conclusions and future directions}
The standard perturbation-theory approach to inflationary non-Gaussianity fails when one is interested in very unlikely events, on the tail of the probability distribution, like for instance in the case of PBHs. We showed that in this case one has to resort to semiclassical methods, approximating the wavefunction of the Universe with a non-linear saddle point, i.e.~a non-linear solution of the (Euclidean) classical equations of motion. In this paper we explained the general logic of the approach and we applied it to a specific interaction of single-field inflation, $\propto \lambda \dot\zeta^4$. One is able to make predictions for arbitrarily large values of $\zeta$ and in particular, with a combination of analytic and numerical analyses, one is able to show that the tail of the probability distribution behaves as $\exp(-\lambda^{-1/4}\zeta^{3/2})$. The non-analytic dependence on $\lambda$ makes clear that this result cannot be reproduced by perturbation theory. 

This paper represents a first step in understanding inflation beyond perturbation theory and many directions remain open. Let us list some of them.
\begin{itemize}
\item On the more phenomenological side, it will be interesting to explore the effects on PBH production. One should first of all understand the impact of the various inflaton operators on the tail of the $\zeta$ distribution. In general the tail will fall slower or faster than in the Gaussian case, but one can also envisage a scenario in which the combination of various operators produces a ``bump" in the distribution for large values of $\zeta$, boosting the PBH abundance. Besides affecting the rate, one expects that going beyond perturbation theory will also change the clustering properties of PBH and therefore their merger rate \cite{Ali-Haimoud:2018dau,Desjacques:2018wuu}. 
\item In a minimal scenario of slow-roll inflation, non-Gaussianities are slow-roll suppressed $f_{\rmS NL} \sim {\cal O(\epsilon, \eta)}$ and the non-perturbative effects we studied are only relevant for $\zeta \gtrsim {\cal O}(\epsilon^{-1}, \eta^{-1})$. This is not relevant for PBHs. However, it would still be interesting to explore the unlikely tail of the wavefunction of the Universe in this minimal scenario. It is important conceptually, since we should learn how to make predictions about the initial conditions of our Universe and it also may have some impact in the study of eternal inflation \cite{Creminelli:2008es,Dubovsky:2008rf,Dubovsky:2011uy}. It is not obvious what is the best strategy to approach the problem, since in this case one has to take into account the modification of the geometry. It looks challenging to derive the full non-linear action of $\zeta$ solving for the constraint variables, like one does in the standard perturbation theory approach, so one may have to resort to a direct solution of Einstein equation with prescribed boundary conditions. The same logic applies to tensor modes: the exploration of the tail of the distribution is clearly not interesting phenomenologically, but it is appealing theoretically since it is fixed by the non-linearities of General Relativity and it is intrinsic of de Sitter space. Exact solutions of gravitational waves in de Sitter \cite{Bicak:1999ha} may be a good starting point for this problem.
\item For the operator we studied in this paper, $\dot\zeta^4$, the Euclidean non-linear solution exists for arbitrarily large values of the $\zeta$. This does not happen for all possible operators (for instance with the interaction $-(\partial_i\zeta)^4$). It is not clear what happens after the solution stops existing. One possibility is that one finds complex saddle solutions that dominate the path integral. In general, starting from the Lorentzian path integral it is a challenging problem to understand which saddles contribute. The scenario at hand, in which one can neglect perturbations of the geometry, may be a good place to understand how to make a more precise sense of the wavefunction of the Universe (for a recent discussion see \cite{Feldbrugge:2017kzv}).
\item Scattering amplitudes and correlation functions in the limit of large number of external legs have been studied using semiclassical methods. (See \cite{Badel:2019khk} for recent studies in flat space and \cite{Panagopoulos:2019ail,Panagopoulos:2020sxp} in the case of inflation.) Naively, one expects that correlation functions with many legs $\langle\zeta^N\rangle$ are related to the behaviour of the probability distribution on the tail. It would be nice to make this connection explicit and relate our approach with the existing literature on the subject.
\item In this paper we studied solutions of the scalar equations of motion in dS with prescribed boundary conditions, without resorting to perturbation theory. The same kind of approach should be possible also in AdS in the context of AdS/CFT. This would correspond to study the dual CFT in the presence of a {\em finite} external source, without treating the source perturbatively.  
\end{itemize}
Work is needed in all directions.

\section*{Acknowledgements}
It is a pleasure to thank D.~Anninos, F.~Benini,  G.~Cabass, L.~Di Pietro, V.~Gorbenko, O.~Janssen, M.~Mirbabayi, J.~E.~Miro, E.~Pajer, G.~Pimentel, R.~Rattazzi, L.~Senatore, E.~Silverstein, M.~Simonovi\'c and G.~Villadoro for useful discussions. G.~T.~acknowledges the support of SISSA in the period when this project started.


\appendix 


\section{Comparison with the WKB approximation}\label{app:WBK_approximation}

The result we obtained for $\Psi_0(x)$ in eq.~\eqref{eq:QM_AHO_ground_state_final} matches with the standard WKB approximation in both the large distance ($\bar x \gg 1$) and small coupling ($\lambda \ll 1$) limit, as we are going to show. From the calculation of the WKB wavefunction we can also appreciate how the prefactor of $\Psi_0$ induces a subleading $x$-dependence with respect to the exponential factor. 

In the WKB approximation, the wavefunction is given by 
\begin{equation}
\Psi_{{\rm \scriptscriptstyle WKB}}(x) = \frac{\EuScript{N}}{\sqrt{p(x)}} \exp\left( \pm i \int_{x_0}^x \frac{p(x')}{\hbar} \, d x'\right)\;,
\end{equation} 
where $\EuScript N$ is again a normalization, $x_0$ is an arbitrary point, $p(x) = \sqrt{2 m (E - V(x))}$ is the momentum of the classical trajectory with energy $E$ and the sign at the exponent is fixed by requiring appropriate boundary conditions at infinity. 
For the WKB approximation to be valid one requires that $ \hbar |p'(x)|\ll p^2(x) $.
Note that in the case of the anharmonic oscillator with potential \eqref{eq:QM_AHO_potential}, this condition is satisfied even for the ground state in the classically-forbidden region $V(x) \gg E$. For small $\lambda$, this point is parametrically smaller than the point where the quartic term starts to dominate the potential ($\bar x \sim 1$). This means that the WKB should match eq.~\eqref{eq:QM_AHO_ground_state_final} even for small $\bar x$.

Let us start from the prefactor. For fixed $\bar x$ and small $\lambda$ we obtain
\begin{equation}\label{eq:QM_AHO_WKB_prefactor}
\frac{\EuScript N}{\sqrt{p(x)}} \simeq \frac{\EuScript N'}{\sqrt{\bar x} \left(1+ \bar x^2\right)^{1/4}}\;.
\end{equation}
Notice that this expression matches with the prefactor obtained in the semiclassical expansion for $\bar x \gg 1$, but for general values differs. Therefore, in order to have a match with the full wavefunction, we expect some correction to come from the exponent.

The exponent can be rewritten as the following integral
\begin{align}\label{eq:QM_AHO_WKB_exponent}
\int^x_{x_0} \frac{p(x')}{\hbar}\, d x' & = i \frac{\sqrt{2 m }}{\hbar}\int^x_{x_0} \sqrt{V(x') - E}\, d x' \nonumber \\
& =  \frac{i }{2 \lambda } \int^{\bar x}_{\bar x_0} \sqrt{y^2 (1+y^2) - \epsilon}\, d y\,, 
\end{align}
where in the second line we defined $y^2 \equiv 2 \lambda x^2 / d^2 $ and $\epsilon \equiv 4\lambda E/(\hbar \omega)$. To perform the integration above, one can either expand the integrand for small $\alpha \equiv \epsilon/(y^2 (1+y^2))$ first and perform the integral after, or evaluate the integral first and expand it for small $\alpha$ after. The latter method is more complicated than the former since the integral will involve Elliptic functions of the first and second kind, so one needs proper care in taking the small $\alpha$ limit. In any case the two ways of performing that calculation must coincide. Let us now proceed with the first method. Expanding the integrand for small $\alpha$ and performing the integral afterwards yields
\begin{align}
\int^x_{x_0} \frac{p(x')}{\hbar}\, dx'  = \frac{i}{6\lambda}\bigg\{ \left[\left(1+ \bar x^2\right)^{3/2} - 1  \right] + \frac{6\lambda E}{\hbar \omega}  \log\left( \frac{1+ \sqrt{1+\bar x^2}}{\bar x} \right) + {\rm const} \bigg\} + \mathcal{O}(\alpha^2)\;,
\end{align}
where the constant terms only depend on $\bar{x}_0$ and can be absorbed into a redefinition of the normalization. Notice that the term $-i/(6\lambda)$ is needed in order to match our result with the wavefunction of the harmonic oscillator when $\lambda$ is taken to be zero. We notice that at order $\sim 1/\lambda$ the exponent matches with the one found from the Euclidean action eq.~\eqref{eq:QM_AHO_action_E}. Moreover, we have corrections at order $\sim  \lambda^0$. The logarithmic term can be important for small $\bar x$ and in fact affects the prefactor of eq.~\eqref{eq:QM_AHO_WKB_prefactor}.

By putting both the prefactor \eqref{eq:QM_AHO_WKB_prefactor} and the exponent \eqref{eq:QM_AHO_WKB_exponent} together (and choosing the appropriate sign) we obtain
\begin{equation}\label{eq:psi_WKB_x}
\Psi_{{\rm \scriptscriptstyle WKB}}(x) \simeq \frac{\EuScript N}{\sqrt{\bar x} (1+ \bar x^2)^{1/4}} \left[ \frac{1+ \sqrt{1+\bar x^2}}{\bar x} \right]^{-E / (\hbar \omega)} \exp \left\{-\frac{1}{6\lambda}\left[ \left(1 + \bar{x}^2 \right)^{3/2} - 1\right] \right\}\;.
\end{equation}
Clearly, by choosing the ground-state energy at leading order $E = \hbar \omega / 2$ we recover the result from semiclassics eq.~\eqref{eq:QM_AHO_ground_state_final}, as expected.


\section{Perturbative check of the PDE}\label{sec:perturb}
As a check, one expects that the numerical result found in Section~\ref{subsec_zeta'^4:PDE} is reduced to the one obtained using perturbation theory when the coupling $\tilde{\lambda}$ is small. More precisely, the check we are going to do will be a comparison between the 4-point coefficient of the WFU derived from perturbation theory and its numerical value evaluated on the classical solutions with the sinusoidal profile. It is more complicated to do a similar check for the Gaussian profile, since one would have to integrate over Fourier space in the perturbative calculation.

Let us start with the perturbative calculation. For simplicity, we only focus on the first order correction in $\tilde{\lambda}$ which corresponds to the first graph in Figure~\ref{fig:TreeWitten_resum}. To compute such a diagram one just needs to know the bulk-to-boundary propagator (\ref{eq:bulk-to-boundary_dS}). Then the 4-point coefficient $\psi^{(4)}$ of the WFU is given by 
\begin{align}
\psi^{(4)}(k_1,k_2,k_3,k_4) = \frac{i\lambda}{P_\zeta^2} \int_{-\infty(1-i\epsilon)}^{\eta_{\rm f}} d\eta~ K'(\eta,\vect{k}_1)K'(\eta,\vect{k}_2)K'(\eta,\vect{k}_3)K'(\eta,\vect{k}_4) \;,
\end{align}
where as usual the $i\epsilon$ prescription has been imposed for the integral to converge. The integral above can be performed analytically so we get
\begin{align}
\psi^{(4)}(k_1,k_2,k_3,k_4) = \frac{24\lambda k_1^2k_2^2k_3^2k_4^2}{P_\zeta^2(k_1 + k_2 + k_3 + k_4)^5} \;.
\end{align}
Note that there is no divergence one has to worry about at late times.  We now want to compare this perturbative result and the numerical one done in Section~\ref{subsec_zeta'^4:PDE}. 

Before that, it is instructive to put the coefficient $\psi^{(4)}$ back in the on-shell action: 
\begin{align}\label{AppB:interacting_action}
i S_{\rm{int}} = \frac{1}{4!} \int \left(\prod_{i = 1}^{4}\frac{d^3k_i}{(2\pi)^3}\right) (2\pi)^3\delta(\sum_{i = 1}^4 \vect{k}_i)~\psi^{(4)}(k_1,k_2,k_3,k_4)~ \zeta_0(\vect{k}_1)\zeta_0(\vect{k}_2)\zeta_0(\vect{k}_3)\zeta_0(\vect{k}_4)  \;.
\end{align}  
This on-shell action, as we have said before, does not capture the loop diagrams shown in Figures~\ref{fig:LoopWitten_resum} and \ref{fig:2LoopWitten_resum}. Apparently, the formula (\ref{AppB:interacting_action}) depends on the late-times boundary condition $\zeta_0(\vect{k})$. One can generally apply this formula to a generic boundary condition at late times, but here we are going to choose a single Fourier mode which is exactly what we considered in Section~\ref{subsec_zeta'^4:PDE}.  

Let us now focus on the single Fourier mode namely, $\zeta(\eta_{\text{f}},x) = \zeta_0 \sin(kx)$. Trivially, the mode $\sin(kx)$ will be converted into the Dirac delta function in $k$-space,
\begin{align}
\zeta_0(\vect{k}) = -\zeta_0 \frac{i}{2}(2\pi)^3 \delta(k_z)\delta(k_y)\bigg[\delta(k_x - k) - \delta(k_x + k)\bigg] \;.
\end{align}
This form of $\zeta_0(\vect{k})$ greatly simplifies the interacting action (\ref{AppB:interacting_action}) into
\begin{align}
iS'_{\rm{int}} = \frac{3 \lambda k^3\zeta_0^4}{8192 P_\zeta^2} = \frac{\zeta_0^2}{P_\zeta} \frac{3 \tilde{\lambda}k^3}{8192} \;, \label{eq:pert_final}
\end{align}
where $iS'$ denotes the action divided by the factor $(2\pi)^3\delta(k_x - k)\delta(k_y)\delta(k_z)$, and we have written in terms of $\tilde{\lambda}$ for the second equality. Again, this is the first order correction in $\tilde{\lambda}$ obtained using perturbation theory. 

Let us turn to the numerical calculation. Let $\Delta S_{\rmS PDE}^{\tilde{\lambda}}$ be the corrections of order $\tilde{\lambda}$ or higher. Then, one way to extract $\Delta S_{\rmS PDE}^{\tilde{\lambda}}$ from (\ref{sec4:action_PDE_sine}) is to subtract the finite part of the free on-shell action, denoted by $\Delta S_{\rmS PDE}^0$:
\begin{align}\label{AppB:num_action}
\Delta S_{\rmS PDE}^{\tilde{\lambda}} = -(\Delta S_{\rmS PDE} - \Delta S_{\rmS PDE}^0) \;. 
\end{align}
The minus sign in front is to make $\Delta S_{\rmS PDE}^{\tilde{\lambda}}$ positive definite. Note that in general this $\Delta S_{\rmS PDE}^{\tilde{\lambda}}$ contains all orders in $\tilde{\lambda}$, but we will show below that for small $\tilde{\lambda}$ it is dominated by the first order corrections (it fits almost perfectly with (\ref{eq:pert_final})). We then numerically evaluate the $\Delta S_{\rmS PDE}^{\tilde{\lambda}}$, divided by the spatial volume $2\pi$, on the classical solution for small $\tilde{\lambda}$ ($\lambda,\zeta_0$ $\ll$ 1). Omitting the common factor $\zeta_0^2/P_\zeta$, we find that $\Delta S_{\rmS PDE}^{\tilde{\lambda}}/2\pi$ matches with the analytic calculation (\ref{eq:pert_final}) for small values of $\tilde{\lambda}$, setting $k = 1$. 

\begin{figure}[t!]
	\centering
	\includegraphics[width=0.6\linewidth]{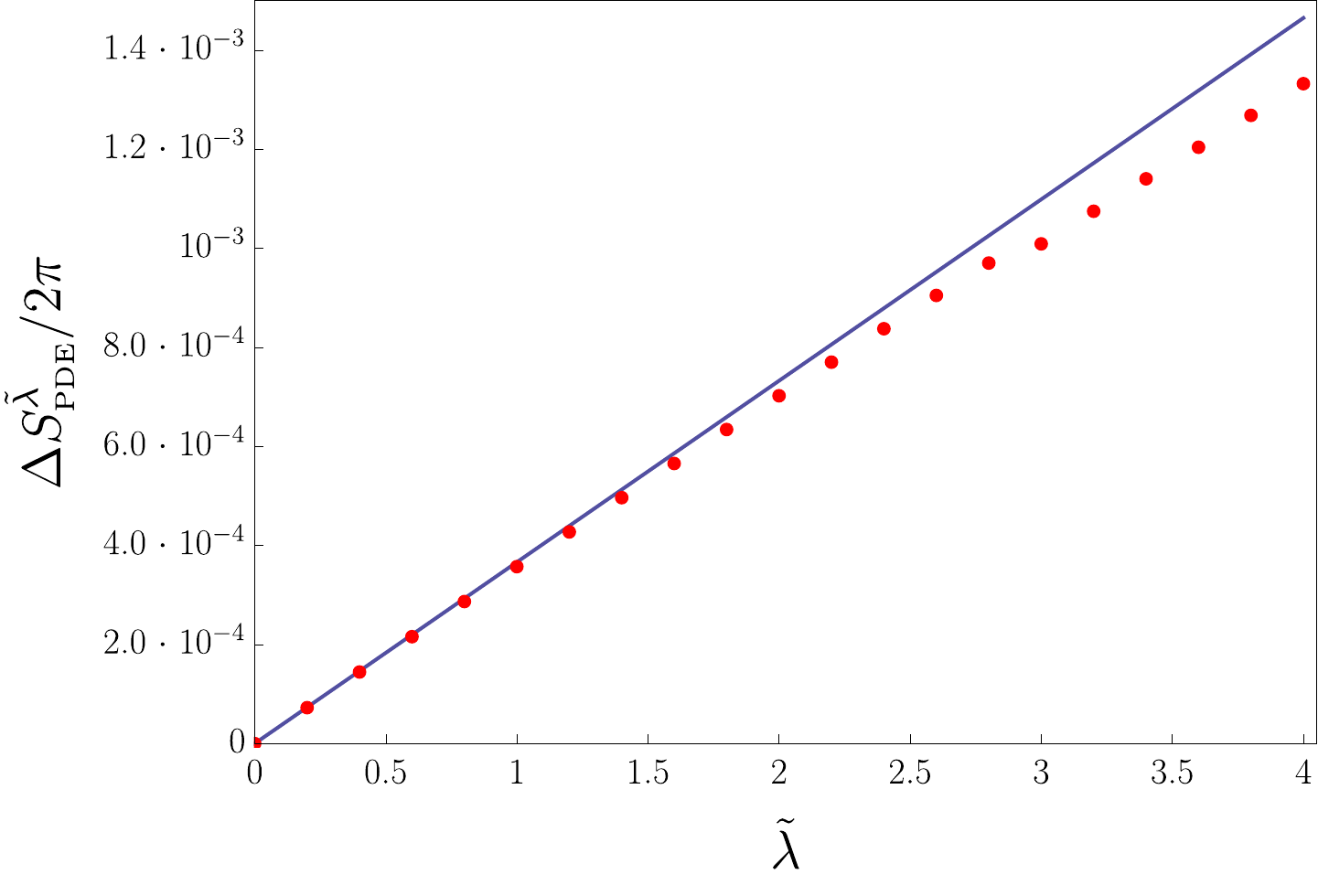}
	\caption{~The blue curve shows the perturbative result (\ref{eq:pert_final}) as a function of $\tilde{\lambda}$. The red points indicate the numerical values of (\ref{AppB:num_action}). As expected, for small $\tilde{\lambda}$ the two approaches coincide, whereas the departure happens around $\tilde{\lambda} \sim \mathcal{O}(1)$.}  
	\label{fig:action_PDE_Sine_zeta4_perturbative}
\end{figure}

Finally, it is also worth checking that for small $\tilde{\lambda}$ (\ref{AppB:num_action}) is dominated by the first order corrections in $\tilde{\lambda}$, as expected in perturbation theory. This result is confirmed in Figure~\ref{fig:action_PDE_Sine_zeta4_perturbative}, where the blue line represents the perturbative result (\ref{eq:pert_final}) and the red points denote the numerical value of $\Delta S_{\rmS PDE}^{\tilde{\lambda}}/2\pi$ for $\tilde{\lambda} \in \{0.2,0.4,\ldots,4\}$. Notice that as $\tilde{\lambda}$ increases, one expects that (\ref{AppB:num_action}) no longer coincides with (\ref{eq:pert_final}) and, indeed, from Figure~\ref{fig:TreeWitten_resum} this departure happens when $\tilde{\lambda}$ is of order unity.

\footnotesize
\bibliographystyle{utphys}
\bibliography{bib_v3}

\providecommand{\href}[2]{#2}\begingroup\raggedright\begin{thebibliography}{10}

\bibitem{Akrami:2019izv}
{\bf Planck} Collaboration, Y.~Akrami {\em et.~al.}, ``{Planck 2018 results.
  IX. Constraints on primordial non-Gaussianity},'' {\em Astron. Astrophys.}
  {\bf 641} (2020) A9, \href{https://arxiv.org/abs/1905.05697}{{\tt
  1905.05697}}.

\bibitem{Maldacena:2002vr}
J.~M. Maldacena, ``{Non-Gaussian features of primordial fluctuations in single
  field inflationary models},'' {\em JHEP} {\bf 0305} (2003) 013,
  \href{https://arxiv.org/abs/astro-ph/0210603}{{\tt astro-ph/0210603}}.

\bibitem{Musco:2020jjb}
I.~Musco, V.~De~Luca, G.~Franciolini, and A.~Riotto, ``{The Threshold for
  Primordial Black Hole Formation: a Simple Analytic Prescription},''
  \href{https://arxiv.org/abs/2011.03014}{{\tt 2011.03014}}.

\bibitem{Franciolini:2018vbk}
G.~Franciolini, A.~Kehagias, S.~Matarrese, and A.~Riotto, ``{Primordial Black
  Holes from Inflation and non-Gaussianity},'' {\em JCAP} {\bf 03} (2018) 016,
  \href{https://arxiv.org/abs/1801.09415}{{\tt 1801.09415}}.

\bibitem{Atal:2018neu}
V.~Atal and C.~Germani, ``{The role of non-gaussianities in Primordial Black
  Hole formation},'' {\em Phys. Dark Univ.} {\bf 24} (2019) 100275,
  \href{https://arxiv.org/abs/1811.07857}{{\tt 1811.07857}}.

\bibitem{Starobinsky:1986fx}
A.~A. Starobinsky, ``{STOCHASTIC DE SITTER (INFLATIONARY) STAGE IN THE EARLY
  UNIVERSE},'' {\em Lect. Notes Phys.} {\bf 246} (1986) 107--126.

\bibitem{Gorbenko:2019rza}
V.~Gorbenko and L.~Senatore, ``{$\lambda \phi^4$ in dS},''
  \href{https://arxiv.org/abs/1911.00022}{{\tt 1911.00022}}.

\bibitem{Pattison_2017}
C.~Pattison, V.~Vennin, H.~Assadullahi, and D.~Wands, ``Quantum diffusion
  during inflation and primordial black holes,'' {\em Journal of Cosmology and
  Astroparticle Physics} {\bf 2017} (oct, 2017) 046--046.

\bibitem{Ezquiaga:2019ftu}
J.~M. Ezquiaga, J.~Garc\'\i{}a-Bellido, and V.~Vennin, ``{The exponential tail
  of inflationary fluctuations: consequences for primordial black holes},''
  {\em JCAP} {\bf 03} (2020) 029, \href{https://arxiv.org/abs/1912.05399}{{\tt
  1912.05399}}.

\bibitem{Chen:2018uul}
X.~Chen, G.~A. Palma, W.~Riquelme, B.~Scheihing~Hitschfeld, and S.~Sypsas,
  ``{Landscape tomography through primordial non-Gaussianity},'' {\em Phys.
  Rev. D} {\bf 98} (2018), no.~8 083528,
  \href{https://arxiv.org/abs/1804.07315}{{\tt 1804.07315}}.

\bibitem{Chen:2018brw}
X.~Chen, G.~A. Palma, B.~Scheihing~Hitschfeld, and S.~Sypsas, ``{Reconstructing
  the Inflationary Landscape with Cosmological Data},'' {\em Phys. Rev. Lett.}
  {\bf 121} (2018), no.~16 161302, \href{https://arxiv.org/abs/1806.05202}{{\tt
  1806.05202}}.

\bibitem{Panagopoulos:2019ail}
G.~Panagopoulos and E.~Silverstein, ``{Primordial Black Holes from non-Gaussian
  tails},'' \href{https://arxiv.org/abs/1906.02827}{{\tt 1906.02827}}.

\bibitem{Panagopoulos:2020sxp}
G.~Panagopoulos and E.~Silverstein, ``{Multipoint correlators in multifield
  cosmology},'' \href{https://arxiv.org/abs/2003.05883}{{\tt 2003.05883}}.

\bibitem{rattazzi2009path}
R.~Rattazzi, ``The path integral approach to quantum mechanics lecture notes
  for quantum mechanics iv.''

\bibitem{Escobar-Ruiz:2017uhx}
M.~Escobar-Ruiz, E.~Shuryak, and A.~Turbiner, ``{Fluctuations in quantum
  mechanics and field theories from a new version of semiclassical theory.
  II},'' {\em Phys. Rev. D} {\bf 96} (2017), no.~4 045005,
  \href{https://arxiv.org/abs/1705.06159}{{\tt 1705.06159}}.

\bibitem{Arkani-Hamed:2015bza}
N.~Arkani-Hamed and J.~Maldacena, ``{Cosmological Collider Physics},''
  \href{https://arxiv.org/abs/1503.08043}{{\tt 1503.08043}}.

\bibitem{Goodhew:2020hob}
H.~Goodhew, S.~Jazayeri, and E.~Pajer, ``{The Cosmological Optical Theorem},''
  \href{https://arxiv.org/abs/2009.02898}{{\tt 2009.02898}}.

\bibitem{Senatore:2010jy}
L.~Senatore and M.~Zaldarriaga, ``{A Naturally Large Four-Point Function in
  Single Field Inflation},'' {\em JCAP} {\bf 01} (2011) 003,
  \href{https://arxiv.org/abs/1004.1201}{{\tt 1004.1201}}.

\bibitem{Cheung:2007st}
C.~Cheung, P.~Creminelli, A.~L. Fitzpatrick, J.~Kaplan, and L.~Senatore, ``{The
  Effective Field Theory of Inflation},'' {\em JHEP} {\bf 0803} (2008) 014,
  \href{https://arxiv.org/abs/0709.0293}{{\tt 0709.0293}}.

\bibitem{Creminelli:2019kjy}
P.~Creminelli, G.~Tambalo, F.~Vernizzi, and V.~Yingcharoenrat, ``{Dark-Energy
  Instabilities induced by Gravitational Waves},'' {\em JCAP} {\bf 05} (2020)
  002, \href{https://arxiv.org/abs/1910.14035}{{\tt 1910.14035}}.

\bibitem{Alishahiha:2004eh}
M.~Alishahiha, E.~Silverstein, and D.~Tong, ``{DBI in the sky},'' {\em Phys.
  Rev. D} {\bf 70} (2004) 123505,
  \href{https://arxiv.org/abs/hep-th/0404084}{{\tt hep-th/0404084}}.

\bibitem{Nicolis:2008in}
A.~Nicolis, R.~Rattazzi, and E.~Trincherini, ``{The Galileon as a local
  modification of gravity},'' {\em Phys. Rev.} {\bf D79} (2009) 064036,
  \href{https://arxiv.org/abs/0811.2197}{{\tt 0811.2197}}.

\bibitem{Anninos:2014lwa}
D.~Anninos, T.~Anous, D.~Z. Freedman, and G.~Konstantinidis, ``{Late-time
  Structure of the Bunch-Davies De Sitter Wavefunction},'' {\em JCAP} {\bf 11}
  (2015) 048, \href{https://arxiv.org/abs/1406.5490}{{\tt 1406.5490}}.

\bibitem{Babich:2004gb}
D.~Babich, P.~Creminelli, and M.~Zaldarriaga, ``{The Shape of
  non-Gaussianities},'' {\em JCAP} {\bf 08} (2004) 009,
  \href{https://arxiv.org/abs/astro-ph/0405356}{{\tt astro-ph/0405356}}.

\bibitem{Hertog:2011ky}
T.~Hertog and J.~Hartle, ``{Holographic No-Boundary Measure},'' {\em JHEP} {\bf
  05} (2012) 095, \href{https://arxiv.org/abs/1111.6090}{{\tt 1111.6090}}.

\bibitem{Maldacena:2019cbz}
J.~Maldacena, G.~J. Turiaci, and Z.~Yang, ``{Two dimensional Nearly de Sitter
  gravity},'' {\em JHEP} {\bf 01} (2021) 139,
  \href{https://arxiv.org/abs/1904.01911}{{\tt 1904.01911}}.

\bibitem{teschl2012ordinary}
G.~Teschl, {\em Ordinary differential equations and dynamical systems},
  vol.~140.
\newblock American Mathematical Soc., 2012.

\bibitem{dyson1952divergence}
F.~J. Dyson, ``Divergence of perturbation theory in quantum electrodynamics,''
  {\em Physical Review} {\bf 85} (1952), no.~4 631.

\bibitem{Harlow:2011ke}
D.~Harlow and D.~Stanford, ``{Operator Dictionaries and Wave Functions in
  AdS/CFT and dS/CFT},'' \href{https://arxiv.org/abs/1104.2621}{{\tt
  1104.2621}}.

\bibitem{Ali-Haimoud:2018dau}
Y.~Ali-Ha\"\i{}moud, ``{Correlation Function of High-Threshold Regions and
  Application to the Initial Small-Scale Clustering of Primordial Black
  Holes},'' {\em Phys. Rev. Lett.} {\bf 121} (2018), no.~8 081304,
  \href{https://arxiv.org/abs/1805.05912}{{\tt 1805.05912}}.

\bibitem{Desjacques:2018wuu}
V.~Desjacques and A.~Riotto, ``{Spatial clustering of primordial black
  holes},'' {\em Phys. Rev. D} {\bf 98} (2018), no.~12 123533,
  \href{https://arxiv.org/abs/1806.10414}{{\tt 1806.10414}}.

\bibitem{Creminelli:2008es}
P.~Creminelli, S.~Dubovsky, A.~Nicolis, L.~Senatore, and M.~Zaldarriaga, ``{The
  Phase Transition to Slow-roll Eternal Inflation},'' {\em JHEP} {\bf 09}
  (2008) 036, \href{https://arxiv.org/abs/0802.1067}{{\tt 0802.1067}}.

\bibitem{Dubovsky:2008rf}
S.~Dubovsky, L.~Senatore, and G.~Villadoro, ``{The Volume of the Universe after
  Inflation and de Sitter Entropy},'' {\em JHEP} {\bf 04} (2009) 118,
  \href{https://arxiv.org/abs/0812.2246}{{\tt 0812.2246}}.

\bibitem{Dubovsky:2011uy}
S.~Dubovsky, L.~Senatore, and G.~Villadoro, ``{Universality of the Volume Bound
  in Slow-Roll Eternal Inflation},'' {\em JHEP} {\bf 05} (2012) 035,
  \href{https://arxiv.org/abs/1111.1725}{{\tt 1111.1725}}.

\bibitem{Bicak:1999ha}
J.~Bicak and J.~Podolsky, ``{Gravitational waves in vacuum space-times with
  cosmological constant. 1. Classification and geometrical properties of
  nontwisting type N solutions},'' {\em J. Math. Phys.} {\bf 40} (1999)
  4495--4505, \href{https://arxiv.org/abs/gr-qc/9907048}{{\tt gr-qc/9907048}}.

\bibitem{Feldbrugge:2017kzv}
J.~Feldbrugge, J.-L. Lehners, and N.~Turok, ``{Lorentzian Quantum Cosmology},''
  {\em Phys. Rev. D} {\bf 95} (2017), no.~10 103508,
  \href{https://arxiv.org/abs/1703.02076}{{\tt 1703.02076}}.

\bibitem{Badel:2019khk}
G.~Badel, G.~Cuomo, A.~Monin, and R.~Rattazzi, ``{Feynman diagrams and the
  large charge expansion in $3-\varepsilon$ dimensions},'' {\em Phys. Lett. B}
  {\bf 802} (2020) 135202, \href{https://arxiv.org/abs/1911.08505}{{\tt
  1911.08505}}.

\end{thebibliography}\endgroup

\end{document}